%% file: hyper-normalisation.tex
\documentclass{CSML}
\pdfoutput=1

\usepackage{lastpage}
\newcommand{\dOi}{13(3:17)2017}
\lmcsheading{}{1--\pageref{LastPage}}{}{}%
{Aug.~22,~2016}{Aug.~29, 2017}{}

\newif\ifignore 
\ignorefalse
\newcommand{\auxproof}[1]{
\ifignore\mbox{}\newline
\textbf{PROOF:} \dotfill\newline
{\it #1}\mbox{}\newline
\textbf{ENDPROOF}\dotfill
\fi}

\usepackage{etex} 
\usepackage{amsmath}
\usepackage{amssymb}
\usepackage{amstext}
\usepackage{mathtools}
\usepackage{mathrsfs}
\usepackage{stmaryrd}
\usepackage{nicefrac}
\usepackage{xspace}
\usepackage{cmll}
\usepackage{hyperref}
\hypersetup{hidelinks}
\usepackage{url}
\usepackage{fancybox}
\usepackage{esvect}
\usepackage{nth}

\usepackage{xypic}
\usepackage[all]{xy}
\xyoption{all}
\xyoption{2cell}
\UseTwocells
\xyoption{tips}
\SelectTips{cm}{}  
\usepackage{tikz}
\usetikzlibrary{circuits.ee.IEC}

\input prooftree

\newtheorem{theorem}{Theorem}
\newtheorem{lemma}[theorem]{Lemma}
\newtheorem{proposition}[theorem]{Proposition}

\newtheorem{definition}[theorem]{Definition}
\newtheorem{example}[theorem]{Example}

\newtheorem{remark}[theorem]{Remark}

\newenvironment{myproof}[1][Proof]%
   { \begin{trivlist}%
     \item[\hskip \labelsep {\bfseries #1}]%
   }%
   { \end{trivlist}%
   }

\newcommand{\QEDbox}{\square}
\newcommand{\QED}{\hspace*{\fill}$\QEDbox$}

\newcommand{\klafter}{\mathrel{\bullet}}
\newcommand{\idmap}[1][]{\ensuremath{\mathrm{id}_{#1}}}
\newcommand{\after}{\mathrel{\circ}}

\newcommand{\st}{\ensuremath{\mathrm{st}}}

\newcommand{\gr}{\ensuremath{\mathrm{gr}}}
\newcommand{\tw}{\ensuremath{\mathrm{tw}}}

\newcommand{\spr}{\ensuremath{\mathrm{spr}}}
\newcommand{\Nrm}{\ensuremath{\mathcal{N}}}

\newcommand{\instr}{\ensuremath{\textrm{instr}}}

\newcommand{\nrm}{\ensuremath{\textrm{nrm}}}
\newcommand{\orthogonal}{\mathrel{\bot}}
\newcommand{\set}[2]{\{#1\;|\;#2\}}
\newcommand{\setin}[3]{\{#1\in#2\;|\;#3\}}
\newcommand{\supp}{\mathrm{supp}}

\newcommand{\tuple}[1]{\langle#1\rangle}
\newcommand{\bang}{\mathord{!}}
\newcommand{\ket}[1]{\ensuremath{|{\kern.1em}#1{\kern.1em}\rangle}}
\newcommand{\bigket}[1]{\ensuremath{\big|{\kern.1em}#1{\kern.1em}\big\rangle}}
\newcommand{\Bigket}[1]{\ensuremath{\Big|{\kern.1em}#1{\kern-.1em}\Big\rangle}}

\newcommand{\one}{\ensuremath{\mathbf{1}}}
\newcommand{\zero}{\ensuremath{\mathbf{0}}}

\newcommand{\defn}[1]{\smash{\stackrel{\text{def}}{#1}}}
\newcommand{\tes}[1]{2_{\scriptscriptstyle #1}} 
\newcommand{\no}[1]{#1^{\scriptscriptstyle \bot}} 
\newcommand{\leftScottint}{[{\kern-.3ex}[}
\newcommand{\rightScottint}{]{\kern-.3ex}]}
\newcommand{\Scottint}[1]{\leftScottint\,#1\,\rightScottint}
\newcommand{\distributionsymbol}{\mathcal{D}}

\newcommand{\multisetsymbol}{\mathcal{M}}

\newcommand{\Dst}{\distributionsymbol}

\newcommand{\Mlt}{\multisetsymbol}
\newcommand{\Giry}{\mathcal{G}}

\newcommand{\UF}{\ensuremath{\mathcal{U}{\kern-.75ex}\mathcal{F}}}

\newcommand{\Cat}[1]{\ensuremath{\mathbf{#1}}\xspace}
\newcommand{\cat}[1]{\Cat{#1}}
\newcommand{\Kl}{\mathcal{K}{\kern-.4ex}\ell}
\newcommand{\EM}{\mathcal{E}{\kern-.4ex}\mathcal{M}}

\newcommand{\Sets}{\Cat{Sets}}

\newcommand{\NNO}{\mathbb{N}}

\newcommand{\cond}[2]{\ensuremath{{#1}|_{#2}}}
\newcommand{\hypcond}[2]{\ensuremath{\mathop{{#1}\!\parallel_{#2}}}}

\newcommand{\Ef}{\ensuremath{\mathcal{E}{\kern-.5ex}f}}
\newcommand{\intd}{{\kern.2em}\mathrm{d}{\kern.03em}}
\newcommand{\indic}[1]{\mathbf{1}_{#1}}

\newcommand{\bigovee}{\mathop{\vphantom{\sum}\mathchoice%
        {\vcenter{\hbox{\huge $\ovee$}}}%
        {\vcenter{\hbox{\Large $\ovee$}}}%
        {\ovee}{\ovee}}\displaylimits}

\newcommand{\OF}{\ensuremath{\mathcal{O}{\kern-.1em}\mathcal{F}}}
\newcommand{\Closed}{\ensuremath{\mathcal{C}{\kern-.05em}\ell}}

\newcommand{\conglongrightarrow}{\mathrel{\smash{\stackrel{
           \raisebox{.5ex}{$\scriptstyle\cong$}}{
           \raisebox{0ex}[0ex][0ex]{$\longrightarrow$}}}}}

\makeatother
 \newdir{ >}{{}*!/-7.5pt/@{>}}
 \newdir{|>}{!/4.5pt/@{|}*:(1,-.2)@^{>}*:(1,+.2)@_{>}}
 \newdir{ |>}{{}*!/-3pt/@{|}*!/-7.5pt/:(1,-.2)@^{>}*!/-7.5pt/:(1,+.2)@_{>}}

\newsavebox\sbground
\savebox\sbground{\begin{tikzpicture}[circuit ee IEC,yscale=0.5,xscale=0.4]
                \draw (0,-2ex) to (0,0) node[ground,rotate=90,xshift=.65ex] {};
                \end{tikzpicture}}

\newsavebox\sbtto
\savebox\sbtto{\begin{tikzpicture}[baseline=-2.5pt]
            \filldraw[fill=black,draw=white] circle (1.4pt);
                \end{tikzpicture}}

\newcommand\klto{\mathrel{\ooalign{$\to$\cr
            \hfil$\raisebox{0.16pt}{\usebox\sbtto\kern0.6pt}$\hfil\cr}}}

\title{Hyper Normalisation and Conditioning for Discrete Probability
  Distributions}

\dedicatory{\color{red}\backslash{}dedicatory\{...\}}
\dedicatory{Dedicated to Ji\v{r}\'\i~Ad\'amek on the occasion of his \nth{70} birthday}

\author{Bart Jacobs}\thanks{The research leading to these results has
    received funding from the European Research Council under the
    European Union's Seventh Framework Programme (FP7/2007-2013) / ERC
    grant agreement nr.~320571}
\address{
Institute for Computing and Information Sciences, 
Radboud University, Nijmegen, The Netherlands. 
}
\urladdr{www.cs.ru.nl/B.Jacobs}
\email{bart@cs.ru.nl}

\date{}

\begin{document}
\maketitle
\begin{abstract}
Normalisation in probability theory turns a subdistribution into a
proper distribution. It is a partial operation, since it is undefined
for the zero subdistribution. This partiality makes it hard to reason
equationally about normalisation. A novel description of normalisation
is given as a mathematically well-behaved total function. The output
of this `hyper' normalisation operation is a distribution of
distributions. It improves reasoning about normalisation.

After developing the basics of this theory of (hyper) normalisation,
it is put to use in a similarly new description of conditioning,
producing a distribution of conditional distributions. This is used to
give a clean abstract reformulation of refinement in quantitative
information flow.
\end{abstract}

\section{Introduction}\label{IntroSec}

\colorlet{ExColor}{red!14!green!28!blue!58!}

We start with the RGB colour model to illustrate normalisation of
distributions.  This model describes each colour as an additive
combination of the primary colours red (R), green (G) and blue (B). It
is standardly used in colour screens and cameras.  We can write a
colour $C$ for instance as sum:
$$\begin{array}{rclr}
C
& = &
\frac{1}{8}\ket{R} + \frac{1}{4}\ket{G} + \frac{1}{2}\ket{B}
\qquad\mbox{which is (colour) printed as}\qquad
&
\mbox{\color{ExColor}{C}}.
\end{array}$$

\noindent The `ket' notation $\ket{-}$ is used as meaningless
syntactic sugar in such formal sums. We see that the three weights add
up to $\frac{1}{8}+\frac{1}{4} + \frac{1}{2} =
\frac{7}{8}$. Normalisation, in its simplest form, re-scales these
weights so that they add up to one. This is done via division by their
sum, as in:
$$\begin{array}{rcccl}
\nrm(C)
& = &
\frac{\nicefrac{1}{8}}{\nicefrac{7}{8}}\ket{R} + 
   \frac{\nicefrac{1}{4}}{\nicefrac{7}{8}}\ket{G} + 
   \frac{\nicefrac{1}{2}}{\nicefrac{7}{8}}\ket{B}
& = &
\frac{1}{7}\ket{R} + \frac{2}{7}\ket{G} + \frac{4}{7}\ket{B}.
\end{array}$$

\noindent We see that in this normalised description $\nrm(C)$, the
relative weights of the values is the same, but their sum has been
adjusted to one. We can understand $\nrm(C)$ as a formal \emph{convex}
sum of $R,G,B$, that is, as a probability distribution over the set
$\{R,G,B\}$. The original colour $C$ is called a subdistribution, since
the sum of its values is below (sub) one.

Normalisation of subdistributions (to distributions) is one of the
fundamental operations in probability theory. It forms the basis of
many other constructions, notably of conditioning, which is so
important in calculating influences in Bayesian
networks~\cite{Barber12}. The problem with normalisation is that it is
a partial operation: it is undefined for the zero subdistribution ---
of the form $0\ket{R} + 0\ket{G} + 0\ket{B}$ in the context of the
above colour example. This partiality makes it difficult to develop an
equational system for normalisation.

The main contribution of this paper is a re-description of
normalisation as a \emph{total} operation that satisfies various
equations. This new, mathematically civilised formulation makes use of
`hyper' distributions, that is, of distributions of
distributions. Hence we often refer to the new formulation as `hyper'
normalisation, in order to distinguish it from traditional
normalisation --- illustrated in the earlier colour example.  Our
hyper normalisation operation $\Nrm$ takes the following form:
$$\xymatrix{
\Dst(n\cdot A)\ar[rr]^-{\Nrm} & & \Dst\big(n\cdot \Dst(A)\big)
}$$

\noindent The set $A$ describes the sample space, and $n$ is a natural
number, used in the copower $n\cdot A$, which produces $n$ copies of
$A$. A distribution $\omega\in\Dst(n\cdot A)$ over the copower $n\cdot
A$ consists of $n$ subdistributions over $A$, over each of these
copies of $A$. The normalisation $\Nrm(\omega)$ produces a
distribution of normalised distributions, by normalising these
subdistributions in parallel, each with weight proportional to the
original subdistribution.  How this works precisely is explained in
Section~\ref{NormSec}, once the notions of distribution and copower
are described in detail.

Applying hyper normalisation in conditioning yields what we call
`hyper' conditioning. It is again a total operation. The use of such
hyper conditioning is briefly illustrated in a Bayesian reasoning
example, and more extensively in a re-description of refinement in
quantitative information flow. Since hyper normalisation satisfies
various equations, for which see Section~\ref{NormSec}, it may be a
useful operation in languages for probabilistic programming and
reasoning; see
\textit{e.g.}~\cite{BorgstromGGMG13,ScibiorGG15,JansenKKOGM15,AdamsJ17,StatonYHKW16,JacobsZ16}.

Actually, the whole idea of describing normalisation in `hyper' form
emerged from the study of the `denotation of a channel' construction
in~\cite{McIverMM10,McIverMSEM14,McIverMT15,McIverMM15}. Normalisation
is an implicit step in this construction, which is defined and
characterised here as a separate, explicit operation. The original
denotation construction in information flow then re-appears as hyper
conditioning. We illustrate the close connection with a new, abstract
proof of a known result from the area (see Theorem~\ref{OrderThm}
below).

In addition, there are two clear points of influence for the current
work.
\begin{itemize}
\item Quantum probability theory. The systematic and formal
  description of aspects of probability theory fits in a wider study
  of quantum foundations and probability
  theory~\cite{JacobsWW15a,Jacobs15d,ChoJWW15b,JacobsZ16}. This
  influence becomes visible here in some of the notation, like the
  kets $\ket{-}$, and in some of the terminology, like tests and
  instruments. However, this quantum background is not needed to
  follow what happens here.

\item Category theory. Several descriptions, constructions and
  calculations in probability theory can be greatly simplified by
  using the categorical notion of monad, concretely in the form of the
  distribution monad $\Dst$ for discrete probability and the Giry
  monad $\Giry$ for continuous probability, see~\cite{Jacobs17a} for
  more information. However, this categorical aspect is deliberately
  suppressed here in order to reach a wider audience: the main ideas
  and constructions of the paper are accessible, hopefully, to readers
  without such categorical background. But the categorical influence
  is manifest, for instance in the frequent use of diagrams to express
  equations.
\end{itemize}

\noindent This paper focuses on normalisation and conditioning in
\emph{discrete} probability. The question immediately arises: what
about continuous or even quantum probability? This matter is postponed
to future work.

\section{Mathematical background}\label{BackgroundSec}

\subsection{Copowers}\label{CoproductSumsec}

For a number $n$ and a set $A$ one commonly writes $A^{n}$ for the
$n$-fold cartesian product $A \times \cdots \times A$ of $A$ with
itself, consisting of all $n$-tuples of elements from $A$. Each
function $f\colon A \rightarrow B$ can be extended to $f^{n} \colon
A^{n} \rightarrow B^{n}$ by $f^{n}(a_{0}, \ldots, a_{n-1}) =
(f(a_{0}), \ldots, f(a_{n-1}))$. More generally, for $n$ different
functions $f_{i}\colon A \rightarrow B$ we can define a map $A^{n}
\rightarrow B^{n}$ that applies $f_i$ to the $i$-th element in a
tuple. This map is written as $n$-tuple
$\tuple{f_{i}\after\pi_{i}}_{i<n} \colon A^{n} \rightarrow B^{n}$,
where the maps $\pi_i$ are \emph{projections} $A^{n} \rightarrow A$.
Finally, there is a diagonal map $\Delta \colon A \rightarrow A^{n}$
sending an element $a\in A$ to the diagonal $n$-tuple
$\tuple{a,\ldots, a}\in A^{n}$.

For $n=0$, the power $A^{n}$ is the singleton set, commonly written as
$1$. It contains only the empty tuple. For each set $B$ there is a
unique function $B \rightarrow 1$, which is written as $\bang$.

These sets $A^{n}$ are called \emph{powers} of $A$. There are also
\emph{copowers} $n\cdot A$, given by the cartesian product
$\{0,\ldots,n-1\}\times A$. Its elements are thus pairs $(i,a)$ where
$0\leq i\leq n-1$ and $a\in A$. We shall use `coprojection'
functions $\kappa_{i}\colon A \rightarrow n\cdot A$, given by
$\kappa_{i}(a) = (i,a)$. As for powers, a function $f\colon A
\rightarrow B$ gives rise to a function $n\cdot f \colon n\cdot A
\rightarrow n\cdot B$, given by $\kappa_{i}a \mapsto \kappa_{i}f(a)$.
For different functions $f_{i}\colon A \rightarrow B$ there is a map
$[\kappa_{i} \after f_{i}]_{i<n}$ mapping $\kappa_{i}a\in n\cdot A$ to
$\kappa_{i}f_{i}(a)\in n\cdot B$. Notice that the empty copower is the
empty set $0$. The analogue of the diagonal map $\Delta\colon A
\rightarrow A^{n}$ is the codiagonal $\nabla \colon n\cdot A
\rightarrow A$ sending each $\kappa_{i}a$ to $a$. Clearly, it removes
all the tags $\kappa_i$. 

In line with these descriptions we write $n$ not only for the natural
number $n\in\NNO$ but also for the $n$-element set $\{0, 1, \ldots,
n-1\}$. Notice that $0$ is then the empty set, $1$ is the singleton
set $\{0\}$, and $2 = \{0,1\}$ is the set of Booleans. We have $2
\cong 2\cdot 1$, and more generally $n \cong n\cdot 1$. When the
copower $n\cdot A$ is read as product $n\times A$, then $\nabla$ is
simply the second projection $\pi_{2} \colon n \times A \rightarrow
A$. We freely switch between these descriptions.

(Categorically, the copower $n\cdot A$ is the $n$-fold coproduct/sum
$A+ \cdots + A$ of sets, just as the power $A^{n}$ is the $n$-fold
product $A\times\cdots\times A$. This coproduct description of
copowers generalises to other categories. The coincidence of copowers
$n\cdot A$ with products $n\times A$ for sets does not work in general
categories.)

\auxproof{
For two sets $A,B$ we write $A+B$ for their disjoint union, given by:
\begin{equation}
\label{CoproductEqn}
\begin{array}{rcl}
A+B
& = &
\set{\kappa_{1}a}{a\in A} \cup \set{\kappa_{2}b}{b\in B}.
\end{array}
\end{equation}

\noindent The $\kappa_{1}$ and $\kappa_{2}$ are just notational tags
that are used to distinguish the elements from $A$ and from $B$ inside
$A+B$, enforcing disjointness.  Hence we assume $\kappa_{1}a \neq
\kappa_{2}b$ for all $a\in A, b\in B$. Moreover, these tags keep
different elements apart: $a\neq a' \Rightarrow \kappa_{1}a\neq
\kappa_{1}a'$, and similarly for $\kappa_2$.

The tags $\kappa_{1}, \kappa_{2}$ give rise to `coprojection'
functions $\kappa_{1} \colon A \rightarrow A+B$ and $\kappa_{2} \colon
B \rightarrow A+B$. By assumption, they are injective and have
disjoint images. This new disjoint union set $A+B$ is also called a
sum, or a coproduct. The latter terminology is often used in category
theory, and is followed here.  In this categorical formulation one
also uses \emph{cotuples} in the following way. For an arbitrary set
$C$ and arbitrary functions $f \colon A \rightarrow C$ and $g\colon B
\rightarrow C$ there is a unique cotuple map $[f,g]\colon A+B
\rightarrow C$ with $[f,g] \after \kappa_{1} = f$ and $[f,g] \after
\kappa_{2} = g$. This cotuple is of course defined by
pattern-matching, as:
$$\begin{array}{rcl}
[f,g](x)
& = &
{\left\{\begin{array}{ll}
f(a) \quad & \mbox{if } x = \kappa_{1}a \\
g(b) & \mbox{if } x = \kappa_{2}b.
\end{array}\right.}
\end{array}$$

\noindent If we have two functions $h\colon A \rightarrow X$ and
$k\colon B \rightarrow Y$ we can now form a map $A+B \rightarrow X+Y$,
namely the map $[\kappa_{1} \after h, \kappa_{2} \after k]$. It sends
$\kappa_{1}a$ to $\kappa_{1}f(a)$ and $\kappa_{2}b$ to
$\kappa_{2}k(b)$. It is convenient to overload notation, and write
$h+k \colon A+B \rightarrow X+Y$ for this function.

Instead of binary coproducts, as in~\eqref{CoproductEqn}, one can also
form $n$-ary coproducts $A_{1} + \cdots + A_{n}$. For $n=0$, this
coproduct is defined as the empty set. If all the sets $A_i$ are the
same, say $A_{i} = A$, then one writes $n\cdot A = A + \cdots +
A$. This $n\cdot A$ is called the \emph{copower}. For a map $f\colon A
\rightarrow X$ we get a function $n\cdot f \colon n\cdot A \rightarrow
n\cdot X$ which sends $\kappa_{i}a$ to $\kappa_{i}f(x)$. More formally
we can describe it as $n$-ary cotuple $[\kappa_{i} \after f]_{i}$.

This copower $n\cdot A = A + \cdots + A$ is formally similar to the
power $A^{n} = A\times \cdots \times A$. Each function $f\colon A
\rightarrow B$ gives rise to a function $f^{n} \colon A^{n}
\rightarrow B^{n}$, with $f^{n}(a_{1}, \ldots, a_{n}) = (f(a_{1}),
\ldots, f(a_{n}))$. More formally, $f^{n}$ is the $n$-ary tuple
$\tuple{f \after \pi_{i}}_{i}$.

For each natural number $n\in\mathbb{N}$ we also write $n$ for the
$n$-element set $\{0, \ldots, n-1\}$. Notice that $0$ is then the
empty set, $1$ is the singleton set $\{0\}$, and $2 = \{0,1\}$ is the
set of Booleans. We have $2 \cong 1+1 = 2\cdot 1$, and more generally
$n \cong 1 + \cdots + 1 = n\cdot 1$. For an arbitrary set $A$ there is
precisely one function $\bang \colon A \rightarrow 1$, namely
$\bang(a) = 0$.
}

\subsection{Probability distributions}\label{DstributionSubsec}

A (discrete) \emph{distribution} over a `sample' set $A$ is a
weigh\-ted combination of elements of $A$, where the weights are
probabilities from the unit interval $[0,1]$ that add up to $1$. Here
we only consider finite combinations and write them as:
\begin{equation}
\label{ConvexFormalSumEqn}
\begin{array}{rclcc}
\omega
& = &
r_{1}\ket{a_{1}} + \cdots + r_{n}\ket{a_{n}}
& \;\mbox{ where }\; &
\left\{\begin{array}{l}
a_{1}, \ldots, a_{n} \in A \\
r_{1}, \ldots, r_{n} \in [0,1] \mbox{ with } \sum_{i}r_{i} = 1.
\end{array}\right.
\end{array}
\end{equation}
\noindent The `ket' notation $\ket{a}$ is syntactic sugar, used to
distinguish elements $a\in A$ from their occurrence in such formal
convex sums. For instance, the uniform distribution of $n$-elements
$a_{1}, \ldots, a_{n}$ is described as $\frac{1}{n}\ket{a_1} + \cdots
+ \frac{1}{n}\ket{a_n}$, or more succinctly as
$\sum_{i}\frac{1}{n}\ket{a_i}$.

We write $\Dst(A)$ for the set of all (finite, discrete) distributions
$\sum_{i}r_{i}\ket{a_i}$ over $A$
from~\eqref{ConvexFormalSumEqn}. Distributions are also called
\emph{states}; they express knowledge, in terms of likelihoods of
occurrence of elements of $A$. Notice that such a state
$\omega\in\Dst(A)$ can be identified with a `probability mass'
function $\omega \colon A \rightarrow [0,1]$ with finite support
$\supp(\omega) = \setin{a}{A}{\omega(a) \neq 0}$ and with $\sum_{a\in
  A}\omega(a) = 1$. This function-description is often more
convenient; we freely switch between this function-description and the
formal convex sum description in~\eqref{ConvexFormalSumEqn}.

In formal convex sums like $\sum_{i}r_{i}\ket{a_i}$
in~\eqref{ConvexFormalSumEqn} we implicitly use equations such as:
$r\ket{a} + s\ket{b} = s\ket{b} + r\ket{a}$, and: $r\ket{a} + s\ket{a}
= (r+s)\ket{a}$. Further, terms $0\ket{a}$ do not contribute to the
sum and are omitted.

The elements of the set $\Dst(n)$ can be identified
with $n$-tuples of non-negative real numbers $(r_{1}, \ldots, r_{n})$
with $\sum_{i}r_{i} = 1$. The set $\Dst(n)$ is called the standard
$n-1$ simplex in topology.

A \emph{hyper distribution}, according
to~\cite{McIverMM10,McIverMM15,McIverMT15,McIverMSEM14}, is a
distribution of distributions, that is, an inhabitant of $\Dst^{2}(A)
= \Dst(\Dst(A))$. There is `multiplication' map $\mu \colon
\Dst^{2}(A) \rightarrow \Dst(A)$ turning a hyper distribution into an
ordinary distribution, via:
\begin{equation}
\label{MuEqn}
\begin{array}{rcl}
\mu\big(\sum_{i}r_{i}\ket{\omega_{i}}\big)
& = &
{\displaystyle\sum}_{a} (\sum_{i} r_{i}\cdot \omega_{i}(a))\Bigket{a}.
\end{array}
\end{equation}

\noindent On the right hand side, the outer sum over $a$ is a formal
convex sum, whereas the inner sum over $i$ is an actual sum, in the
unit interval $[0,1]$. In this equation~\eqref{MuEqn}, the formal
convex sum and the function notation are mixed. We shall use the term
`hyper distribution' in `tagged' form, as distribution on a copower
$n\cdot\Dst(A)$ of distributions, that is, as inhabitant of
$\Dst(n\cdot\Dst(A))$.

The mapping $A \mapsto \Dst(A)$ is \emph{functorial}: it does not only
work on sets, but also on functions. Each function $f\colon A
\rightarrow B$ gives rise to a function $\Dst(A) \rightarrow \Dst(B)$,
for which we use the overloaded notation $\Dst(f)$. It is given in
the obvious way, like map-list in functional programming:
\begin{equation}
\label{DstFunEqn}
\begin{array}{rcl}
\Dst(f)\Big(\sum_{i}r_{i}\ket{a_i}\Big)
& = &
\sum_{i}r_{i}\ket{f(a_{i})}.
\end{array}
\end{equation}

\noindent The result is sometimes called the \emph{push-forward}
distribution. The sum on the right hand side may involve fewer items
than the original sum $\sum_{i}r_{i}\ket{a_i}$, when $f(a_{i}) =
f(a_{j})$ for certain indices $i\neq j$. It is not hard to see that
identity functions and compositions are preserved: $\Dst(\idmap) =
\idmap$ and $\Dst(g \after f) = \Dst(g) \after \Dst(f)$.

Marginalisation can be described via functoriality of $\Dst$. For a
distribution $\omega\in\Dst(A\times B)$ on a product set, the
marginalisations of $\omega$ are obtained as $\Dst(\pi_{1})(\omega)
\in\Dst(A)$ and $\Dst(\pi_{2})(\omega)\in\Dst(B)$, via the two
projections $A \stackrel{\pi_1}{\longleftarrow} A\times B
\stackrel{\pi_2}{\longrightarrow} B$. Explicitly:
$$\begin{array}{rclcrcl}
\Dst(\pi_{1})(\omega)
& = &
{\displaystyle\sum}_{a} \big(\sum_{b}\omega(a,b)\big)\Bigket{a}
& \qquad\mbox{and}\qquad &
\Dst(\pi_{2})(\omega)
& = &
{\displaystyle\sum}_{b} \big(\sum_{a}\omega(a,b)\big)\Bigket{b}.
\end{array}$$

\subsection{Kleisli maps and Kleisli composition}\label{KleisliSubsec}


The mapping $A \mapsto \Dst(A)$ is an instance of the categorical
notion of monad. We shall suppress the categorical perspective, and
stick to rather concrete descriptions. It is not hard to see that a
map of the form $n\rightarrow\Dst(m)$ corresponds to an $m\times n$
stochastic matrix, with $n$ columns of $m$ entries adding up to
$1$. Matrix composition corresponds to a special form of function
composition, which we shall write as $\klafter$.  We often call
$\klafter$ Kleisli composition, since it is composition in the
so-called Kleisli category associated with $\Dst$, as monad.

We shall write $f \colon A \klto B$ to express that $f$ is a function
$A \rightarrow \Dst(B)$. Such a map is sometimes called a
\emph{conditional distribution}, or just a \emph{conditional}, since
one can understand $f(a)(b) \in [0,1]$ as the conditional probability
$P(b\mid a)$. The point of the notation $A\klto B$ is that the letter
`$\Dst$' can be suppressed in the codomain. A bit formally, we can
write a state $\omega\in\Dst(A)$ as a Kleisli map $\omega\colon 1
\klto A$, where $1 = \{0\}$ is the singleton set, as above.  This
arrow formulation is useful in diagrams.

If we have two such Kleisli maps $A \klto B$ and $B\klto C$, given by
functions $f\colon A \rightarrow \Dst(B)$ and $g\colon B \klto
\Dst(C)$, then we write $g\klafter f = g_{*} \after f \colon A \klto
C$, where $\after$ is ordinary composition, and $g_{*} \colon \Dst(B)
\rightarrow \Dst(C)$ is the `Kleisli lifting' function defined by:
\begin{equation}
\label{KlLiftEqn}
\begin{array}{rcl}
g_{*}\Big(\sum_{i} r_{i}\ket{b_i}\Big)
& = &
{\displaystyle\sum}_{c\in C} (\sum_{i}r_{i}\cdot g(b_{i})(c))\Bigket{c}.
\end{array}
\end{equation}

\noindent Abstractly, we can write $g_{*} = \mu \after \Dst(g)$.

When $f,g$ are seen as matrices, then $\klafter$ is matrix
composition. It is not hard to see that Kleisli composition $\klafter$
is associative. Its unit is the `Dirac' map $\eta \colon A \rightarrow
\Dst(A)$ given by `point' distributions $\eta(a) = 1\ket{a}$. In
various calculations we shall use the following basic equations about
Kleisli extension $(-)_{*}$; they hold for monads in general. Proofs
are left to the interested reader.

\begin{lemma}
\label{KleisliLem}
The above definition~\eqref{KlLiftEqn} satisfies:
\begin{enumerate}
\item $g_{*} \after \eta = g$;

\item $\eta_{*} = \idmap$;

\item $(\eta \after f)_{*} = \Dst(f)$;

\item $\Dst(h) \after g_{*} = (\Dst(h) \after g)_{*}$;

\item $g_{*} \after f_{*} = (g \klafter f)_{*}$. \QED
\end{enumerate}
\end{lemma}

The following special maps play an important role in the sequel.

\begin{definition}
\label{StrengthGraphDef}
Let $A,B$ be arbitrary sets. There are two \emph{strength} functions:
\begin{equation}
\label{StrengthEqn}
\vcenter{\xymatrix@R-2pc@C-.5pc{
\Dst(A)\times B\ar[r]^-{\st_1} & \Dst(A\times B)
& &
A\times \Dst(B)\ar[r]^-{\st_2} & \Dst(A\times B)
\\
\st_{1}(\sum_{i}r_{i}\ket{a_i}, b)\ar@{|->}[r] & \sum_{i} r_{i}\ket{a_{i},b}
& &
\st_{2}(a, \sum_{i}r_{i}\ket{b_i})\ar@{|->}[r] & \sum_{i} r_{i}\ket{a,b_{i}}
}}
\end{equation}

\noindent For a function $f\colon A \rightarrow \Dst(B)$ there is
a \emph{graph} function:
\begin{equation}
\label{GraphEqn}
\vcenter{\xymatrix@C+.5pc{
A\ar[r]^-{\gr(f)} & \Dst(B\times A)
}}
\qquad\mbox{via}\qquad
\begin{array}{rcl}
\gr(f)(a)
& = &
\sum_{b} f(a)(b)\ket{b,a}.
\end{array}
\end{equation}
\end{definition}

If we write $\tw = \tuple{\pi_{2}, \pi_{1}} \colon A\times B
\conglongrightarrow B\times A$ for the `twist' map, then we see that
the two strength maps are related via $\st_{2} \after \tw = \Dst(\tw)
\after \st_{1}$. These strength functions $\st_i$ make $\Dst$ a
`strong monad', a basic notion in functional programming. The graph
map can be defined abstractly as $\gr(f) = \st_{1} \after
\tuple{f,\idmap}$, as in~\cite{FurberJ15b}.


We need some basic results about how strength and graph interact with
marginalisation, as succinctly expressed in the following diagrams.
\begin{equation}
\label{StrengthGraphDiag}
\vcenter{\xymatrix@C-0pc{
\Dst(A) & \Dst(A\times B)\ar[l]_-{\Dst(\pi_{1})}
   \ar[r]^-{\Dst(\pi_{2})} & \Dst(B)
&
\Dst(B) & \Dst(B\times A)\ar[l]_-{\Dst(\pi_{1})}
   \ar[r]^-{\Dst(\pi_{2})} & \Dst(A)
\\
& \Dst(A)\times B\ar[u]_{\st_1}\ar[r]_-{\pi_2}\ar[ul]^-{\pi_1} & B\ar[u]_{\eta} 
&
& A\ar[u]_{\gr(f)}\ar[ul]^{f}\ar[ur]_{\eta} & 
}}
\end{equation}

\noindent These results are easily verified. On an abstract level, the
rectangle on the left follows from the fact that $\Dst$ is a `strongly
affine' monad; see~\cite{Jacobs16a,Jacobs17a}.




\subsection{Normalisation, traditionally}\label{TradNormSubsec}

In~\eqref{ConvexFormalSumEqn} we have seen that in a distribution
$\sum_{i}r_{i}\ket{a_i}$ the weights $r_{i} \in [0,1]$ add up to one.
We speak of a \emph{subdistribution} when the sum is below one, that
is, when $\sum_{i}r_{i} \leq 1$. What we call \emph{normalisation}, in
the traditional sense, is the process of turning a subdistribution
into a proper distribution by adjusting the weights so that they add
up to one --- as illustrated in the RGB example at the very beginning
of this article. Normalisation is a partial operation that can be
described as follows. If $\omega = \sum_{i}r_{i}\ket{a_i}$ is a
subdistribution we first take the sum $r = \sum_{i}r_{i}$ of all
weights; if $r\neq 0$, then we can readjust the original weights to
form a proper distribution:
\begin{equation}
\label{TradNormEqn}
\begin{array}{rcl}
\nrm(\omega)
& = &
\sum_{i} \frac{r_i}{r} \ket{a_i}.
\end{array}
\end{equation}

\noindent By construction $\nrm(\omega)$ is a distribution since its
weights add up to one: $\sum_{i}\frac{r_i}{r} = \frac{r}{r} = 1$.

Via the graph construction in Definition~\ref{StrengthGraphDef} one
can produce a joint distribution on a set $B\times A$ from a Kleisli
map (conditional) $A \rightarrow \Dst(B)$. The reverse process is
sometimes called \emph{disintegration}. We shall concentrate on the
special case of joint distributions on copowers $n\cdot A = n\times
A$.

If have a `joint' distribution $\Omega \in \Dst(n\cdot A)$ on a
copower $n\cdot A$ we obtain for each element $a\in A$ a
subdistribution on $n$, namely:
\begin{equation}
\label{JointSubdistrEqn}
\begin{array}{rcl}
\Omega_{a}
& = &
\sum_{i} \Omega(\kappa_{i}a)\ket{i}.
\end{array}
\end{equation}

\noindent Normalisation of these subdistributions is what we call
\emph{pointwise} normalisation. It is crucial in the following result
showing how a \emph{conditional} $A \rightarrow \Dst(n)$ can be
associated with a joint distribution on $n\cdot A$.  It can be seen as
a discrete version of \textit{e.g.}~\cite[Prop.~3.3]{Fong12}
and~\cite[Prop.~6.7]{Panangaden09}. The existence of such `regular
conditional probability' in (continuous) measure theory is a
consequence of the Radon-Nikodym Theorem. Here, in the discrete
setting, things are much simpler.

\begin{proposition}
\label{PointNormProp}
There is a bijective correspondence between $\Omega$ above the double
lines and pairs $(f,\omega)$ below, in:
$$\begin{prooftree}
\Omega\in\Dst(n\cdot A) \quad\mbox{with}\quad
   \supp\big(\Dst(\pi_{2})(\Omega)\big) = A
\Justifies
\xymatrix{A\ar[r]_-{f} & \Dst(n) \quad\mbox{ and }\quad
   \omega\in\Dst(A) \quad\mbox{with}\quad \supp(\omega) = A}
\end{prooftree}$$
\end{proposition}

The side-condition $r\neq 0$ in normalisation~\eqref{TradNormEqn}
translates in this pointwise formalisation into the requirement that
the support of the relevant distributions is the whole set $A$.

\begin{myproof}
In the upward direction we define $\Omega(\kappa_{i}(a)) =
\omega(a)\cdot f(a)(i)$. More formally, we first take the graph
$\gr(f) \colon A \rightarrow \Dst(n\cdot A)$ from
Definition~\ref{StrengthGraphDef} and then obtain a joint distribution
by applying its Kleisli extension to $\omega\in\Dst(A)$, as in:
\begin{equation}
\label{PointNormUp}
\begin{array}{rcccl}
\Omega
& = &
\gr(f)_{*}(\omega)
& = &
{\displaystyle\sum}_{i,a}\, \omega(a)\cdot f(a)(i)\Bigket{\kappa_{i}a}
\;\in\;\Dst(n\cdot A).
\end{array}
\end{equation}

\noindent We show that $\Dst(\pi_{2})(\Omega) = \omega$ in two ways. First
we reason with distributions:
$$\begin{array}{rcll}
\Dst(\pi_{2})(\Omega)
& = &
{\displaystyle\sum}_{a}\, \big(\sum_{i}\omega(a)\cdot f(a)(i)\big)
   \bigket{a} \quad \\
& = &
{\displaystyle\sum}_{a}\, \big(\omega(a)\cdot \sum_{i}f(a)(i)\big)\bigket{a}
\\
& = &
{\displaystyle\sum}_{a}\, \big(\omega(a)\cdot 1\big)\bigket{a}
   & \mbox{since $f(a) \in \Dst(n)$} 
\\
& = &
\omega.
\end{array}$$

\noindent A more abstract proof uses Lemma~\ref{KleisliLem} and
Diagram~\eqref{StrengthGraphDiag}:
$$\begin{array}{rcccccccl}
\Dst(\pi_{2})(\Omega)
& = &
\big(\Dst(\pi_{2}) \after \gr(f)_{*}\big)(\omega)
& = &
\big(\Dst(\pi_{2}) \after \gr(f)\big)_{*}(\omega)
& = &
\eta_{*}(\omega) 
& = &
\omega.
\end{array}$$

\noindent Hence $\supp\big(\Dst(\pi_{2})(\Omega)\big) = \supp(\omega)
= A$.

In the other direction, given $\Omega \in\Dst(n\cdot A)$ we take
$\omega = \Dst(\pi_{2})(\Omega) = \Dst(\nabla)(\Omega) \in \Dst(A)$
and use the subdistribution~\eqref{JointSubdistrEqn} to define a
function $f\colon A \rightarrow \Dst(n)$ via normalisation:
\begin{equation}
\label{PointNormDown}
\begin{array}{rcccccl}
f(a)
& = &
\nrm(\Omega_{a})
& = &
{\displaystyle\sum}_{i} 
   \frac{\Omega(\kappa_{i}a)}{\sum_{i}\Omega(\kappa_{i}a)}\Bigket{i}
& = &
{\displaystyle\sum}_{i} \frac{\Omega(\kappa_{i}a)}{\omega(a)}\Bigket{i}.
\end{array}
\end{equation}

\noindent This is well-defined since $\supp(\omega) = A$, so that
$\omega(a) = \sum_{i}\Omega(\kappa_{i}a) \neq 0$, for each $a\in A$.

We show that $\Omega$ re-appears via the
formula~\eqref{PointNormUp}:
$$\begin{array}{rcccccl}
\gr(f)_{*}(\omega)
& = &
{\displaystyle\sum}_{i,a} \omega(a)\cdot f(a)(i)\bigket{\kappa_{i}a}
& = &
{\displaystyle\sum}_{i,a} \Omega(\kappa_{i}a)\bigket{\kappa_{i}a}
& = &
\Omega.
\end{array}$$

\noindent We leave it to the interested reader to show that first
applying~\eqref{PointNormUp} to $f,\omega$ and
then~\eqref{PointNormDown} yields the original pair $f,\omega$. \QED

\auxproof{
We already know that $\Dst(\pi_{1})(\omega) = \omega$ for $\Omega$
defined in~\eqref{PointNormUp}. The resulting function applied to
$a\in A$ is:
$$\begin{array}{rcl}
\nrm(\Omega_{a})
& = &
{\displaystyle\sum}_{b} \frac{\Omega(b,a)}{\omega(a)}\Bigket{b} \\
& = &
{\displaystyle\sum}_{b} \frac{\omega(a)\cdot f(a)(b)}{\omega(a)}\Bigket{b} \\
& = &
{\displaystyle\sum}_{b} f(a)(b)\Bigket{b} \\
& = &
f(a).
\end{array}$$
}
\end{myproof}


Since distributions in the current setting always have finite support,
the assumptions $\supp\big(\Dst(\pi_{2})(\Omega)\big) = A$ and
$\supp(\omega) = A$ in Proposition~\ref{PointNormProp} imply that $A$
must be a finite set. Hence we could identify $A$ with a finite set
$m$.


\section{Hyper normalisation}\label{NormSec}

Having seen these preliminary definitions and results, we can turn to
our new description of normalisation in `hyper' form. It will be a
function $\Nrm$ of the following type.
$$\vcenter{\xymatrix{
\Dst\big(n\cdot A\big)\ar[rr]^-{\Nrm} & & \Dst\big(n\cdot\Dst(A)\big)
}}$$

\noindent This normalisation map $\Nrm$ thus sends a distribution over
a copower of a set $A$ to a distribution over a copower of
distributions over $A$. Before defining the map $\Nrm$ in full
generality we give an illustration of how it works.

Consider a finite set $A = \{a, b, c, d\}$ and number $n=3$. Let's
start from the distribution $\omega \in \Dst(3\cdot A)$ given by:
$$\begin{array}{rcl}
\omega
& = &
\frac{1}{8}\ket{\kappa_{0}a} + \frac{1}{4}\ket{\kappa_{0}b} +
   \frac{1}{2}\ket{\kappa_{1}c} + \frac{1}{8}\ket{\kappa_{1}d}.
\end{array}$$

\noindent This distribution contains elements $a,b\in A$ from the
first sum component in the copower $3\cdot A = A+A+A$, and elements
$c,d\in A$ from the second component, and nothing from the third
component. There are associated subdistributions $\omega_{i}$, for
$i\in 3$, are given by:
$$\begin{array}{rclcrclcrcl}
\omega_{0}
& = &
\frac{1}{8}\ket{a} + \frac{1}{4}\ket{b}
& \qquad &
\omega_{1}
& = &
\frac{1}{2}\ket{c} + \frac{1}{8}\ket{d}
& \qquad &
\omega_{2}
& = &
0.
\end{array}$$

\noindent We see that these subdistributions $\omega_{i}$ aggregate
the items in $\omega$ from the same component --- \textit{i.e.}~with
the same coprojection $\kappa_i$. Normalisation turns these
subdistributions $\omega_{i}$ into proper `inner' distributions in
$\Nrm(\omega)$ via normalisation as in~\eqref{TradNormEqn}, while
keeping track of their origin.  That is, $\Nrm(\omega) \in
\Dst\big(3\cdot\Dst(A)\big) = \Dst\big(\Dst(A) + \Dst(A) +
\Dst(A)\big)$ is given by:
$$\begin{array}{rcl}
\Nrm(\omega)
& = &
\frac{3}{8}\Bigket{\kappa_{0}(\nrm(\omega_{0}))} \;+\;
   \frac{5}{8}\Bigket{\kappa_{1}(\nrm(\omega_{1}))} \;+\;
   0\Bigket{\kappa_{1}(\nrm(\omega_{2}))}
\\[.6em]
& = &
\frac{3}{8}\Bigket{\kappa_{0}(\frac{\nicefrac{1}{8}}{\nicefrac{3}{8}}\ket{a}+
   \frac{\nicefrac{1}{4}}{\nicefrac{3}{8}}\ket{b})} \;+\;
   \frac{5}{8}\Bigket{\kappa_{1}(\frac{{\nicefrac{1}{2}}}{\nicefrac{5}{8}}\ket{c} + 
   \frac{{\nicefrac{1}{8}}}{{\nicefrac{5}{8}}}\ket{d})}.
\\[.6em]
& = &
\frac{3}{8}\Bigket{\kappa_{0}(\frac{1}{3}\ket{a}+\frac{2}{3}\ket{b})} \;+\;
   \frac{5}{8}\Bigket{\kappa_{1}(\frac{4}{5}\ket{c} + \frac{1}{5}\ket{d})}.
\end{array}$$

\noindent The outer distribution is a convex combination
$\frac{3}{8}\ket{-} + \frac{5}{8}\ket{-} + 0\ket{-}$ of inner
distributions, where the weights $\frac{3}{8} =
\frac{1}{8}+\frac{1}{4}$ and $\frac{5}{8} = \frac{1}{2} + \frac{1}{8}$
and $0$ are the normalisation factors for $\omega_{0}$ and
$\omega_{1}$ and $\omega_{2}$. Notice that the third term
$0\bigket{\kappa_{1}(\nrm(\omega_{2}))}$ in the above first line of
$\Nrm(\omega)$ disappears because of the weight $0$ upfront. This is
good news, because normalisation of the zero subdistribution
$\omega_2$ is not defined. Hence the hyper formulation deals with
undefinedness in a natural way: it disappears automatically.

We are now ready for the general description of hyper normalisation.

\begin{definition}
\label{NrmDef}
Let $A$ be a set, and $n$ be a natural number.  The hyper
normalisation map $\Nrm \colon \Dst\big(n\cdot A\big) \rightarrow
\Dst\big(n\cdot\Dst(A)\big)$ is defined as:
\begin{equation}
\label{NrmEqn}
\begin{array}{rcl}
\Nrm(\omega)
& = &
\displaystyle \sum_{\stackrel{\scriptstyle 0\leq i \leq n-1}{\omega[i]\neq 0}}
   \omega[i]\Bigket{\kappa_{i}\big(\textstyle
      \sum_{a\in A} \frac{\omega(\kappa_{i}a)}{\omega[i]}\ket{a}\big)}
\end{array}
\end{equation}

\noindent where:
$$\begin{array}{rclcrcl}
\omega[i]
& = &
\sum_{a}\omega(\kappa_{i}a)
& \qquad\mbox{so that}\qquad &
\sum_{i}\omega[i]
& = &
1.
\end{array}$$
\end{definition}

\noindent Notice that each inner distribution $\sum_{a\in A}
\frac{\omega(\kappa_{i}a)}{\omega[i]}\ket{a}$ in $\Dst(A)$ is the
normalisation~\eqref{TradNormEqn} of the subdistribution $\omega_{i} =
\sum_{a} \omega(\kappa_{i}a)\ket{a}$. It is well-defined, since
$\omega[i] = \sum_{a} \omega(\kappa_{i}a) \neq 0$ in the above formal
convex sum~\eqref{NrmEqn} and:
$$\begin{array}{rcccccl}
\sum_{a} \frac{\omega(\kappa_{i}a)}{\omega[i]}
& = &
\frac{\sum_{a} \omega(\kappa_{i}a)}{\omega[i]}
& = &
\frac{\omega[i]}{\omega[i]}
& = &
1.
\end{array}$$

\noindent For $n\leq 1$ the map $\Nrm \colon \Dst(n\cdot A)
\rightarrow \Dst(n\cdot \Dst(A))$ is trivial: if $n=0$, then $n\cdot A
= 0 = n\cdot \Dst(A)$, so that $\Nrm$ is the identity map on the empty
set $0 = \Dst(0)$. For $n=1$ we have $1 \cdot A \cong A$ and $1 \cdot
\Dst(A) \cong \Dst(A)$, so that the map $\Nrm \colon \Dst(1\cdot A)
\rightarrow \Dst(1\cdot \Dst(A))$ can be identified with the unit /
Dirac map $\eta\colon \Dst(A) \rightarrow \Dst(\Dst(A))$, sending
$\omega$ to $1\ket{\omega}$. We prefer not to exclude these trivial
border cases, to avoid unnecessary side conditions.

One can call a distribution $\omega\in\Dst(n\cdot A)$
\emph{normalised} if each $\kappa_{i}$ occurs at most once in
$\omega$. More formally, this can be expressed as $\Nrm(\omega) =
\Dst(n\cdot\eta)(\omega)$, so that $\Nrm(\omega)$ consists of point
distributions $r_{i}\ket{\kappa_{i}\eta(a)}$, for subexpressions
$r_{i}\ket{\kappa_{i}a}$ in $\omega$. The fact that $\Nrm(\omega)$ is
itself normalised occurs in point~\eqref{NrmLemIdempotent} below.

The hyper normalisation map $\Nrm$ is mathematically quite civilised:
it satisfies some basic equations, listed below. These equations are
formulated --- in categorical style --- in terms of commuting
diagrams, so that the relevant types are clearly visible.

\begin{lemma}
\label{NrmLem}
The hyper normalisation map $\Nrm$ from Definition~\ref{NrmDef} makes
the diagrams below commute.
\begin{enumerate}
\item \label{NrmLemInput} Normalising trivial input gives trivial
  output:
\begin{equation}
\label{NrmInputDiag}
\vcenter{\xymatrix{
\Dst(A)\ar[r]^-{\Dst(\kappa_{i})}\ar[d]_{\kappa_i} & \Dst(n\cdot A)\ar[d]^{\Nrm}
&
\Dst(n)\times A\ar[r]^-{\st_1}\ar[d]_{\idmap\times\eta} & \Dst(n\cdot A)\ar[d]^{\Nrm}
\\
n\cdot\Dst(A)\ar[r]_-{\eta} & \Dst(n\cdot\Dst(A))
&
\Dst(n)\times\Dst(A)\ar[r]_-{\st_1} & \Dst(n\cdot\Dst(A))
}}
\end{equation}

\item \label{NrmLemOutput} Destroying the output structure destroys
  normalisation:
\begin{equation}
\label{NrmOutputDiag}
\qquad\vcenter{\xymatrix{
\Dst(n\cdot A)\ar[r]^-{\Nrm}\ar[d]|{\Dst(n\cdot\bang) = \Dst(\pi_{1})} & 
   \Dst(n\cdot\Dst(A))\ar[d]|{\Dst(n\cdot\bang) = \Dst(\pi_{1})}
&
\Dst(n\cdot A)\ar[r]^-{\Nrm}\ar[d]|{\Dst(\nabla) = \Dst(\pi_{2})} & 
   \Dst(n\cdot\Dst(A))\ar[r]^-{\Dst(\nabla)} & \Dst(\Dst(A))\ar[d]^{\mu}
\\
\Dst(n)\ar@{=}[r] & \Dst(n)
&
\Dst(A)\ar@{=}[rr] & & \Dst(A)
}}
\end{equation}

\item \label{NrmLemIdempotent} Normalisation is idempotent:
\begin{equation}
\label{NrmIdempotentEtaDiag}
\vcenter{\xymatrix@C+.5pc{
\Dst(n\cdot A)\ar[r]^-{\Nrm}\ar[d]_{\Nrm} & 
   \Dst(n\cdot\Dst(A))\ar[d]^-{\Nrm}
\\
\Dst(n\cdot\Dst(A))\ar[r]_-{\Dst(n\cdot\eta)} &
   \Dst(n\cdot\Dst^{2}(A))
}}
\end{equation}

\noindent And thus:
\begin{equation}
\label{NrmIdempotentMuDiag}
\vcenter{\xymatrix{
\Dst(n\cdot A)\ar[r]^-{\Nrm}\ar[d]_{\Nrm} & 
   \Dst(n\cdot\Dst(A))\ar[r]^-{\Nrm} & 
   \Dst(n\cdot\Dst^{2}(A))\ar[d]^{\Dst(n\cdot\mu)}
\\
\Dst(n\cdot\Dst(A))\ar@{=}[rr] & & \Dst(n\cdot\Dst(A))
}}
\end{equation}

\item \label{NrmLemMonic} Normalisation can be undone: it has a left
  inverse (is a split mono):
\begin{equation}
\label{NrmMonicDiag}
\vcenter{\xymatrix@C+1pc{
\Dst(n\cdot A)\ar[r]^-{\Nrm}\ar@{=}[dr] & 
   \Dst(n\cdot\Dst(A))\ar[d]^-{(\st_{2})_{*}}
\\
& \Dst(n\cdot A)
}}
\end{equation}

\item \label{NrmLemNat} Normalisation is natural both for ordinary
  functions and for Kleisli maps: for all functions $f\colon A
  \rightarrow B$ and $g\colon A \rightarrow \Dst(B)$ the following
  diagram commutes.
\begin{equation}
\label{NrmNatDiag}
\vcenter{\xymatrix{
\Dst(n\cdot A)\ar[r]^-{\Nrm}\ar[d]_{\Dst(n\cdot f)} & 
   \Dst(n\cdot\Dst(A))\ar[d]^{\Dst(n\cdot \Dst(f))}
& &
\Dst(n\cdot A)\ar[r]^-{\Nrm}\ar[d]_{\qquad (n\cdot g)_{*}} & 
   \Dst(n\cdot\Dst(A))\ar[d]^{\Dst(n\cdot g_{*})}
\\
\Dst(n\cdot B)\ar[r]^-{\Nrm} & \Dst(n\cdot\Dst(B))
& &
\Dst(n\cdot B)\ar[r]^-{\Nrm} & \Dst(n\cdot\Dst(B))
}}
\end{equation}

\noindent We write $n\cdot g \colon n\cdot A \rightarrow \Dst(n\cdot
B)$ for the function $\kappa_{i}a \mapsto
\sum_{b}g(a)(b)\ket{\kappa_{i}b}$, that is, $n\cdot g = \st_{2} \after
(\idmap[n]\times g)$. Commutation of the first rectangle
in~\eqref{NrmNatDiag} follows from commutation of the second one, for
$g = \eta \after f$. But we prefer to make this special (first) case
explicit.
\end{enumerate}
\end{lemma}

\auxproof{
Abstractly, we have $n\cdot g = \st_{2} \after (\idmap\times g)$, so
that $n\cdot (\eta\after f) = \st_{2} \after (\idmap\times\eta) \after
(\idmap\times f) = \eta \after (\idmap\times f)$. Hence $\big(n\cdot
(\eta\after f)\big)_{*} = \Dst(n\cdot f)$ as in the diagram on the
left. Clearly we have $\Dst(n\cdot (\eta\after f)_{*}) =
\Dst(n\cdot\Dst(f))$.
}

\begin{myproof}
\begin{enumerate}
\item For the first diagram in~\eqref{NrmInputDiag} we have
for $\varphi\in\Dst(A)$,
$$\begin{array}{rcl}
\big(\Nrm \after \Dst(\kappa_{i})\big)(\varphi)
& = &
\Nrm\Big(\Dst(\kappa_{i})\big(\sum_{a}\varphi(a)\ket{a}\big)\Big) \\
& = &
\Nrm\Big(\sum_{a}\varphi(a)\ket{\kappa_{i}a}\Big) \\
& = &
1\bigket{\kappa_{i}(\sum_{a}\varphi(a)\ket{a})} \\
& = &
1\bigket{\kappa_{i}\varphi} \\
& = &
\big(\eta \after \kappa_{i}\big)(\varphi).
\end{array}$$

\noindent Commutation of the second diagram is obtained via:
$$\begin{array}{rcl}
\big(\Nrm \after \st_{1}\big)(\sum_{i}r_{i}\ket{i}, a)
& = &
\Nrm\big(\sum_{i} r_{i}\ket{\kappa_{i}a}\big) \\
& = &
\sum_{i}r_{i}\bigket{\kappa_{i}(1\ket{a})} \\
& = &
\st_{1}(\sum_{i}r_{i}\ket{i}, 1\ket{a}) \\
& = &
\big(\st_{1} \after (\idmap\times\eta)\big)(\sum_{i}r_{i}\ket{i}, a).
\end{array}$$

\item For the first diagram in~\eqref{NrmOutputDiag} we first note that:
$$\begin{array}{rcl}
\Dst(\pi_{1})(\omega)
& = &
\Dst(\pi_{1})\big(\sum_{i,a}\omega(\kappa_{i}a)\ket{\kappa_{i}a}\big) \\
& = &
\sum_{i,a}\omega(\kappa_{i}a)\ket{\pi_{1}(\kappa_{i}a)} \\
& = &
\sum_{i,a}\omega(\kappa_{i}a)\ket{i} \\
& = &
\sum_{i}(\sum_{a}\omega(\kappa_{i}a))\ket{i} \\
& = &
\sum_{i} \omega[i]\ket{i}.
\end{array}$$

\noindent Via a similar but simpler calculation one also gets
$\Dst(\pi_{1})\big(\Nrm(\omega)\big) = \sum_{i} \omega[i]\ket{i}$.

For the second diagram in~\eqref{NrmOutputDiag} we have:
$$\begin{array}{rcl}
\big(\mu \after \Dst(\pi_{2}) \after \Nrm\big)(\omega)
& = &
\big(\mu \after \Dst(\pi_{2})\big)\big(
   \sum_{i}\omega[i]\bigket{\kappa_{i}\big(\textstyle
      \sum_{a\in A} \frac{\omega(\kappa_{i}a)}{\omega[i]}\ket{a}\big)}\big) \\
& = &
\mu\big(\sum_{i}\omega[i]\bigket{\sum_{a\in A} 
   \frac{\omega(\kappa_{i}a)}{\omega[i]}\ket{a}}\big) \\
& \smash{\stackrel{\eqref{MuEqn}}{=}} &
\sum_{a} (\sum_{i} \omega[i]\cdot \frac{\omega(\kappa_{i}a)}{\omega[i]})
   \ket{a} \\
& = &
\sum_{a} (\sum_{i} \omega(\kappa_{i}a)) \ket{a} \\
& = &
\Dst(\pi_{2})\big(\sum_{a, i} \omega(\kappa_{i}a)\ket{\kappa_{i}a}\big) \\
& = &
\Dst(\pi_{2})(\omega).
\end{array}$$

\item Normalisation is idempotent since for $\omega \in \Dst(n\cdot
  A)$,
$$\begin{array}{rcl}
\big(\Nrm \after \Nrm\big)(\omega)
& = &
\Nrm\Big({\displaystyle\sum}_{i} \omega[i]\bigket{\kappa_{i}\big(
   \sum_{a} \frac{\omega(\kappa_{i}a)}{\omega[i]}\ket{a}\big)}\Big)
\\
& = &
{\displaystyle\sum}_{i} \omega[i]\Bigket{\kappa_{i}\big(1\bigket{
   \sum_{a} \frac{\omega(\kappa_{i}a)}{\omega[i]}\ket{a}}\big)}
\\
& = &
\Dst(n\cdot\eta)\Big({\displaystyle\sum}_{i} \omega[i]\bigket{\kappa_{i}\big(
   \sum_{a} \frac{\omega(\kappa_{i}a)}{\omega[i]}\ket{a}\big)}\Big)
\\
& = &
\big(\Dst(n\cdot\eta) \after \Nrm\big)(\omega).
\end{array}$$

\noindent Commutation of~\eqref{NrmIdempotentMuDiag} is now easy:
$$\begin{array}{rcl}
\Dst(n\cdot\mu) \after \Nrm \after \Nrm
& \smash{\stackrel{\eqref{NrmIdempotentEtaDiag}}{=}} &
\Dst(n\cdot\mu) \after \Dst(n\cdot\eta) \after \Nrm
\\
& = &
\Dst(n\cdot(\mu \after \eta)) \after \Nrm
\\
& = &
\Nrm
\end{array}$$

\item Recall from Definition~\ref{StrengthGraphDef} that
  $\st_{2}(\kappa_{i}\varphi) = \sum_{a}
  \varphi(a)\ket{\kappa_{i}a}$. Hence:
$$\begin{array}{rcl}
\big((\st_{2})_{*} \after \Nrm\big)(\omega)
& = &
\big(\mu \after \Dst(\st_{2})\big)\Big(
   {\displaystyle\sum}_{i} \omega[i]\bigket{\kappa_{i}\big(
   \sum_{a} \frac{\omega(\kappa_{i}a)}{\omega[i]}\ket{a}\big)}\Big)
\\
& = &
\mu\Big({\displaystyle\sum}_{i} \omega[i]\bigket{
   \sum_{a} \frac{\omega(\kappa_{i}a)}{\omega[i]}\ket{\kappa_{i}a}}\Big)
\\
& = &
{\displaystyle\sum}_{i,a} 
  \big(\omega[i]\cdot\frac{\omega(\kappa_{i}a)}{\omega[i]}\big)
  \ket{\kappa_{i}a}
\\
& = &
{\displaystyle\sum}_{i,a} \omega(\kappa_{i}a)\ket{\kappa_{i}a}
\\
& = &
\omega.
\end{array}$$

\item We only (need to) prove commutation of the diagram on the right
  in~\eqref{NrmNatDiag}. So let $g\colon A \rightarrow \Dst(B)$ be
  given. Then, for $\omega\in\Dst(n\cdot A)$,
$$\begin{array}[b]{rcl}
\big(\Nrm \after (n\cdot g)_{*}\big)(\omega)
& = &
\Nrm\big(\sum_{i, b} (\sum_{a} \omega(\kappa_{i}a)\cdot g(a)(b))
   \bigket{\kappa_{i}b}\big)
\\
& = &
{\displaystyle\sum}_{i} \omega[i]\Bigket{\kappa_{i}\big(
   \sum_{b}\frac{\sum_{a} \omega(\kappa_{i}a)\cdot g(a)(b)}{\omega[i]}
   \ket{b}\big)}
\\
& & \qquad\mbox{since } \begin{array}[t]{rcl}
   \lefteqn{\textstyle\sum_{b,a}\omega(\kappa_{i}a)\cdot g(a)(b)} \\
   & = &
   \sum_{a}\omega(\kappa_{i}a)\cdot (\sum_{b}g(a)(b)) \\
   & = &
   \sum_{a}\omega(\kappa_{i}a) \\
   & = &
   \omega[i]
   \end{array}
\\
& = &
{\displaystyle\sum}_{i} \omega[i]\Bigket{\kappa_{i}\big(g_{*}\big(
   \sum_{a}\frac{\omega(\kappa_{i}a)}{\omega[i]}\ket{a}\big)\big)}
\\
& = &
\Dst(n\cdot g_{*})\Big({\displaystyle\sum}_{i} \omega[i]\Bigket{
   \kappa_{i}\big(\sum_{a}\frac{\omega(\kappa_{i}a)}{\omega[i]}\ket{a}\big)}\Big)
\\
& = &
\big(\Dst(n\cdot g_{*}) \after \Nrm\big)(\omega).
\end{array}\eqno{\QEDbox}$$

\auxproof{
For naturality, let $f\colon A \rightarrow B$ be an arbitrary
  function. For $\omega \in \Dst(n\cdot A)$ we have:
$$\begin{array}[b]{rcl}
\big(\Dst(n\cdot\Dst(f)) \after \Nrm\big)(\omega)
& = &
\Dst(n\cdot\Dst(f))\big(
   \sum_{i}\omega[i]\bigket{\kappa_{i}\big(\textstyle
      \sum_{a\in A} \frac{\omega(\kappa_{i}a)}{\omega[i]}\ket{a}\big)}\big) \\
& = &
\sum_{i}\omega[i]\bigket{\kappa_{i}\big(\textstyle
      \sum_{a\in A} \frac{\omega(\kappa_{i}a)}{\omega[i]}\ket{f(a)}\big)} \\
& = &
\Nrm\big(\sum_{a,i}\omega(\kappa_{i}a)\ket{\kappa_{1}f(a)}\big) \\
& = &
\big(\Nrm \after \Dst(n\cdot f)\big)(\omega).
\end{array}\eqno{\QEDbox}$$
}
\end{enumerate}
\end{myproof}

We need the following auxiliary map for the subsequent next 
result about hyper normalisation.

\begin{definition}
\label{SprinkleDef}
Let $A$ be a set and $n\in\NNO$ an arbitrary number. We define a
`sprinkle' function
$$\vcenter{\xymatrix{
\Dst(n)\times \Dst(A)^{n}\ar[r]^-{\spr} & \Dst(A)
}}$$

\noindent by:
$$\begin{array}{rcccl}
\spr\big((r_{1}, \ldots, r_{n}), (\varphi_{1}, \ldots, \varphi_{n})\big)
& = &
\sum_{i}r_{i}\varphi_{i}
& = &
{\displaystyle\sum}_{a} (\sum_{i}r_{i}\cdot \varphi_{i}(a))\bigket{a}.
\end{array}$$
\end{definition}

This function $\spr$ thus sprinkles the convex $n$-tuple $r_{1},
\ldots, r_{n}$ of probabilities over the $n$-tuple of distributions
$\varphi_{1}, \ldots, \varphi_{n} \in \Dst(A)$, and produces a new
distribution over $A$, namely the convex sum of the $\varphi_i$.  This
works because the set $\Dst(A)$ is a convex set, in which such convex
sums $\sum_{i}r_{i}\varphi_{i}$ exist. More abstractly, the sprinkle
map can be obtained from strength followed by evaluation and
multiplication: $\Dst(n)\times \Dst(A)^{n} \rightarrow \Dst(n\times
\Dst(A)^{n}) \rightarrow \Dst(\Dst(A)) \rightarrow \Dst(A)$.

Our next result about normalisation is an equational
characterisation. It says that $\Nrm$ is the unique function
satisfying $\Nrm\big(\sum_{i}r_{i}\Dst(\kappa_{i})(\varphi_{i})\big) =
\sum_{i} r_{i}\ket{\kappa_{i}(\varphi_{i})}$. As before, this equation
is expressed in diagrammatic form.

\begin{theorem}
\label{NrmThm}
For each set $A$ and number $n$, the normalisation map $\Nrm$ is the
unique map $h\colon \Dst(n\cdot A) \rightarrow \Dst(n\cdot\Dst(A))$
making the following diagram commute.
$$\xymatrix@C+1pc{
\Dst(n)\times\Dst(A)^{n}
      \ar[r]^-{\idmap\times\tuple{\Dst(\kappa_{i})\after\pi_{i}}_{i}}
      \ar[d]_{\idmap\times\tuple{\kappa_{i}\after\pi_{i}}_i} &
   \Dst(n)\times\Dst(n\cdot A)^{n}\ar[r]^-{\spr} &
   \Dst(n\cdot A)\ar[d]^{h}
\\
\Dst(n)\times (n\cdot\Dst(A))^{n}\ar[r]_-{\idmap\times\eta^{n}} &
   \Dst(n)\times \Dst\big((n\cdot\Dst(A))\big)^{n}\ar[r]_-{\spr} & 
   \Dst(n\cdot\Dst(A))
}$$

\noindent where $\spr$ is the sprinkle map from
Definition~\ref{SprinkleDef}.
\end{theorem}

\begin{myproof}
We first show that the map $\Nrm$ as introduced in
Definition~\ref{NrmDef} makes the above rectangle commute.
$$\begin{array}{rcl}
\lefteqn{\big(\Nrm \after \spr \after 
      (\idmap\times\tuple{\Dst(\kappa_{i})\after\pi_{i}}_{i})\big)
   ((r_{1}, \ldots, r_{n}), (\varphi_{1}, \ldots, \varphi_{n}))} \\
& = &
\big(\Nrm \after \spr\big)((r_{1}, \ldots, r_{n}), 
   (\Dst(\kappa_{1})(\varphi_{1}), \ldots, \Dst(\kappa_{n})(\varphi_{n}))) \\
& = &
\Nrm\big(\sum_{i}r_{i}(\sum_{a}\varphi_{i}(a)\ket{\kappa_{i}a})\big) \\
& = &
\Nrm\big(\sum_{i,a}r_{i}\cdot\varphi_{i}(a)\ket{\kappa_{i}a}\big) \\
& = &
\sum_{i} r_{i}\bigket{\kappa_{i}(\sum_{a}\varphi_{i}(a)\ket{a})} \\
& = &
\sum_{i} r_{i}\bigket{\kappa_{i}(\varphi_{i})} \\
& = &
\spr\big((r_{1}, \ldots, r_{n}), 
   (1\ket{\kappa_{1}(\varphi_{1})}, \ldots, 1\ket{\kappa_{n}(\varphi_{n})})) \\
& = &
\big(\spr \after (\idmap\times\eta^{n}) \after 
   (\idmap\times\tuple{\kappa_{i}\after\pi_{i}}_{i})\big)
   ((r_{1}, \ldots, r_{n}), (\varphi_{1}, \ldots, \varphi_{n})).
\end{array}$$

Next, let $h\colon \Dst(n\cdot A) \rightarrow \Dst(n\cdot\Dst(A))$ make
the above diagram commute. Then, for $\omega\in\Dst(n\cdot A)$,
$$\begin{array}[b]{rcl}
h(\omega)
& = &
h\big(\sum_{a,i} \omega(\kappa_{i}a)\ket{\kappa_{i}a}\big) \\
& = &
h\big(\sum_{i, \omega[i]\neq 0} 
  \omega[i](\sum_{a}\frac{\omega(\kappa_{i}a)}{\omega[i]}\ket{\kappa_{i}a})\big) \\
& = &
h\big(\sum_{i, \omega[i]\neq 0} 
  \omega[i]\Dst(\kappa_{i})(\sum_{a}
    \frac{\omega(\kappa_{i}a)}{\omega[i]}\ket{a})\big) \\
& = &
\big(h \after \spr \after 
   (\idmap\times\tuple{\Dst(\kappa_{i})\after\pi_{i}}_{i})\big)
   ((\omega[1], \ldots, \omega[n]), (\varphi_{1}, \ldots, \varphi_{n})) \\
& & \qquad \mbox{where } {\begin{array}{rcl}
   \varphi_{i} & = & \left\{\begin{array}{ll}
     \sum_{a}\frac{\omega(\kappa_{i}a)}{\omega[i]}\ket{a} \quad &
         \mbox{if } \omega[i] \neq 0 \\
     \mbox{arbitrary} & \mbox{otherwise} \end{array}\right.\end{array}}
\\
& = &
\big(\spr \after (\idmap\times\eta) \after 
   (\idmap\times\tuple{\kappa_{i}\after\pi_{i}}_{i})\big)
   ((\omega[1], \ldots, \omega[n]), (\varphi_{1}, \ldots, \varphi_{n})) \\
& = &
\spr\big((\omega[1], \ldots, \omega[n]), 
   (1\ket{\kappa_{1}(\varphi_{1})}, \ldots, 1\ket{\kappa_{n}(\varphi_{n})})\big) \\
& = &
\sum_{i} \omega[i]\bigket{\kappa_{i}\varphi_{i}} \\
& = &
\sum_{i} \omega[i]\bigket{\kappa_{i}(\sum_{a}
   \frac{\omega(\kappa_{i}a)}{\omega[i]}\ket{a})} \\
& = &
\Nrm(\omega).
\end{array}\eqno{\QEDbox}$$
\end{myproof}

In Lemma~\ref{NrmLem}~\eqref{NrmLemNat} we have seen naturality of the
normalisation map $\Nrm \colon \Dst(n\cdot A) \rightarrow
\Dst(n\cdot\Dst(A))$ in the parameter $A$. But what about naturality
in the other parameter $n$? This also exists, but in more complicated
form.

\begin{lemma}
\label{NrmLemNNat}
For a Kleisli map $h\colon n \rightarrow \Dst(m)$ write $h\cdot A =
\st_{1} \after (h\times\idmap) \colon n\cdot A \rightarrow \Dst(m\cdot
A)$ for the map $\kappa_{i}a \mapsto \sum_{j}
h(i)(j)\ket{\kappa_{j}a}$. The following diagram then commutes.
\begin{equation}
\label{NrmNNatDiag}
\vcenter{\xymatrix@C-.5pc{
\Dst(n\cdot A)\ar[d]_{(h\cdot A)_{*}}\ar[r]^-{\Nrm} &
   \Dst(n\cdot\Dst(A))\ar[rr]^-{(h\cdot\Dst(A))_{*}} & &
   \Dst(m\cdot\Dst(A))\ar[r]^-{\Nrm} &
   \Dst(m\cdot\Dst^{2}(A))\ar[d]^-{\Dst(m\cdot\mu)}
\\
\Dst(m\cdot A)\ar[rrrr]_-{\Nrm} & & & & \Dst(m\cdot\Dst(A))
}}
\end{equation}

\auxproof{
\noindent We now write $\gr(h_{1}) \colon n\cdot A \rightarrow
\Dst(m\cdot (n\cdot A))$, where $h_{1} = h \after \pi_{1} \colon
n\cdot A \rightarrow \Dst(m)$, for the map $\kappa_{i}a \mapsto
\sum_{j} f(i)(j) \ket{\kappa_{j}\kappa_{i}a}$. The following diagram
then also commutes.
\begin{equation}
\label{NrmGrNatDiag}
\vcenter{\xymatrix@C-1.1pc@R-.5pc{
\Dst(n\cdot A)\ar[dd]_{\gr(h_{1})_{*}}\ar[r]^-{\Nrm} &
   \Dst(n\cdot\Dst(A))\ar[rr]^-{\gr(h_{1})_{*}} & &
   \Dst(m\cdot(n\cdot\Dst(A)))\ar[rrr]^-{\Dst(m\cdot [\Dst(\kappa_{i})]_{i})} 
& & &
   \Dst(m\cdot\Dst(n\cdot A))\ar[d]^-{\Nrm}
\\
& & & & & & \Dst(m\cdot\Dst^{2}(n\cdot A))\ar[d]^-{\Dst(m\cdot\mu)}
\\
\Dst(m\cdot (n\cdot A))\ar[rrrrrr]_-{\Nrm} & & & & & & 
   \Dst(m\cdot\Dst(n\cdot A))
}}
\end{equation}
}
\end{lemma}

\begin{myproof}
For $\omega\in\Dst(n\cdot A)$ we compute:
$$\begin{array}[b]{rcl}
\lefteqn{\big(\Dst(m\cdot\mu) \after \Nrm \after (h\cdot\Dst(A))_{*}
   \after \Nrm\big)(\omega)} 
\\
& = &
\big(\Dst(m\cdot\mu) \after \Nrm \after (h\cdot\Dst(A))_{*}\big)\Big(
   {\displaystyle\sum}_{i} \omega[i]\bigket{\kappa_{i}\big(
      \sum_{a}\frac{\omega(\kappa_{i}a)}{\omega[i]}\ket{a}\big)}\Big)
\\
& = &
\big(\Dst(m\cdot\mu) \after \Nrm\big)\Big(
   {\displaystyle\sum}_{i,j} h(i)(j)\cdot\omega[i]
      \bigket{\kappa_{j}\big(
      \sum_{a}\frac{\omega(\kappa_{i}a)}{\omega[i]}\ket{a}\big)}\Big)
\\
& = &
\Dst(m\cdot\mu)\Big(
   {\displaystyle\sum}_{j} \big(\sum_{i}h(i)(j)\cdot\omega[i]\big)
      \bigket{\kappa_{j}\big(
      \sum_{i}\frac{h(i)(j)\cdot\omega[i]}{\sum_{i}h(i)(j)\cdot\omega[i]}
      \cdot\big(\sum_{a}\frac{\omega(\kappa_{i}a)}{\omega[i]}\ket{a}\big)\big)}\Big)
\\
& = &
{\displaystyle\sum}_{j} \big(\sum_{i}h(i)(j)\cdot\omega[i]\big)
      \bigket{\kappa_{j}\big(
      \sum_{a}\big(\sum_{i}\frac{h(i)(j)\cdot\omega[i]}{\sum_{i}h(i)(j)\cdot\omega[i]}
      \cdot\frac{\omega(\kappa_{i}a)}{\omega[i]}\big)\ket{a}\big)}
\\
& = &
{\displaystyle\sum}_{j} \big(\sum_{i}h(i)(j)\cdot\omega[i]\big)
      \bigket{\kappa_{j}\big(
      \sum_{a}\frac{\sum_{i}h(i)(j)\cdot\omega(\kappa_{i}a)}
                  {\sum_{i}h(i)(j)\cdot\omega[i]}\ket{a}\big)}
\\
& = &
{\displaystyle\sum}_{j} \big(\sum_{i,a} h(i)(j)\cdot \omega(\kappa_{i}a)\big)
   \bigket{\kappa_{j}\big(\sum_{a} 
   \frac{\sum_{i} h(i)(j)\cdot \omega(\kappa_{i}a)}
        {\sum_{i,a} h(i)(j)\cdot \omega(\kappa_{i}a)}\ket{a}\big)}
\\
& = &
\Nrm\big(\sum_{j,a} \big(\sum_{i}h(i)(j)\cdot \omega(\kappa_{i}a)\big)
   \bigket{\kappa_{j}a}\big)
\\
& = &
\big(\Nrm \after (h\cdot A)_{*}\big)(\omega).
\end{array}\eqno{\QEDbox}$$

\auxproof{
\noindent Commutation of the second rectangle~\eqref{NrmGrNatDiag} is
obtained in a similar way:
$$\begin{array}[b]{rcl}
\lefteqn{\big(\Dst(m\cdot\mu) \after \Nrm \after 
   \Dst(m\cdot [\Dst(\kappa_{i})]_{i}) \after \gr(h_{1})_{*} \after \Nrm\big)
   (\omega)} 
\\
& = &
\big(\Dst(m\cdot\mu) \after \Nrm \after 
   \Dst(m\cdot [\Dst(\kappa_{i})]_{i}) \after \gr(h_{1})_{*}\big)\Big(
   {\displaystyle\sum}_{i} \omega[i]\bigket{\kappa_{i}\big(
      \sum_{a}\frac{\omega(\kappa_{i}a)}{\omega[i]}\ket{a}\big)}\Big)
\\
& = &
\big(\Dst(m\cdot\mu) \after \Nrm \after 
   \Dst(m\cdot [\Dst(\kappa_{i})]_{i})\big)\Big(
   {\displaystyle\sum}_{i,j} h(i)(j)\cdot\omega[i]
      \bigket{\kappa_{j}\kappa_{i}\big(
      \sum_{a}\frac{\omega(\kappa_{i}a)}{\omega[i]}\ket{a}\big)}\Big)
\\
& = &
\big(\Dst(m\cdot\mu) \after \Nrm\big)\Big(
   {\displaystyle\sum}_{i,j} h(i)(j)\cdot\omega[i]\bigket{\kappa_{j}\big(
      \sum_{a}\frac{\omega(\kappa_{i}a)}{\omega[i]}\ket{\kappa_{i}a}\big)}\Big)
\\
& = &
\Dst(m\cdot\mu)\Big(
   {\displaystyle\sum}_{j} \big(\sum_{i}h(i)(j)\cdot\omega[i]\big)
      \bigket{\kappa_{j}\big(\sum_{i} 
      \frac{h(i)(j)\cdot\omega[i]}{\sum_{i}h(i)(j)\cdot\omega[i]}
      \sum_{a}\frac{\omega(\kappa_{i}a)}{\omega[i]}\ket{\kappa_{i}a}\big)}\Big)
\\
& = &
{\displaystyle\sum}_{j} \big(\sum_{i}h(i)(j)\cdot\omega[i]\big)
      \bigket{\kappa_{j}\big(\sum_{i,a}  
      \frac{h(i)(j)\cdot\omega[i]}{\sum_{i}h(i)(j)\cdot\omega[i]}
      \cdot\frac{\omega(\kappa_{i}a)}{\omega[i]}\ket{\kappa_{i}a}\big)}\Big)
\\
& = &
{\displaystyle\sum}_{j} \big(\sum_{i}h(i)(j)\cdot\omega[i]\big)
      \bigket{\kappa_{j}\big(\sum_{i,a}  
      \frac{h(i)(j)\cdot\omega(\kappa_{i}a)}{\sum_{i}h(i)(j)\cdot\omega[i]}
      \ket{\kappa_{i}a}\big)}\Big)
\\
& = &
\Nrm\big(\sum_{i,j,a} h(i)(j)\cdot \omega(\kappa_{i}a)
   \ket{\kappa_{j}\kappa_{i}a}\big)
\\
& = &
\big(\Nrm \after \gr(h_{1})_{*}\big)(\omega).
\end{array}\eqno{\QEDbox}$$
}
\end{myproof}

\begin{remark}
\label{NrmNonAffineRem}
Normalisation $\Nrm \colon \Dst(n\cdot A) \rightarrow
\Dst(n\cdot\Dst(A))$ is \emph{not} an affine map, that is, it does not
preserve convex combinations. We describe a simple counterexample, for
$A = \{a,b\}$ and $n=2$.
$$\begin{array}{rcl}
\frac{1}{4}\Nrm(1\ket{\kappa_{0}a}) + \frac{3}{4}\Nrm(1\ket{\kappa_{0}b})
& = &
\frac{1}{4}\big(1\ket{\kappa_{0}(1\ket{a})}\big) + 
   \frac{3}{4}\big(1\ket{\kappa_{0}(1\ket{b})}\big)
\\
& = &
\frac{1}{4}\ket{\kappa_{0}(1\ket{a})} + \frac{3}{4}\ket{\kappa_{0}(1\ket{b})}
\\
\Nrm\big(\frac{1}{4}(1\ket{\kappa_{0}a}) + \frac{3}{4}(1\ket{\kappa_{0}b})\big)
& = &
\Nrm\big(\frac{1}{4}\ket{\kappa_{0}a} + \frac{3}{4}\ket{\kappa_{0}b}\big)
\\
& = &
1\ket{\kappa_{0}(\frac{1}{4}\ket{a} + \frac{3}{4}\ket{b})}.
\end{array}$$
\end{remark}

One may ask how pointwise normalisation from
Proposition~\ref{PointNormProp} and hyper normalisation are
related. This requires some preparatory work, where we use a `twisted'
version of Proposition~\ref{PointNormProp}, using the twist map $\tw =
\tuple{\pi_{2}, \pi_{1}}$. For a distribution $\omega\in\Dst(n\cdot
A)$ we write $\omega_{\tw} = \Dst(\tw)(\omega) \in \Dst(A\times n)$
and assume that $\omega_{1} = \Dst(\pi_{1})(\omega) =
\Dst(\pi_{2})(\omega_{\tw}) \in \Dst(n)$ satisfies $\supp(\omega_{1})
= n$. Notice that $\omega_{1}(i) = \sum_{a}\omega(\kappa_{i}a) =
\omega[i]$, as introduced in Definition~\ref{NrmDef}. Via
Proposition~\ref{PointNormProp} we can write $\omega_{\tw} =
\gr(f)_{*}(\omega_{1})$, for the unique conditional $f\colon n
\rightarrow \Dst(A)$. This map $f$ is, basically as described
in~\eqref{PointNormDown}:
\begin{equation}
\label{HyperPointCondEqn}
\begin{array}{rcccccl}
f(i)
& = &
\nrm(\omega_{i})
& = &
{\displaystyle\sum}_{a} \frac{\omega(\kappa_{i}a)}{\sum_{a}\omega(\kappa_{i}a)}
   \bigket{a} 
& = &
{\displaystyle\sum}_{a} \frac{\omega(\kappa_{i}a)}{\omega[i]}\bigket{a} 
\end{array}
\end{equation}

\noindent We see that these $f(i)$'s are the normalised `inner'
distributions occurring in the formula for $\Nrm(\omega)$ in
Definition~\ref{NrmDef}. The next result describes this situation in a
precise manner.

\begin{proposition}
\label{HyperPointProp}
Let $\omega\in\Dst(n\cdot A)$ be a distribution whose twisted version
$\omega_{\tw} = \Dst(\tw)(\omega) \in \Dst(A\times n)$ has
conditional $f\colon n \rightarrow \Dst(A)$, so that:
$$\begin{array}{rclcrcl}
\omega_{\tw}
& = &
\gr(f)_{*}(\omega_{1})
& \qquad\mbox{where}\qquad &
\omega_{1}
& = &
\Dst(\pi_{1})(\omega).
\end{array}$$

\noindent The hyper normalisation $\Nrm(\omega) \in
\Dst(n\cdot\Dst(A))$ of $\omega$ can then be described via the adapted
conditional $\eta \after f\colon n \rightarrow \Dst(\Dst(A))$ as:
$$\begin{array}{rcl}
\Nrm(\omega)_{\tw}
& = &
\gr(\eta \after f)_{*}(\omega_{1}) \,\in\, \Dst(\Dst(A)\times n).
\end{array}$$
\end{proposition}

\begin{myproof}
We first notice that the graph function $\gr(\eta \after f) \colon n
\rightarrow \Dst(\Dst(A)\cdot n)$ is given by $\gr(\eta \after f)(i) =
1\ket{f(i),i}$, see Definition~\ref{StrengthGraphDef}. Then:
$$\begin{array}[b]{rcl}
\gr(\eta \after f)_{*}(\omega_{1})
& \smash{\stackrel{\eqref{KlLiftEqn}}{=}} &
{\displaystyle\sum}_{i,\varphi}\,
   \big(\sum_{j} \omega_{1}(j) \cdot 
      \gr(\eta\after f)(j)(\varphi,i)\big)\bigket{\kappa_{i}\varphi}
\\
& = &
{\displaystyle\sum}_{i,\varphi}\, \omega[i] \bigket{f(i),i}
\\
& \smash{\stackrel{\eqref{HyperPointCondEqn}}{=}} &
{\displaystyle\sum}_{i}\, \omega[i] \bigket{\tw\big(\kappa_{i}\big(
   \sum_{a} \frac{\omega(\kappa_{i}a)}{\omega[i]}\bigket{a}\big)\big)}
\\
& \smash{\stackrel{\eqref{NrmEqn}}{=}} &
\Dst(\tw)\big(\Nrm(\omega)\big)
\\
& = &
\Nrm(\omega)_{\tw}.
\end{array}\eqno{\QEDbox}$$
\end{myproof}

\subsection{Comparison to other formulations of normalisation}\label{NrmCompareSubsec}

We briefly compare our `hyper' approach to normalisation to other
approaches. First, in~\cite{JacobsWW15a} normalisation is defined for
non-zero subdistributions. A subdistribution on $A$ is a subconvex
combination $\sum_{i}r_{i}\ket{a_i}$ with $\sum_{i}r_{i} \leq 1$. It
may be identified with a distribution $\omega\in\Dst(A+1)$ on $A+1$,
of the form $\sum_{i\leq n}r_{i}\ket{\kappa_{1}a_i} +
r_{n+1}\ket{\kappa_{2}0}$, where $r_{n+1} = 1 - (\sum_{i}r_{i})$ is
the `one-deficit' capturing the probability of non-termination; see
also~\cite{McIverMS96,McIverM04}.  This $\omega$ is non-zero if
$r_{n+1}\neq 1$. In that case we can normalise it to $\sum_{i}
\frac{r_{i}}{1-r_{n+1}}\ket{a_i}$. This process is described
abstractly in~\cite{JacobsWW15a}.

We sketch how it fits in the current setting. We first map
$\omega\in\Dst(A+1)$ to the distribution $\omega' \in
\Dst\big((A+1)+(A+1)\big)$, given by $\omega' =
\Dst(\kappa_{1}+\kappa_{2})(\omega)$. Applying our hyper normalisation
operation $\Nrm \colon \Dst\big((A+1)+(A+1)\big) \rightarrow \Dst\big(
\Dst(A+1)+\Dst(A+1)\big)$ yields $\Nrm(\omega')$ of the form:
$$\textstyle
   (1-r_{n+1})\Bigket{\kappa_{1}\big(\frac{r_1}{1-r_{n+1}}\ket{\kappa_{1}a_{1}} 
   + \cdots +
   \frac{r_n}{1-r_{n+1}}\ket{\kappa_{1}a_{n}}\big)} 
   + r_{n+1}\Bigket{\kappa_{2}\big(1\ket{\kappa_{2}0}\big)}.$$

\noindent The normalised distribution now appears as the first inner
component.

Second, in~\cite{StatonYHKW16} normalisation is defined wrt.\ a
`score'.  We slightly adapt its description, so that it fits in the
current setting. Normalisation like in~\cite{StatonYHKW16} can then be
described as a partial function $\Dst([0,1]\times A) \rightarrow
\Dst(A)$.  The number in $[0,1]$ in the input type $\Dst([0,1]\times
A)$ is called the score. It is a non-negative real number
in~\cite{StatonYHKW16}, but here we restrict it to the unit interval.
It allows us to massage the input type via a strength map, so that it
becomes a subdistribution that can be normalised, as above. We use
that $[0,1] \cong \Dst(2)$ in:
$$\xymatrix@R-1.5pc@C+.5pc{
\Dst\big([0,1]\times A\big)\ar@{=}[d]^{\wr}
\\
\Dst\big(\Dst(2)\times A\big)\ar[r]^-{\Dst(\st_{1})} &
   \Dst(\Dst(2\times A)\big)\ar@{=}[d]^{\wr}
\\
& \Dst\big(\Dst(A+A)\big)\ar[r]^-{\mu} & 
   \Dst(A+A)\ar[r]^-{\Dst(\idmap+\bang)} & \Dst(A+1)
}$$

\noindent This form of normalisation sends a distribution
$\sum_{i}r_{i}\ket{(s_{i}, a_{i})} \in \Dst\big([0,1]\times A\big)$,
with scores $s_i$, to $\sum_{i} \frac{r_{i}\cdot
  s_{i}}{\sum_{i}r_{i}\cdot s_{i}}\ket{a_i} \in \Dst(A)$. It is only
defined if $\sum_{i}r_{i}\cdot s_{i} \neq 0$.

\auxproof{
Explicitly, these steps do the following.
$$\begin{array}{rcl}
\sum_{i}r_{i}\ket{(s_{i}, a_{i})}
& \longmapsto &
\sum_{i}r_{i}\ket{(s_{i}\ket{1} + (1-s_{i})\ket{0}, a_{i})} \\
& \longmapsto &
\sum_{i}r_{i}\bigket{s_{i}\ket{1,a_{i}} + (1-s_{i})\ket{0, a_{i}}} \\
& \longmapsto &
\sum_{i}r_{i}\bigket{s_{i}\ket{\kappa_{1}a_{i}} + (1-s_{i})\ket{\kappa_{2}a_{i}}} \\
& \longmapsto &
\sum_{i}r_{i}\cdot s_{i}\ket{\kappa_{1}a_{i}} + 
   \sum_{i}r_{i}\cdot (1-s_{i})\ket{\kappa_{2}a_{i}} \\
& \longmapsto &
\sum_{i}r_{i}\cdot s_{i}\ket{\kappa_{1}a_{i}} + 
   (\sum_{i}r_{i}\cdot (1-s_{i}))\ket{\kappa_{2}0}.
\end{array}$$

\noindent Normalisation, if succesful then yields the distribution
$\sum_{i} \frac{r_{i}\cdot s_{i}}{1 - \sum_{i}r_{i}\cdot
  (1-s_{i})}\ket{a_i}$. Notice that the divisor satisfies:
$$\begin{array}{rcccl}
1 - \sum_{i}r_{i}\cdot (1-s_{i})
& = &
1 - \sum_{i}r_{i} + \sum_{i}r_{i}\cdot s_{i}
& = &
\sum_{i}r_{i}\cdot s_{i}.
\end{array}$$
}

\section{Normalisation as distributive law}\label{DistributiveSec}

This section is meant for the categorically proficient reader, knowing
about (co)monads and distributive laws --- see
\textit{e.g.}~\cite{Jacobs14d} for more information. It can be skipped
safely, since it presents only a categorical curiosity. This section
shows that hyper normalisation $\Nrm$ forms a distributive law,
between a comonad and a functor. It is not a distributive law between
two comonads, since one of the counit laws fails to hold --- whereas
the corresponding comultiplication law does hold.

The standard adjunction $\Kl(\Dst) \rightleftarrows \Sets$ between a
Kleisli category and its underlying category induces a comonad on
$\Kl(\Dst)$, which shall write as $\overline{\Dst}$.  On objects it is
given by $X \mapsto \Dst(X)$. It sends a map $f\colon X \rightarrow
\Dst(Y)$ to $\overline{\Dst}(f) = \eta \after f_{*} = \eta \after \mu
\after \Dst(f) \colon \Dst(X) \rightarrow \Dst^{2}(Y)$. The counit
$\varepsilon \colon \overline{\Dst}(X) \klto X$ is the identity map
$\Dst(X) \rightarrow \Dst(X)$. The comultiplication $\delta \colon
\overline{\Dst}(X) \klto \overline{\Dst}^{2}(X)$ is $\eta\after\eta =
\Dst(\eta) \after \eta$.

\auxproof{
We have:
$$\begin{array}{rcl}
\overline{\Dst}(\idmap)
& = &
\eta \after \mu \after \Dst(\eta)
\\
& = &
\eta
\\
& = &
\idmap
\\
\overline{\Dst}(g) \klafter \overline{\Dst}(f)
& = &
\mu \after \Dst(\eta \after \mu \after \Dst(g)) \after \eta \after \mu
   \after \Dst(f)
\\
& = &
\mu \after \eta \after \eta \after \mu \after \Dst(g) \after \mu
   \after \Dst(f)
\\
& = &
\eta \after \mu \after \mu \after \Dst^{2}(g) \after \Dst(f)
\\
& = &
\eta \after \mu \after \Dst(\mu \after \Dst(g) \after f)
\\
& = &
\overline{\Dst}(g \klafter f)
\end{array}$$
}

For each $n\in\mathbb{N}$ the $n$-fold copower $n\cdot (-)$ is a
comonad on a category with finite coproducts. This is also the case on
the Kleisli category $\Kl(\Dst)$. In this case we describe it with a
star $n\ast (-)$, to distinguish it from $n\cdot (-)$ on $\Sets$. For
a map $f\colon X \rightarrow \Dst(Y)$ we get $n\ast f \colon n\ast X
\rightarrow \Dst(n\ast Y)$ given by $\st_{2} \after (\idmap[n]\times
f)$. Explicitly, $(n\ast f)(\kappa_{i}x) = \sum_{y}
f(x)(y)\ket{\kappa_{i}y}$. The counit $\varepsilon \colon n\ast X
\klto X$ is $\eta \after \nabla$. The comultiplication $\delta \colon
n\ast X \rightarrow n\ast (n\ast X)$ is the map $\eta \after
(\kappa_{1}+\cdots+\kappa_{n}) = [\eta \after \kappa_{i} \after
  \kappa_{i}]_{i}$.

Thus we are looking at a situation:
\begin{equation}
\label{ComonadDiag}
\vcenter{\xymatrix@R-.5pc@C-1pc{
& & & \\
& \Kl(\Dst)\ar[d]_{\dashv}
   \ar `u[r] `[rd] `[u]^{n\ast(-)} []
   \ar `u[l] `[ld] `[u]_{\overline{\Dst}} [] & \\
& \Sets\ar@/^3ex/[u]
   \ar `l[ld] `[u]_{\Dst} `[lr] [] 
   & \\
& & &
}}\quad\mbox{where}\qquad
\xymatrix{
\overline{\Dst}(n\ast(-))\ar@{=>}[r]^-{\Nrm} & n\ast\overline{\Dst}(-)
}
\end{equation}

The normalisation operation $\Nrm$ is a map $\Nrm_{A} \colon
\overline{\Dst}(n\ast A) \klto n\ast\overline{\Dst}(A)$ in
$\Kl(\Dst)$. It is natural by~\eqref{NrmNatDiag}, since for a map
$f\colon A \klto B$ in $\Kl(\Dst)$,
$$\begin{array}{rcl}
(n\ast \overline{\Dst})(f) \klafter \Nrm_{A}
& = &
\mu \after \Dst(\st_{2} \after (\idmap\times \overline{\Dst}(f))) \after \Nrm
\\
& = &
\mu \after \Dst(\st_{2} \after (\idmap\times \eta) \after
   (\idmap\times f_{*})) \after \Nrm
\\
& = &
\mu \after \Dst(\eta \after  (\idmap\times f_{*})) \after \Nrm
\\
& = &
\Dst(\idmap\times f_{*}) \after \Nrm
\\
& \smash{\stackrel{\eqref{NrmNatDiag}}{=}} &
\Nrm_{B} \after (n\ast f)_{*}
\\
& = &
\mu \after \eta \after \Nrm_{B} \after (n\ast f)_{*}
\\
& = &
\mu \after \Dst(\Nrm_{B}) \after \eta \after (n\ast f)_{*}
\\
& = &
\Nrm_{B} \klafter \overline{\Dst}(n\ast f).
\end{array}$$

The normalisation map $\Nrm$ commutes appropriately with the
comultiplication maps of the two comonads, as expressed in
the two rectangles:
$$\xymatrix{
\overline{\Dst}(n\ast A)\ar[r]|-{\bullet}^-{\delta_{n\ast A}}
   \ar[dd]|-{\bullet}_{\Nrm_A}
   & \overline{\Dst}^{2}(n\ast A)
     \ar[d]|-{\bullet}^{\overline{\Dst}(\Nrm_{A})}
& &
\overline{\Dst}(n\ast A)\ar[r]|-{\bullet}^-{\overline{\Dst}(\delta_{A})}
   \ar[dd]|-{\bullet}_{\Nrm_A}
   & \overline{\Dst}(n\ast (n\ast A))\ar[d]|-{\bullet}^{\Nrm_{n\ast A}}
\\
& \overline{\Dst}(n\ast \overline{\Dst}(A))
   \ar[d]|-{\bullet}^{\Nrm_{\overline{\Dst}(A)}}
& &
& n\ast \overline{\Dst}(n\ast A)\ar[d]|-{\bullet}^{n\ast\Nrm_{A}}
\\
n\ast\overline{\Dst}(A)\ar[r]|-{\bullet}_-{n\ast\delta_{A}} &
  n\ast\overline{\Dst}^{2}(A)
& &
n\ast\overline{\Dst}(A)\ar[r]|-{\bullet}_-{\delta_{\overline{\Dst}(A)}} &
   n\ast (n\ast\overline{\Dst}(A))
}$$

\noindent Commutation of the diagram on the left follows
from~\eqref{NrmIdempotentEtaDiag}:
$$\begin{array}{rcl}
\lefteqn{\Nrm_{\overline{\Dst}(A)} \klafter \overline{\Dst}(\Nrm_{A}) \klafter
    \delta_{n\ast A}}
\\
& = &
\mu \after \Dst(\Nrm) \after \mu \after \Dst(\eta \after \mu \after
   \Dst(\Nrm)) \after \eta \after \eta
\\
& = &
\mu \after \Dst(\Nrm \after \mu \after \Dst(\Nrm)) \after \eta \after \eta
\\
& = &
\mu \after \eta \after \Nrm \after \mu \after \Dst(\Nrm) \after \eta
\\
& = &
\Nrm \after \mu \after \eta \after \Nrm
\\
& = &
\Nrm \after \Nrm
\\
& \smash{\stackrel{\eqref{NrmIdempotentEtaDiag}}{=}} &
\Dst(\idmap\times \eta) \after \Nrm
\\
& = &
\mu \after \Dst(\eta \after (\idmap\times\eta)) \after \Nrm
\\
& = &
\mu \after \Dst(\st_{2} \after (\idmap\times(\eta\after\eta))) \after \Nrm
\\
& = &
(n\ast\delta_{A}) \klafter \Nrm_{A}.
\end{array}$$

\noindent The above diagram on the right requires more work:
$$\begin{array}{rcl}
\lefteqn{\big(n\ast\Nrm_{A} \klafter \Nrm_{n\ast A} \klafter 
   \overline{\Dst}(\delta_{A})\big)} \\
& = &
\mu \after \Dst(n\ast\Nrm_{A}) \after \mu \after \Dst(\Nrm_{n\ast A})
   \after \eta \after \mu \after 
   \Dst(\eta \after [\kappa_{i}\after\kappa_{i}]_{i}) 
\\
& = &
\mu \after \Dst(n\ast\Nrm_{A}) \after \mu \after 
   \eta \after \Nrm_{n\ast A} \after \Dst([\kappa_{i}\after\kappa_{i}]_{i})
\\
& = &
\mu \after \Dst(n\ast\Nrm_{A}) \after \Nrm_{n\ast A} \after 
   \Dst([\kappa_{i}\after\kappa_{i}]_{i})
\\
& \smash{\stackrel{(*)}{=}} &
\Dst([\kappa_{i}\after\kappa_{i}]_{i}) \after \Nrm_{A}
\\
& = &
\mu \after \Dst([\eta \after \kappa_{i}\after\kappa_{i}]_{i}) \after \Nrm_{A}
\\
& = &
\delta_{\overline{\Dst}(A)} \klafter \Nrm_{A}
\end{array}$$

\noindent We explicitly prove the marked equation:
$$\begin{array}{rcl}
\lefteqn{\big(\mu \after \Dst(n\ast\Nrm_{A}) \after \Nrm_{n\ast A} \after 
   \Dst([\kappa_{i}\after\kappa_{i}]_{i})\big)(\omega)}
\\
& = &
\big(\mu \after \Dst(n\ast\Nrm_{A}) \after \Nrm_{n\ast A}\big)\big(
   \sum_{i,a} \omega(\kappa_{i}a)\ket{\kappa_{i}\kappa_{i}a}\big)
\\
& = &
\big(\mu \after \Dst(n\ast\Nrm_{A})\big)\Big(
   {\displaystyle\sum}_{i} \omega[i]\bigket{\kappa_{i}\big(
      \sum_{a}\frac{\omega(\kappa_{i}a)}{\omega[i]}\ket{\kappa_{i}a}\big)}\Big)
\\
& = &
\mu\Big({\displaystyle\sum}_{i} \omega[i]\bigket{\sum_{\varphi}
    \Nrm_{A}\big(\sum_{a}\frac{\omega(\kappa_{i}a)}{\omega[i]}
       \ket{\kappa_{i}a}\big)(\varphi)\ket{\kappa_{i}\varphi}}\Big)
\\
& = &
\mu\Big({\displaystyle\sum}_{i} \omega[i]\bigket{1\bigket{\kappa_{i}
    \kappa_{i}\big(\sum_{a}\frac{\omega(\kappa_{i}a)}{\omega[i]}
       \ket{a}\big)}}\Big)
\\
& = &
{\displaystyle\sum}_{i} \omega[i]\bigket{\kappa_{i}
    \kappa_{i}\big(\sum_{a}\frac{\omega(\kappa_{i}a)}{\omega[i]}
       \ket{a}\big)}
\\
& = &
\Dst([\kappa_{i}\after\kappa_{i}]_{i})\Big(
   {\displaystyle\sum}_{i} \omega[i]\bigket{\kappa_{i}\big(
      \sum_{a}\frac{\omega(\kappa_{i}a)}{\omega[i]}\ket{a}\big)}\Big)
\\
& = &
\big(\Dst([\kappa_{i}\after\kappa_{i}]_{i}) \after \Nrm_{A}\big)(\omega).
\end{array}$$

Commutation of $\Nrm$ with the two counits is expressed in the
diagrams:
$$\xymatrix{
\overline{\Dst}(n\ast A)\ar[r]|-{\bullet}^-{\Nrm}
   \ar[d]|-{\bullet}_{\varepsilon_{n\ast A}}
   & n\ast\overline{\Dst}(A)\ar[d]|-{\bullet}^{n\ast\varepsilon_{A}}
& &
\overline{\Dst}(n\ast A)\ar[r]|-{\bullet}^-{\Nrm}
   \ar[d]|-{\bullet}_{\overline{\Dst}(\varepsilon_{A})}
   & n\ast\overline{\Dst}(A)\ar[d]|-{\bullet}^{\varepsilon_{\overline{\Dst}(A)}}
\\
n\ast A\ar@{=}[r] & n\ast A
& &
\overline{\Dst}(A)\ar@{=}[r] & \overline{\Dst}(A)
}$$

\noindent The diagram on the left commutes:
$$\begin{array}{rcl}
n\ast\varepsilon_{A} \klafter \Nrm
& = &
\mu \after \Dst(\st_{2} \after (\idmap\times\idmap)) \after \Nrm
\\
& = &
(\st_{2})_{*}(\Nrm)
\\
& \smash{\stackrel{\eqref{NrmMonicDiag}}{=}} &
\idmap
\\
& = &
\varepsilon_{n\ast A}.
\end{array}$$

\noindent Somewhat surprisingly, the above rectangle on the right does
not commute, despite~\eqref{NrmOutputDiag}. The latter diagram
translates into the following diagram in $\Kl(\Dst)$.
$$\xymatrix{
\overline{\Dst}(n\ast A)\ar[r]|-{\bullet}^-{\Nrm}
   \ar[d]|-{\bullet}_{\overline{\Dst}(\varepsilon_{A})}
   & n\ast\overline{\Dst}(A)\ar[d]|-{\bullet}^{\varepsilon_{\overline{\Dst}(A)}}
\\
\overline{\Dst}(A)\ar@{=}[r]\ar[d]|-{\bullet}_{\varepsilon_{A}} & 
   \overline{\Dst}(A)\ar[d]|-{\bullet}^{\varepsilon_{A}}
\\
A\ar@{=}[r] & A
}$$

\noindent This non-standard diagram does commute in $\Kl(\Dst)$, because:
$$\begin{array}{rcl}
\varepsilon_{A} \klafter \varepsilon_{\overline{\Dst}(A)} \klafter \Nrm
& = &
\mu \after \Dst(\idmap) \after \mu \after \Dst(\eta \after \nabla) \after \Nrm
\\
& = &
\mu \after \Dst(\nabla) \after \Nrm
\\
& \smash{\stackrel{\eqref{NrmOutputDiag}}{=}} &
\Dst(\nabla)
\\
& = &
\mu \after \Dst(\eta \after \nabla)
\\
& = &
(\eta \after \nabla)_{*}
\\
& = &
\mu \after \Dst(\idmap) \after \eta \after (\varepsilon_{A})_{*}
\\
& = &
\varepsilon_{A} \klafter \overline{\Dst}(\varepsilon_{A})
\end{array}$$

\auxproof{
We check where the computation stops --- Fabio has constructed a
counterexample.
$$\begin{array}{rcl}
\varepsilon_{\overline{\Dst}(A)} \klafter \Nrm
& = &
\mu \after \Dst(\eta \after \nabla) \after \Nrm
\\
& = &
\Dst(\nabla) \after \Nrm
\\
& = &
\cdots \\
& = &
\eta \after \Dst(\nabla)
\\
& = &
\eta \after (\eta \after \nabla)_{*}
\\
& = &
\overline{\Dst}(\varepsilon_{A})
\end{array}$$
}

\auxproof{
More abstractly, let $S,T\colon \cat{C} \rightarrow \cat{C}$ be
comonads, with a law $\sigma\colon ST \rightarrow TS$ commuting with
$\delta^{S}, \delta^{T}, \varepsilon^{S}$, and with $\varepsilon^{S}$
only via the diagram:
$$\xymatrix{
ST\ar[r]^-{\sigma}\ar[d]_{S(\varepsilon^{T})} & TS\ar[d]^{\varepsilon^T}
\\
S\ar[d]_{\varepsilon^S} & S\ar[d]^{\varepsilon_S}
\\
\idmap\ar@{=}[r] & \idmap
}$$

\noindent This is \emph{not} enough to prove that $ST$ a comonad, via
the usual comultiplication and counit:
$$\begin{array}{rcl}
\delta
& = &
\xymatrix{
\Big(ST\ar[r]^-{\delta^{S}T} & S^{2}T\ar[r]^-{S^{2}\delta^{T}} &
   S^{2}T^{2}\ar[r]^-{S(\sigma T)} & STST\Big)
}
\\
& = &
\xymatrix{
\Big(ST\ar[r]^-{S\delta^{T}} & ST^{2}\ar[r]^-{\delta^{S}T^{2}} &
   S^{2}T^{2}\ar[r]^-{S(\sigma T)} & STST\Big)
}
\\
\varepsilon
& = &
\xymatrix{
\Big(ST\ar[r]^-{\varepsilon^{S}T} & T\ar[r]^-{\varepsilon^T} & \idmap\Big)
}
\\
& = &
\xymatrix{
\Big(ST\ar[r]^-{S(\varepsilon^{T})} & S\ar[r]^-{\varepsilon^S} & \idmap\Big)
}
\end{array}$$

\noindent Then:
$$\begin{array}{rcl}
\varepsilon \after \delta
& = &
\varepsilon^{S} \after S(\varepsilon^{T}) \after S(\sigma) \after
   S^{2}(\delta^{T}) \after \delta^{S}
\\
& = &
\varepsilon^{T} \after \sigma \after
   S(\delta^{T}) \after \varepsilon^{S} \after \delta^{S}
\\
& = &
\varepsilon^{T} \after \sigma \after S(\delta^{T})
\\
& \neq &
S(\varepsilon^{T}) \after S(\delta^{T})
\\
& = &
\idmap
\\
ST(\varepsilon) \after \delta
& = &
ST(\varepsilon^{S}) \after STS(\varepsilon^{T}) \after S(\sigma) \after
   S^{2}(\delta^{T}) \after \delta^{S}
\\
& = &
ST(\varepsilon^{S}) \after S(\sigma) \after S^{2}T(\varepsilon^{T}) \after 
   S^{2}(\delta^{T}) \after \delta^{S}
\\
& = &
S(\varepsilon^{S}) \after \delta^{S}
\\
& = &
\idmap.
\end{array}$$
}

We summarise what we have described above.

\begin{proposition}
In the situation~\eqref{ComonadDiag} the hyper normalisation map
$\Nrm$ is a distributive law of the functor $n\ast (-)$ over the
comonad $\overline{\Dst}$. It commutes with the
$n\ast(-)$-comultiplication, but not with the $n\ast(-)$-counit. \QED
\end{proposition}

\section{Predicates}\label{IntermezzoSec}

We continue the main line of our story by using the new `hyper' form
of normalisation to describe conditioning.  Our description of
conditioning makes crucial uses of predicates.  Hence we first have to
explain what predicates in a (discrete) probabilistic setting are, and
how they are used as `evidence'. The current section provides the
required background information on predicates, which is used in the
next section to describe `hyper' conditioning.

\subsection{Events and predicates}\label{PredSubsec}

Let $A$ be an arbitrary set, seen as `sample space'. An \emph{event}
is a subset $E\subseteq A$ of the sample space.  These events are
traditionally used as predicates on $A$. We need to use a more
general `fuzzy' kind of predicate, namely functions $p \colon A
\rightarrow [0,1]$, where $[0,1] \subseteq \mathbb{R}$ is the unit
interval. An event $E\subseteq A$ can be identified with a `sharp'
predicate $A \rightarrow \{0,1\} \subseteq [0,1]$, taking values in
the subset $\{0,1\}$ of Booleans. For an event $E$ we write $\indic{E}
\in [0,1]^{A}$ for the associated sharp predicate, given by the
indicator function $\indic{E}$, defined by $\indic{E}(a) = 1$ if $a\in
E$ and $\indic{E}(a) = 0$ if $a\not\in E$.

Sharp predicates (subsets) on $A$ form a Boolean algebra. The set
$[0,1]^{A}$ of (non-sharp, fuzzy) predicates over $A$ however is an
`effect module', see~\cite{JacobsM12b,Jacobs15d,ChoJWW15b}. We briefly
describe the relevant structure, without going into the details of
what an effect module precisely is. There are truth and falsity
predicates $\one, \zero \in [0,1]^{A}$ which map each element $a\in A$
to $1$, or to $0$ respectively. Given two predicates $p,q\in
[0,1]^{A}$ we say that they are orthogonal, written as $p\orthogonal
q$, if $p(a) + q(a) \leq 1$, for all $a\in A$. In that case we write
$p \ovee q \in [0,1]^{A}$ for the pointwise sum: $(p \ovee q)(a) =
p(a) + q(a)$. These $(\ovee, \zero)$ make $[0,1]^{A}$ a partially
commutative monoid. There is also a `negation', usually written as
orthosupplement $p^{\bot}$, with $(p^{\bot})(a) = 1 - p(a)$.  Notice
that $p^{\bot\bot} = p$ and $p \ovee p^{\bot} = \one$. Moreover,
$(\indic{E})^{\bot} = \indic{\neg E}$, where $\neg E =
\setin{a}{A}{a\not\in E}$. Finally, for a scalar $s\in [0,1]$ and a
predicate $p\in [0,1]^{A}$ there is a `scaled' predicate $s\cdot p \in
[0,1]^{A}$ given by $(s\cdot p)(a) = s\cdot p(a)$.

An $n$-tuple of predicates $p_{1}, \ldots, p_{n} \in [0,1]^{A}$ is
called a \emph{test} --- or an $n$-test, to be more specific --- if
$p_{1} \ovee \cdots \ovee p_{n} = \one$. This terminology comes from
quantum theory, see \textit{e.g.}~\cite{Jacobs15d,ChoJWW15b}. This
means that these predicates $p_i$ add up to one, pointwise. When we
write such sum expressions, we implicitly assume that the relevant
predicates are orthogonal.


Notice that $\Dst(1) \cong 1$ and $\Dst(2) \cong [0,1]$. Hence we can
identify predicates on $A$ with maps $A \rightarrow \Dst(2)$. It takes
a bit more effort to see that $n$-tests on $A$ can be identified with
functions $p \colon A \rightarrow \Dst(n)$, that is, with Kleisli maps
$A \klto n$. Indeed, the $i$-th predicate $p_{i}\in [0,1]^{A}$ can be
extracted from $p$ as $p_{i}(a) = p(a)(i)$, using the functional
notation for distributions. We thus see that a test can be seen as a
probabilistic partition. After all, an ordinary, partition of a set
$A$ into $n$-parts can be identified with a function $A \rightarrow
n$, see also Section~\ref{FlowSec}. A predicate $p\in [0,1]^{A}$ can
be identified with a 2-test, consisting of $p$ itself and and its
orthosupplement $p^\bot$.

For a Kleisli map $f\colon A \klto B$ and a predicate $q\in [0,1]^{B}$
there is a (weakest precondition) predicate $f^{*}(q)$ on $A$ defined
by:
\begin{equation}
\label{SubstEqn}
\begin{array}{rcl}
f^{*}(q)(a)
& = &
\sum_{b\in B} f(a)(b)\cdot q(b).
\end{array}
\end{equation}

\noindent It is not hard to see that this map $f^{*} \colon [0,1]^{B}
\rightarrow [0,1]^{A}$ preserves the effect module structure described
in Subsection~\ref{PredSubsec}. In this way an $n$-test $q_{1},
\ldots, q_{n}$ can be turned into an $n$-test $f^{*}(q_{1}), \ldots,
f^{*}(q_{n})$ on $A$.

\subsection{Validity}\label{ValiditySubsec}

Given an event $E\subseteq A$ on a sample space $A$ we often like to
know its probability wrt.\ a distribution on $A$. This probability is
commonly written as $P(E)$. In order to make the underlying
distribution $\omega\in \Dst(A)$ explicit we prefer to write
$P_{\omega}(E)$ instead of just $P(E)$.  This probability is defined
as $P_{\omega}(E) = \sum_{a\in E}\omega(a)$. Notice that this is a
finite sum, in $[0,1]$, since the distribution $\omega$ has finite
support.


More generally, for a not necessarily sharp predicate $p\in [0,1]^A$
and a distribution $\omega\in\Dst(A)$ we define the validity (expected
value) $\omega\models p$ in $[0,1]$ as:
\begin{equation}
\label{DstValidityEqn}
\begin{array}{rclcrcl}
\omega\models p
& \;\defn{=}\; &
\sum_{a\in A} \omega(a)\cdot p(a)
& \qquad\mbox{so that}\qquad &
\omega\models\indic{E}
& \;=\; &
P_{\omega}(E).
\end{array}
\end{equation}

\noindent It is easy to see that $(\omega\models\one) = 1$ and
$(\omega\models\zero) = 0$. Moreover, $(\omega\models p^{\bot}) = 1 -
(\omega\models p)$ and $(\omega\models p\ovee q) = (\omega\models p) +
(\omega\models q)$.

\subsection{Conditionals, traditionally}\label{ConditionalSubsec}

For a predicate $p\in [0,1]^{A}$ and a distribution $\omega\in\Dst(A)$
with $\omega\models p \neq 0$ we describe a \emph{conditional
  distribution} $\cond{\omega}{p} \in \Dst(A)$, pronounced as
``$\omega$ given $p$'', and defined as:
\begin{equation}
\label{DstCondDistEqn}
\begin{array}{rcl}
\cond{\omega}{p}
& = &
{\displaystyle\sum}_{a\in A}\frac{\omega(a)\cdot p(a)}{\omega\models p}
   \bigket{a}.
\end{array}
\end{equation}

\noindent The big nuisance with these `traditional' conditionals
$\cond{\omega}{p}$ is that they are not always defined: they involve
division by the validity $\omega\models p\in [0,1]$, which should thus
be non-zero. The key improvement in our novel `hyper' description of
conditioning (in the next section) is that it is a total operation
which does not require such side-conditions --- like hyper
normalisation.


As illustration of validity and conditioning, consider a distribution
$\omega = \frac{1}{4}\ket{a} + \frac{1}{3}\ket{b} +
\frac{5}{12}\ket{c}$ on a set $A = \{a, b, c\}$, an event $E = \{a,
c\} \subseteq A$ and a predicate $p\in [0,1]^{A}$ with $p(a) =
\frac{1}{2}, p(b) = \frac{1}{4}, p(c) = 1$. Then:
$$\begin{array}{rclcrclcrcl}
\omega\models p 
& = &
\frac{5}{8} 
& \qquad &
\cond{\omega}{p}
& = &
\frac{1}{5}\ket{a} + \frac{2}{15}\ket{b} + \frac{2}{3}\ket{c}
& \qquad &
P_{\omega}(E)
& = &
\omega\models\indic{E}
\hspace*{\arraycolsep}=\hspace*{\arraycolsep}
\frac{2}{3} 
\\
\omega \models p^{\bot}
& = &
\frac{3}{8}
& \quad & 
\cond{\omega}{p^\bot}
& = &
\frac{1}{3}\ket{a} + \frac{2}{3}\ket{b}
& \quad &
\omega|_{\indic{E}}
& = &
\frac{3}{8}\ket{a} + \frac{5}{8}\ket{c}.
\end{array}$$

\auxproof{
\noindent Notice that the original state $\omega$ can be recovered as
convex combination of the two conditionals $\cond{\omega}{p}$ and
$\cond{\omega}{p^\bot}$, using the validities $\omega\models p$ and
$\omega\models p^\bot$ as scalars:
$$\begin{array}{rcl}
(\omega\models p)\cdot \cond{\omega}{p} + 
   (\omega\models p^{\bot})\cdot \cond{\omega}{p^\bot}
& = &
\frac{5}{8}\cdot \big(\frac{1}{5}\ket{a} + \frac{2}{15}\ket{b} + 
   \frac{2}{3}\ket{c}\big) +
\frac{3}{8}\cdot \big(\frac{1}{3}\ket{a} + \frac{2}{3}\ket{b}\big) \\
& = &
\frac{1}{4}\ket{a} + \frac{1}{3}\ket{b} + \frac{5}{12}\ket{c}
\hspace*{\arraycolsep}=\hspace*{\arraycolsep}
\omega.
\end{array}$$

\noindent This can also be proven in general, using the
formulas~\eqref{DstValidityEqn} and~\eqref{DstCondDistEqn}. It will
play an important role in the sequel.
}

\section{Hyper conditioning}\label{HyperConditioningSec}

We shall formulate our characterisation of conditionals for tests and
not for predicates. As noted before, predicates are subsumed by tests,
as 2-tests. Tests not only provide greater generality, but also better
capture the underlying idea. They lead to what may be called
`parallel' conditioning. Therefor we use the sign $\|$, commonly used
for a parallel processes in concurrency theory.

\begin{definition}
\label{HypCondDef}
Let $t\colon A \rightarrow \Dst(n)$ be an $n$-test on a set $A$, and
$\omega\in\Dst(A)$ be distribution on $A$. The `hyper' conditional
$\hypcond{\omega}{t} \in \Dst(n\cdot\Dst(A))$ is defined as:
$$\begin{array}{rcl}
\hypcond{\omega}{t}
& \;\defn{=}\; &
\Nrm\Big(\gr(t)_{*}(\omega)\Big).
\end{array}$$

\noindent Thus, the hyper conditioning is a function:
\begin{equation}
\label{HypCondMapDiag}
\vcenter{\xymatrix@C-.5pc{
\hypcond{(-)}{t} = \Big(\Dst(A)\ar[rr]^-{\gr(t)_{*}} & &
   \Dst(n\cdot A)\ar[r]^-{\Nrm} & \Dst(n\cdot\Dst(A))\Big)
}}
\end{equation}
\end{definition}

The map~\eqref{HypCondMapDiag} is a Kleisli map $\Dst(A) \klto
n\cdot\Dst(A)$. It is called an `abstract channel'
in~\cite{McIverMT15,McIverMSEM14}, where it is claimed that such
abstract channels capture the essence of leakages in quantitative
information flow. This will be elaborated in Section~\ref{FlowSec}.

The map $\gr(t) \colon A \rightarrow \Dst(n\cdot A)$ is called the
\emph{instrument} associated with the test $t\colon A \rightarrow n$,
in the sense of~\cite{Jacobs15d,Jacobs16a,Jacobs17a}.

This definition of $\hypcond{\omega}{t}$ is quite abstract, so we give
a more concrete illustration. We re-use the example from the end of
Subsection~\ref{ConditionalSubsec}, with $A = \{a, b, c\}$, and $p\in
[0,1]^{A}$ given by $p(a) = \frac{1}{2}, p(b) = \frac{1}{4}, p(c) =
1$. We identify the predicate $p$ with the 2-test $t = (p, p^{\bot})$,
giving a graph map $\gr(t) \colon A \rightarrow \Dst(2\cdot A)$
defined by $\gr(t)(a) = p(x)\ket{\kappa_{0}x} +
p^{\bot}(x)\ket{\kappa_{1}x}$.

The distribution $\omega = \frac{1}{4}\ket{a} + \frac{1}{3}\ket{b} +
\frac{5}{12}\ket{c}$ on $A$ gives rise to:
$$\begin{array}{rcl}
\gr(t)_{*}(\omega)
& \smash{\stackrel{\eqref{KlLiftEqn}}{=}} &
{\displaystyle\sum}_{z\in A+A} (\sum_{x\in A}\omega(x)\cdot \gr(t)(x)(z))
   \bigket{z} \\
& = &
\omega(a)\cdot p(a)\bigket{\kappa_{0}a} +
   \omega(b)\cdot p(b)\bigket{\kappa_{0}b} +
   \omega(c)\cdot p(c)\bigket{\kappa_{0}c} \\
& & \qquad
\omega(a)\cdot p^{\bot}(a)\bigket{\kappa_{1}a} +
   \omega(b)\cdot p^{\bot}(b)\bigket{\kappa_{1}b} +
   \omega(c)\cdot p^{\bot}(c)\bigket{\kappa_{1}c} \\
& = &
\frac{1}{8}\bigket{\kappa_{0}a} +
   \frac{1}{12}\bigket{\kappa_{0}b} +
   \frac{5}{12}\bigket{\kappa_{0}c} +
   \frac{1}{8}\bigket{\kappa_{1}a} +
   \frac{1}{4}\bigket{\kappa_{1}b}.
\end{array}$$

\noindent Let's use the short name $\rho = \gr(t)_{*}(\omega)$ for
the latter distribution. Then, according to Definition~\ref{NrmDef},
$$\begin{array}{rcccccccl}
\rho[0]
& = &
\sum_{x\in A} \rho(\kappa_{0}x)
& = &
\frac{1}{8} + \frac{1}{12} + \frac{5}{12}
& = &
\frac{5}{8}
& = &
\omega\models p.
\end{array}$$

\noindent Similarly, we have $\rho[1] = \frac{3}{8} = \omega\models
p^{\bot}$. We can now describe the hyper conditional more concretely:
$$\begin{array}{rcl}
\hypcond{\omega}{t}
& = &
\Nrm(\rho) \qquad\qquad \mbox{since we abbreviate } 
   \rho = \gr(t)_{*}(\omega) \\
& \smash{\stackrel{\eqref{NrmEqn}}{=}} &
\rho[0]\Bigket{\kappa_{0}(\sum_{x}
   \frac{\rho(\kappa_{0}x)}{\rho[0]}\ket{x}\big)} \;+\;
   \rho[1]\Bigket{\kappa_{1}(\sum_{x}
   \frac{\rho(\kappa_{1}x)}{\rho[1]}\ket{x}\big)} \\
& = &
(\omega\models p)\Bigket{\kappa_{0}(\sum_{x}
   \frac{\omega(x)\cdot p(x)}{\omega\models p}\ket{x})} \;+\;
   (\omega\models p^{\bot})\Bigket{\kappa_{1}(\sum_{x}
   \frac{\omega(x)\cdot p^{\bot}(x)}{\omega\models p^{\bot}}\ket{x})} \\
& = &
(\omega\models p)\bigket{\kappa_{0}(\cond{\omega}{p})} \;+\;
   (\omega\models p^{\bot})\bigket{\kappa_{1}(\cond{\omega}{p^\bot})} \\
& = &
\frac{5}{8}\bigket{\kappa_{0}(\frac{1}{5}\ket{a} + \frac{2}{15}\ket{b} + 
   \frac{2}{3}\ket{c})} \;+\;
   \frac{3}{8}\bigket{\kappa_{1}(\frac{1}{3}\ket{a} + \frac{2}{3}\ket{b})}.
\end{array}$$

Generalising this example we get the following formulation of hyper
conditioning in terms of traditional conditioning.

\begin{lemma}
\label{CondFormLem}
For an $n$-test $t = (p_{1}, \ldots, p_{n})$ of predicates $p_i$ and a
state $\omega$ we have:
$$\begin{array}{rcl}
\hypcond{\omega}{t}
& = &
\!\!\displaystyle \sum_{\stackrel{\scriptstyle 1 \leq i \leq n}{\omega\models p_{i} \neq 0}} 
   (\omega\models p_{i})\Bigket{\kappa_{i}(\cond{\omega}{p_i})}.
\end{array}\eqno{\QEDbox}$$
\end{lemma}

Notice that the problem that traditional conditionals
$\cond{\omega}{p_i}$ are not defined if $\omega\models p_{i} = 0$
(again) disappears in this `hyper' formulation, since the entries with
$\omega\models p_{i} = 0$ do not show up in the above formal convex
sum.

It turns out that under the distribution $\omega$ and test $t$ can be
recovered from a hyper conditional $\Omega = \hypcond{\omega}{t}$, via
the bijective correspondence of Proposition~\ref{PointNormProp}.

\begin{proposition}
\label{HypCondRecoverProp}
Let $\omega\in\Dst(A)$ be a distribution with $\supp(\omega) = A$, and
let $t\colon A \rightarrow \Dst(n)$ be an $n$-test. Then both $\omega$
and $t$ can be recovered from the hyper conditional
$\hypcond{\omega}{t} \in \Dst(n\cdot\Dst(A))$, namely via:
\begin{itemize}
\item $\omega = (\pi_{2})_{*}\big(\hypcond{\omega}{t}\big)$;

\item $t\colon A \rightarrow \Dst(n)$ is the map determined by the
  distribution $\Omega = (\st_{2})_{*}(\hypcond{\omega}{t}) \in
  \Dst(n\cdot A)$, as in Proposition~\ref{PointNormProp}.
\end{itemize}
\end{proposition}

\begin{myproof}
The first bullet point is easy:
$$\begin{array}{rcccccccl}
(\pi_{2})_{*}\big(\hypcond{\omega}{t}\big)
& \smash{\stackrel{\eqref{NrmOutputDiag}}{=}} &
\Dst(\pi_{2})\big(\gr(t)_{*}(\omega)\big) 
& = & 
\big(\Dst(\pi_{2}) \after \gr(t)\big)_{*}(\omega) 
& \smash{\stackrel{\eqref{StrengthGraphDiag}}{=}} &
\eta_{*}(\omega) 
& = &
\omega.
\end{array}$$

\noindent For the second bullet point, we write $\Omega =
(\st_{2})_{*}(\hypcond{\omega}{t}) \in \Dst(n\cdot A)$.  We first show
that $\Dst(\pi_{2})(\Omega) = \Dst(\nabla)(\Omega) = \omega$, using
what we have just proven:
$$\begin{array}{rcccccccl}
\Dst(\pi_{2})(\Omega)
& = &
\big(\Dst(\pi_{2}) \after (\st_{2})_{*}\big)(\hypcond{\omega}{t})
& = &
(\Dst(\pi_{2}) \after \st_{2})_{*}(\hypcond{\omega}{t})
& \smash{\stackrel{\eqref{StrengthGraphDiag}}{=}} &
(\pi_{2})_{*}(\hypcond{\omega}{t})
& = &
\omega.
\end{array}$$

\noindent Hence the side-condition in the bijective correspondence of
Proposition~\ref{PointNormProp} is satisfied for $\Omega$. Thus, we
can write $\Omega = \gr(f)_{*}(\omega)$, as in~\eqref{PointNormUp},
for a unique map $f\colon A \rightarrow \Dst(n)$. We have to show that
$f = t$, the original test. But this follows from:
$$\begin{array}{rcccccl}
\Omega
& = &
(\st_{2})_{*}(\hypcond{\omega}{t})
& = &
\big((\st_{2})_{*} \after \Nrm\big)\big(\gr(t)_{*}(\omega)\big)
& \smash{\stackrel{\eqref{NrmMonicDiag}}{=}} &
\gr(t)_{*}(\omega).
\end{array}\eqno{\QEDbox}$$
\end{myproof}

\begin{example}
\label{HypCondRecoverEx}
We illustrate how the distribution and test can be recovered for a
hyper distribution $\Phi\in\Dst(2\cdot\Dst(\{H,T\}))$ over the 2-element
set $\{H,T\}$ of `head' and `tail' outcomes.
$$\begin{array}{rcl}
\Phi
& = &
\frac{1}{2}\bigket{\kappa_{0}(\frac{2}{3}\ket{H} + \frac{1}{3}\ket{T})} +
   \frac{1}{2}\bigket{\kappa_{1}(\frac{1}{3}\ket{H} + \frac{2}{3}\ket{T})}
\end{array}$$

\noindent The first bullet in Proposition~\ref{HypCondRecoverProp}
says that we can obtain the underlying distribution
$\omega\in\Dst(\{H,T\})$ as:
$$\begin{array}{rcl}
\omega
\hspace*{\arraycolsep}=\hspace*{\arraycolsep}
(\pi_{2})_{*}(\Phi)
& = &
\mu\big(\Dst(\pi_{2})\big(\frac{1}{2}\bigket{\kappa_{0}(\frac{2}{3}\ket{H} + 
   \frac{1}{3}\ket{T})} +
   \frac{1}{2}\bigket{\kappa_{1}(\frac{1}{3}\ket{H} + \frac{2}{3}\ket{T})}\big)
\\
& = &
\mu\big(\frac{1}{2}\bigket{\frac{2}{3}\ket{H} + \frac{1}{3}\ket{T}} +
   \frac{1}{2}\bigket{\frac{1}{3}\ket{H} + \frac{2}{3}\ket{T}}\big)
\\
& = &
\frac{1}{2}\cdot\frac{2}{3}\ket{H} + \frac{1}{2}\cdot\frac{1}{3}\ket{T} +
   \frac{1}{2}\cdot\frac{1}{3}\ket{H} + \frac{1}{2}\cdot\frac{2}{3}\ket{T}
\\
& = &
\frac{1}{2}\ket{H} + \frac{1}{2}\ket{T}.
\end{array}$$

\noindent For the second bullet we compute:
$$\begin{array}{rcl}
(\st_{2})_{*}(\Psi)
& = &
\mu\big(\Dst(\st_{2})\big(\frac{1}{2}\bigket{\kappa_{0}(\frac{2}{3}\ket{H} + 
   \frac{1}{3}\ket{T})} +
   \frac{1}{2}\bigket{\kappa_{1}(\frac{1}{3}\ket{H} + \frac{2}{3}\ket{T})}\big)
\\
& = &
\mu\big(\frac{1}{2}\bigket{\frac{2}{3}\ket{\kappa_{0}H} + 
      \frac{1}{3}\ket{\kappa_{0}T}} +
   \frac{1}{2}\bigket{\frac{1}{3}\ket{\kappa_{1}H} + 
      \frac{2}{3}\ket{\kappa_{1}T}}\big)
\\
& = &
\frac{1}{3}\ket{\kappa_{0}H} + \frac{1}{6}\ket{\kappa_{0}T} +
   \frac{1}{6}\ket{\kappa_{1}H} + \frac{1}{3}\ket{\kappa_{1}T}.
\end{array}$$

\noindent The recipe~\eqref{PointNormUp} now gives a test function
$s\colon \{H, T\} \rightarrow \Dst(2)$, namely:
$$\begin{array}{rccclcrcccl}
s(H)
& = &
\frac{\nicefrac{1}{3}}{\nicefrac{1}{2}}\ket{0} +
   \frac{\nicefrac{1}{6}}{\nicefrac{1}{2}}\ket{1}
& = &
\frac{2}{3}\ket{0} + \frac{1}{3}\ket{1}
& \qquad &
s(T)
& = &
\frac{\nicefrac{1}{6}}{\nicefrac{1}{2}}\ket{0} +
   \frac{\nicefrac{1}{3}}{\nicefrac{1}{2}}\ket{1}
& = &
\frac{1}{3}\ket{0} + \frac{2}{3}\ket{1}.
\end{array}$$

\noindent Then indeed, $\Phi = \hypcond{\omega}{s}$, as can be checked
easily.

\auxproof{
We compute consecutively:
$$\begin{array}{rcl}
\gr(s)(H)
& = &
\frac{2}{3}\ket{\kappa_{0}H} + \frac{1}{3}\ket{\kappa_{1}H}
\\
\gr(s)(T)
& = &
\frac{1}{3}\ket{\kappa_{0}T} + \frac{2}{3}\ket{\kappa_{1}T}
\\
\gr(s)_{*}(\omega)
& = &
\frac{1}{2}\cdot\frac{2}{3}\ket{\kappa_{0}H} + 
   \frac{1}{2}\cdot\frac{1}{3}\ket{\kappa_{1}H} +
   \frac{1}{2}\cdot\frac{1}{3}\ket{\kappa_{0}T} + 
   \frac{1}{2}\cdot\frac{2}{3}\ket{\kappa_{1}T}
\\
& = &
\frac{1}{3}\ket{\kappa_{0}H} + \frac{1}{6}\ket{\kappa_{1}H} +
   \frac{1}{6}\ket{\kappa_{0}T} + \frac{1}{3}\ket{\kappa_{1}T}
\\
\hypcond{\omega}{s}
& = &
\Nrm\big(\gr(s)_{*}(\omega)\big)
\\
& = &
\Nrm\big(\frac{1}{3}\ket{\kappa_{0}H} + \frac{1}{6}\ket{\kappa_{1}H} +
   \frac{1}{6}\ket{\kappa_{0}T} + \frac{1}{3}\ket{\kappa_{1}T}\big)
\\
& = &
\frac{1}{2}\bigket{\kappa_{0}\big(\frac{\nicefrac{1}{3}}{\nicefrac{1}{2}}\ket{H}
   + \frac{\nicefrac{1}{6}}{\nicefrac{1}{2}}\ket{T}\big)} +
\frac{1}{2}\bigket{\kappa_{0}\big(\frac{\nicefrac{1}{6}}{\nicefrac{1}{2}}\ket{H}
   + \frac{\nicefrac{1}{3}}{\nicefrac{1}{2}}\ket{T}\big)}
\\
& = &
\frac{1}{2}\bigket{\kappa_{0}(\frac{2}{3}\ket{H} + \frac{1}{3}\ket{T})} +
   \frac{1}{2}\bigket{\kappa_{1}(\frac{1}{3}\ket{H} + \frac{2}{3}\ket{T})}
\\
& = &
\Phi.
\end{array}$$
}

The reader may wish to do a similar computation for the hyper
distribution $\Psi\in\Dst(3\cdot\Dst(\{H,T\}))$ given by:
$$\begin{array}{rcl}
\Psi
& = &
\frac{1}{3}\bigket{\kappa_{0}(\frac{2}{3}\ket{H} + \frac{1}{3}\ket{T})} +
   \frac{1}{3}\bigket{\kappa_{1}(\frac{1}{2}\ket{H} + \frac{1}{2}\ket{T})} +
   \frac{1}{3}\bigket{\kappa_{2}(\frac{1}{3}\ket{H} + \frac{2}{3}\ket{T})}
\end{array}$$

\noindent The answer appears in Example~\ref{OrderEx}.

\auxproof{
The underlying distribution is the same $\omega$ as before:
$$\begin{array}{rcl}
(\pi_{2})_{*}(\Psi)
& = &
\mu\big(\Dst(\pi_{2})(\Psi)\big)
\\
& = &
(\frac{1}{3}\cdot\frac{2}{3} + \frac{1}{3}\cdot\frac{1}{2}
    + \frac{1}{3}\cdot\frac{1}{3})\ket{H} + 
   (\frac{1}{3}\cdot\frac{1}{3} + \frac{1}{2}\cdot\frac{1}{2}  
    + \frac{1}{3}\cdot\frac{2}{3})\ket{T}
\\
& = &
\frac{1}{2}\ket{H} + \frac{1}{2}\ket{T}
\\
(\st_{2})_{*}(\Psi)
& = &
\mu\big(\Dst(\st_{2})(\Psi)\big)
\\
& = &
\mu\big(\frac{1}{3}\bigket{\frac{2}{3}\ket{\kappa_{0}H} + 
   \frac{1}{3}\ket{\kappa_{0}T}} + 
   \frac{1}{3}\bigket{\frac{1}{2}\ket{\kappa_{1}H} + 
   \frac{1}{2}\ket{\kappa_{1}T}} +
   \frac{1}{3}\bigket{\frac{1}{3}\ket{\kappa_{2}H} + 
   \frac{2}{3}\ket{\kappa_{2}T}}\big)
\\
& = &
\frac{2}{9}\ket{\kappa_{0}H} + \frac{1}{9}\ket{\kappa_{0}T} + 
   \frac{1}{6}\ket{\kappa_{1}H} + \frac{1}{6}\ket{\kappa_{1}T} +
   \frac{1}{9}\ket{\kappa_{2}H} + \frac{2}{9}\ket{\kappa_{2}T}.
\end{array}$$

\noindent From this we obtain the test map $t\colon \{T,H\}
\rightarrow \Dst(3)$ given by:
$$\begin{array}{rcl}
t(H)
& = &
\frac{\nicefrac{2}{9}}{\nicefrac{1}{2}}\ket{0} +
   \frac{\nicefrac{1}{6}}{\nicefrac{1}{2}}\ket{1} +
   \frac{\nicefrac{1}{9}}{\nicefrac{1}{2}}\ket{2}
\hspace*{\arraycolsep}=\hspace*{\arraycolsep}
\frac{4}{9}\ket{0} + \frac{1}{3}\ket{1} + \frac{2}{9}\ket{2}
\\
t(T)
& = &
\frac{\nicefrac{1}{9}}{\nicefrac{1}{2}}\ket{0} +
   \frac{\nicefrac{1}{6}}{\nicefrac{1}{2}}\ket{1} +
   \frac{\nicefrac{2}{9}}{\nicefrac{1}{2}}\ket{2}
\hspace*{\arraycolsep}=\hspace*{\arraycolsep}
\frac{2}{9}\ket{0} + \frac{1}{3}\ket{1} + \frac{4}{9}\ket{2}.
\end{array}$$

\noindent We now compute:
$$\begin{array}{rcl}
\gr(t)(H)
& = &
\frac{4}{9}\ket{\kappa_{0}H} + \frac{1}{3}\ket{\kappa_{1}H} + 
   \frac{2}{9}\ket{\kappa_{2}H}
\\
\gr(t)(T)
& = &
\frac{2}{9}\ket{\kappa_{0}T} + \frac{1}{3}\ket{\kappa_{1}T} + 
   \frac{1}{9}\ket{\kappa_{2}T}
\\
\gr(t)_{*}(\omega)
& = &
\frac{1}{2}\cdot\frac{4}{9}\ket{\kappa_{0}H} + 
   \frac{1}{2}\cdot\frac{1}{3}\ket{\kappa_{1}H} + 
   \frac{1}{2}\cdot\frac{2}{9}\ket{\kappa_{2}H} +
\frac{1}{2}\cdot\frac{2}{9}\ket{\kappa_{0}T} + 
   \frac{1}{2}\cdot\frac{1}{3}\ket{\kappa_{1}T} + 
   \frac{1}{2}\cdot\frac{4}{9}\ket{\kappa_{2}T}
\\
& = &
\frac{2}{9}\ket{\kappa_{0}H} + \frac{1}{6}\ket{\kappa_{1}H} + 
   \frac{1}{9}\ket{\kappa_{2}H} +
\frac{1}{9}\ket{\kappa_{0}T} + \frac{1}{6}\ket{\kappa_{1}T} + 
   \frac{2}{9}\ket{\kappa_{2}T}
\\
\hypcond{\omega}{t}
& = &
\Nrm\big(\gr(t)_{*}(\omega)\big)
\\
& = &
\Nrm\big(\frac{2}{9}\ket{\kappa_{0}H} + \frac{1}{6}\ket{\kappa_{1}H} + 
   \frac{1}{9}\ket{\kappa_{2}H} +
\frac{1}{9}\ket{\kappa_{0}T} + \frac{1}{6}\ket{\kappa_{1}T} + 
   \frac{2}{9}\ket{\kappa_{2}T}\big)
\\
& = &
\frac{1}{3}\bigket{\kappa_{0}\big(
   \frac{\nicefrac{2}{9}}{\nicefrac{1}{3}}\ket{H} +
   \frac{\nicefrac{1}{9}}{\nicefrac{1}{3}}\ket{T}\big)} +
\frac{1}{3}\bigket{\kappa_{1}\big(
   \frac{\nicefrac{1}{6}}{\nicefrac{1}{3}}\ket{H} +
   \frac{\nicefrac{1}{6}}{\nicefrac{1}{3}}\ket{T}\big)} +
\frac{1}{3}\bigket{\kappa_{2}\big(
   \frac{\nicefrac{1}{9}}{\nicefrac{1}{3}}\ket{H} +
   \frac{\nicefrac{2}{9}}{\nicefrac{1}{3}}\ket{T}\big)}
\\
& = &
\frac{1}{3}\bigket{\kappa_{0}(\frac{2}{3}\ket{H} + \frac{1}{3}\ket{T})} +
   \frac{1}{3}\bigket{\kappa_{1}(\frac{1}{2}\ket{H} + \frac{1}{2}\ket{T})} +
   \frac{1}{3}\bigket{\kappa_{2}(\frac{1}{3}\ket{H} + \frac{2}{3}\ket{T})}
\\
& = &
\Psi.
\end{array}$$
}


\end{example}

We illustrate how the `hyper' approach works in Bayesian reasoning,
for a standard medical examination example copied from~\cite{JacobsZ16}.

\begin{example}
\label{MedTestEx}
Write $\tes{D} = \{d, \no{d}\}$ and $\tes{T} = \{t, \no{t}\}$ for two
2-element sets, where $d$ represents `disease' and $\no{d}$ represents
`no disease'. Similarly, the element $t$ represents a positive test
(examination outcome), and $\no{t}$ a negative outcome. Consider the
following simple Bayesian network, described as Kleisli maps (as
in~\cite{JacobsZ16}):
$$\vcenter{\xymatrix{
1 \ar[r]|-{\bullet}^-{\omega} & \tes{D} \ar[r]|-{\bullet}^-{s} & \tes{T}
}}
\qquad\mbox{with}\qquad
\left\{\begin{array}{rcl}
\omega
& = &
\frac{1}{100}\ket{d} + \frac{99}{100}\ket{\no{d}}
\\
s(d)
& = &
\frac{9}{10}\ket{t} + \frac{1}{10}\ket{\no{t}}
\\
s(\no{d})
& = &
\frac{1}{20}\ket{t} + \frac{19}{20}\ket{\no{t}}.
\end{array}\right.$$

\noindent The state $\omega$ captures the prior probability of $1\%$
of having the disease. The function $s \colon \tes{D} \rightarrow
\Dst(\tes{T})$ describes the sensitivity of the test.

We write $T? \colon \tes{T} \rightarrow [0,1]$ for the
(sharp) predicate given by $T?(t) = 1$ and $T?(\no{t}) = 0$. Together
with $T?^{\bot}$ it forms a 2-test $T! = (T?, T?^{\bot})$ on
$\tes{T}$. It gives rise to a 2-test $s^{*}(T!) = (s^{*}(T?),
s^{*}(T?^{\bot}))$ on $\tes{D}$ via~\eqref{SubstEqn}, given by:
$$\begin{array}{rclcrclcrclcrcl}
s^{*}(T?)(d)
& = &
\frac{9}{10}
& \qquad &
s^{*}(T?^{\bot})(d)
& = &
\frac{1}{10}
& \qquad &
s^{*}(T?)(\no{d})
& = &
\frac{1}{20}
& \qquad &
s^{*}(T?^{\bot})(\no{d})
& = &
\frac{19}{20}.
\end{array}$$

\noindent The associated instrument map $\gr(s^{*}(T!)) \colon \tes{D}
\rightarrow \Dst(2\cdot \tes{D})$ is:
$$\begin{array}{rclcrcl}
\gr(s^{*}(T!))(d)
& = &
\frac{9}{10}\ket{\kappa_{0}d} + \frac{1}{10}\ket{\kappa_{1}d}
& \quad &
\gr(s^{*}(T!))(\no{d})
& = &
\frac{1}{20}\ket{\kappa_{0}\no{d}} + \frac{19}{20}\ket{\kappa_{1}\no{d}}
\end{array}$$

\noindent When applied to the (prior) state $\omega$ it gives:
$$\begin{array}{rcl}
\gr(s^{*}(T!))_{*}(\omega)
& = &
\frac{9}{1000}\ket{\kappa_{0}d} + \frac{1}{1000}\ket{\kappa_{1}d}
+
\frac{99}{2000}\ket{\kappa_{0}\no{d}} + \frac{1881}{2000}\ket{\kappa_{1}\no{d}}

\end{array}$$

\noindent The resulting hyper conditional $\hypcond{\omega}{s^{*}(T!)}
= \Nrm\big(\gr(s^{*}(T!))_{*}(\omega)\big)$ is then:
$$\begin{array}{rcl}
\hypcond{\omega}{s^{*}(T!)}
& = &
\frac{117}{2000}\Bigket{\kappa_{0}(\frac{18}{117}\ket{d} + 
   \frac{99}{117}\ket{\no{d}})} +
   \frac{1883}{2000}\Bigket{\kappa_{1}(\frac{2}{1883}\ket{d} + 
   \frac{1881}{1883}\ket{\no{d}})} 
\end{array}$$

\noindent This hyper distribution $\hypcond{\omega}{s^{*}(T!)} \in
\Dst(2\cdot\Dst(\tes{D}))$ is obtained by backward learning, from the
$2$-test $T!$. It is given by a convex combination of two conditional
(normalised) inner distributions. The left inner distribution
describes the probability $\frac{18}{117} \sim 15\%$ of having the
disease after a positive test outcome $T?$, whereas the right inner
distribution gives the probability $\frac{2}{1883} \sim 0.1\%$ of
having the disease after a negative outcome $T?^{\bot}$. One could say
that the parallel conditioning that happens in a hyper conditional
$\hypcond{\omega}{t}$ corresponds to a many worlds view --- as is
sometimes used, for instance, in counter factual
reasoning~\cite{Pearl09}.

The hyper approach does not give direct access to these inner
distributions. But further calculations can be done with this hyper
distribution. If one is not interested in the second inner
distribution it can be removed via a $\bang$ map to the final
(singleton) set $1$, leading to a distribution:
$$\textstyle\frac{117}{2000}\Bigket{\kappa_{0}(\frac{18}{117}\ket{d} + 
   \frac{99}{117}\ket{\no{d}})} +
   \frac{1883}{2000}\Bigket{\kappa_{1}0}
\;\in\;
\Dst\big(\Dst(\tes{D}) + 1\big).$$

\noindent Via multiplication it can be further reduced to a distribution
in $\Dst(\tes{D}+1)$, but then one loses the conditional, as in:
$$\textstyle\frac{18}{2000}\ket{\kappa_{0}d} + 
   \frac{99}{2000}\ket{\kappa_{0}\no{d}} + 
   \frac{1883}{2000}\ket{\kappa_{1}0}
\;\in\;
\Dst\big(\tes{D} + 1\big).$$
\end{example}

\section{Applications in quantitative information flow}\label{FlowSec}

The hyper conditional construction $\hypcond{\omega}{t}$ that we use
here --- see Definition~\ref{HypCondDef} --- is inspired by a
`denotation of a channel' construction in quantitative information
flow, see~\cite{McIverMM10,McIverMSEM14,McIverMT15,McIverMM15}. This
will be sketched first.  Subsequently we describe how tests and hyper
distributions are ordered, and how these orders are related.


An abstract channel
in~\cite{McIverMM10,McIverMSEM14,McIverMT15,McIverMM15} from a set $X$
to set $Y$ is what we call a Kleisli map $c\colon X \klto Y$, that is,
a function $c\colon X \rightarrow \Dst(Y)$. The sets $X,Y$ used in
this context are finite, so we can replace them by numbers, and write
a channel as Kleisli map $n \klto m$. As noted in
Subsection~\ref{KleisliSubsec} such a channel gives an $m$-test on
$n$.

The \emph{denotation} of a channel $c\colon n \klto m$ is defined
in~\cite{McIverMSEM14,McIverMT15} as a function $\Scottint{c} \colon
\Dst(n) \rightarrow \Dst^{2}(n)$. It uses conditional distributions,
via normalisation. We redescribe this denotation via the notation from
this paper. Let channel $c\colon n \klto m$ correspond to $m$-test
$c_{i}\in [0,1]^{n}$ given by $c_{i}(j) = c(j)(i)$. The denotation
$\Scottint{c}(\omega) \in \Dst(\Dst(n))$ is defined for
$\omega\in\Dst(n)$ as:
\begin{equation}
\label{DenotationEqn}
\begin{array}{rcl}
\Scottint{c}(\omega)
& = &
\displaystyle\sum_{\stackrel{\scriptstyle 0\leq i\leq m-1}{\omega\models c_{i} \neq 0}} 
   \big(\omega\models c_{i}\big)\Bigket{\cond{\omega}{c_i}}.
\end{array}
\end{equation}

\noindent There is an obvious similarity with `our' formula for hyper
conditioning in Lemma~\ref{CondFormLem}. The difference is that we use
an inner copower $\Dst(m\cdot\Dst(n))$ instead of $\Dst(\Dst(n))$,
with corresponding coprojections $\kappa_{j}$, to keep the inner
conditional distributions $\cond{\omega}{c_i}$ separate.

It is not hard to see that the above formulation~\eqref{DenotationEqn}
can be obtained from ours as $\Scottint{c}(\omega) =
\Dst(\nabla)(\hypcond{\omega}{c})$, by removing the coprojections, via
the codiagonal $\nabla\colon m\cdot\Dst(n) \rightarrow
\Dst(n)$. In~\cite{McIverMT15} it is observed\footnote{See after
  Defn.~7 in~\cite{McIverMT15}, where multiplication $\mu$ is called
  average. We add that the construction of $\Scottint{c}(\omega)$
  in~\cite{McIverMSEM14,McIverMT15} requires some \textit{ad hoc}
  `removal' and `renaming' of redundant data that happens
  automatically in the current situation by the formal convex sum
  formalism from Subsection~\ref{DstributionSubsec}.} that applying
multiplication $\mu$ to $\Scottint{c}(\omega)$ yields the original
distribution $\omega$. In our case this follows directly from the
first bullet in Proposition~\ref{HypCondRecoverProp}.

Denotations $\Scottint{c} \colon \Dst(n) \rightarrow \Dst^{2}(n)$ are
instances of Hidden Markov Models in~\cite{McIverMT15}, whose action
on `uncertainty measures' is characterised in terms of uncertainty
transformers. Here we zoom in on the order theoretic aspects.

\auxproof{
A hidden Markov model (HMM) according to~\cite{McIverMM15} is given by
a state space $X$ of hidden variables together with a set $A$ of
visible variables, and a coalgebra $X \rightarrow
\Dst(A)\times\Dst(X)$. It gives rise to a map $X \rightarrow
\Dst(A\times\Dst(X))$ via strength $\st_1$.

The order considered on $\Dst(A\times\Dst(X))$ is a combination of
$\preceq$ (written as $\sqsubseteq$ below), and the pointwise order
$\leq$. Then: $\Phi \sqsubseteq \Psi$ iff there is a $\Omega$ with
$(\pi_{2})_{*}(\Omega) = \Phi$ and $\Dst(m\cdot(\pi_{2})_{*})(\Omega)
\leq \Psi$. This order is apparently well-behaved.
}

\subsection{Refinements}
We continue with refinements of partitions (tests), and start with the
ordinary (non-probabilistic) case. Let $(S_{i})_{i\in n}$ be a
partition of a set $A$. That means $S_{i} \subseteq A$ with
$\bigcup_{i}S_{i} = A$, and $S_{i} \cap S_{i'} = \emptyset$ for $i\neq
i'$. Thus, each element $a\in A$ can be mapped to a unique element
$i\in n$ with $a\in S_{i}$. Hence the partition $(S_{i})$ can be
identified with a function $s \colon A \rightarrow n$, where $S_{i} =
s^{-1}(i)$.

If we have two partitions $(S_{i})_{i\in n}$ and $(T_{j})_{j\in m}$ of
the same set $A$ we can say that $(S_{i}) \sqsubseteq (T_{j})$ if for
each $i\in n$ there is a $j\in m$ with $S_{i} \subseteq T_{j}$. This
means that the $S$-partition is more refined than the $T$-partition,
since each subsets $S_i$ fits in some $T_j$.

There is particularly simple way to express this refinement relation
when we switch to the description in terms of functions. Let $s\colon A
\rightarrow n$ and $t\colon A \rightarrow m$ be the functions corresponding
to the partitions $(S_{i})$ and $(T_{j})$. Then it is not hard to see:
$$\begin{array}{rcl}
(S_{i}) \sqsubseteq (T_{j})
& \Longleftrightarrow &
\mbox{there is a $h\colon n \rightarrow m$ with }
\vcenter{\xymatrix@R-1.5pc@C+1pc{
& n\ar[dd]^{h}
\\
A\ar[ur]^-{s}\ar[dr]_{t} & 
\\
& m
}}
\end{array}$$

\noindent Here we have to assume that $S_{i} \neq \emptyset$, for each
$i\in n$. Then we can define $h(i) = j$ iff $S_{i} \subseteq T_{j}$.
This yields what is sometimes called the lattice of
information~\cite{LandauerR93}.

\auxproof{
\begin{itemize}
\item[$\Rightarrow$] Pick $a\in A$, and write $i = s(a) \in n$. Then
  $a\in S_{i} \subseteq T_{h(i)}$. Hence $a \in t^{-1}(h(i))$, that
  is, $t(a) = h(i) = h(s(a))$. Hence $h \after s = t$.

\item[$\Leftarrow$] Pick $i\in n$ and find an element $a\in
  S_{i}$. Then $s(a) = i$, so that $t(a) = h(s(a)) = h(i)$, so that $a
  \in t^{-1}(h(i)) = T_{h(i)}$. Hence $S_{i} \subseteq T_{h(i)}$.
\end{itemize}
}

This functional description of refinement can be translated very
easily to a probabilistic setting, simply by using Kleisli maps
instead of ordinary functions. This done in the first item below. The
second item givens an alternative formulation of refinement on hyper
distributions, used in quantitative information flow, see
\textit{e.g.}~\cite{McIverMM10,McIverMSEM14,McIverMT15,McIverMM15}. We
slightly adapt it to the current setting.

\begin{definition}
\label{OrderDef}
Let $A$ be a set and $n,m$ be natural numbers.
\begin{enumerate}
\item For two tests $s\colon A \rightarrow \Dst(n)$ and $t\colon A
  \rightarrow \Dst(m)$ on $A$ one defines:
$$\begin{array}{rclcrcl}
s & \sqsubseteq & t
& \quad\mbox{iff}\quad &
\mbox{there is a function $h\colon n \rightarrow \Dst(m)$ with 
  $h\klafter s = t$, as in: }
\vcenter{\xymatrix@R-1.5pc@C+1pc{
& n\ar[dd]|-{\bullet}^{h}
\\
A\ar[ur]|-{\bullet}^-{s}\ar[dr]|-{\bullet}_{t} & 
\\
& m
}}
\end{array}$$

\item For two hyper distributions $\Phi\in \Dst(n\cdot \Dst(A))$ and
  $\Psi\in\Dst(m\cdot \Dst(A))$ we put:
$$\begin{array}{rclcrcl}
\Phi & \sqsubseteq & \Psi
& \quad\mbox{iff}\quad &
\left\{\begin{array}{l}
\mbox{there is an $\Omega\in\Dst(m\cdot \Dst(n\cdot \Dst(A)))$ with} \\
(\pi_{2})_{*}(\Omega) = \Phi
\mbox{ and }
\Dst\big(m\cdot (\pi_{2})_{*}\big)(\Omega) = \Psi.
\end{array}\right.
\end{array}$$
\end{enumerate}
\end{definition}

The theorem below is a basic result in quantitative information flow,
see~\cite{McIverMM10,McIverMSEM14,McIverMT15,McIverMM15}. Our aim is
to illustrate how our approach to normalisation and conditioning can
be used, by giving abstract proof constructions.

\begin{theorem}
\label{OrderThm}
In the situation of Definition~\ref{OrderDef},
\begin{enumerate}
\item if $s \sqsubseteq t$ then $\hypcond{\omega}{s} \sqsubseteq
\hypcond{\omega}{t}$ for each $\omega\in\Dst(A)$;

\item if $\hypcond{\omega}{s} \sqsubseteq \hypcond{\omega}{t}$ for
  some $\omega\in\Dst(A)$ with $\supp(\omega) = A$ and
  $\supp\big(s_{*}(\omega)\big) = n$, then $s \sqsubseteq t$.
\end{enumerate}
\end{theorem}

\begin{myproof}
Let $s \sqsubseteq t$ via Kleisli map $h\colon n \rightarrow \Dst(m)$,
so that $h \klafter s = t$. We write $h_{1} = h \after \pi_{1} \colon
n\cdot\Dst(A) \rightarrow \Dst(m)$, with associated graph map
$\gr(h_{1}) \colon n\cdot\Dst(A) \rightarrow
\Dst(m\cdot(n\cdot\Dst(A)))$. For an arbitrary distribution $\omega
\in\Dst(A)$ we take:
\begin{equation}
\label{OrderThmPrfOmega}
\begin{array}{rcccl}
\Omega
& \defn{=} &
\hypcond{\big(\hypcond{\omega}{s}\big)}{h_1}
& = &
\Nrm\Big(\gr(h_{1})_{*}\big(\hypcond{\omega}{s}\big)\Big) \;\in\;
   \Dst\Big(m\cdot\Dst\big(n\cdot\Dst(A)\big)\Big).
\end{array}
\end{equation}

\noindent By construction, $(\pi_{2})_{*}(\Omega) =
\hypcond{\omega}{s}$, see Proposition~\ref{HypCondRecoverProp}. The
proofs of the following two auxiliary equations are easily obtained.
\begin{equation}
\label{OrderThmPrfGivenh}
\begin{array}{rclcrcl}
\Dst(m\cdot\pi_{2}) \after \gr(h_{1})
& = &
h\cdot\Dst(A)
& \qquad &
(h\cdot A) \klafter \gr(s)
& = &
\gr(h \klafter s)
\end{array}
\end{equation}

\auxproof{
Abstractly,
$$\begin{array}{rcl}
\Dst(m\cdot\pi_{2}) \after \gr(h_{1})
& = &
\Dst(\idmap\times\pi_{2}) \after \st_{1} \after \tuple{h \after \pi_{1},\idmap}
\\
& = &
\st_{1} \after (\idmap\times\pi_{2}) \after \tuple{h \after \pi_{1},\idmap}
\\
& = &
\st_{1} \after (h\times\idmap)
\\
& = & 
h\cdot\Dst(A)
\end{array}$$

\noindent Similarly,
$$\begin{array}{rcl}
(h\cdot A) \klafter \gr(s)
& = &
\mu \after \Dst(\st_{1} \after (h\times\idmap)) \after \st_{1} \after 
   \tuple{s, \idmap}
\\
& = &
\mu \after \Dst(\st_{1}) \after \st_{1} \after (\Dst(h)\times\idmap)) \after 
   \tuple{s, \idmap}
\\
& = &
\st_{1} \after (\mu\times\idmap) \after \tuple{\Dst(h) \after s, \idmap}
\\
& = &
\st_{1} \after \tuple{h \klafter s, \idmap}
\\
& = &
\gr(h \klafter s)
\end{array}$$
}

\noindent Then $\hypcond{\omega}{s} \sqsubseteq \hypcond{\omega}{t}$
via $\Omega$ follows from:
$$\begin{array}{rcl}
\Dst\big(m\cdot (\pi_{2})_{*})\big)(\Omega)
& = &
\big(\Dst(m\cdot\mu) \after \Dst(m\cdot\Dst(\pi_{2})) \after \Nrm \after
   \gr(h_{1})_{*} \after \Nrm \after \gr(s)_{*}\big)(\omega)
\\
& \smash{\stackrel{\eqref{NrmNatDiag}}{=}} &
\big(\Dst(m\cdot\mu) \after \Nrm \after \Dst(m\cdot\pi_{2}) \after 
   \gr(h_{1})_{*} \after \Nrm \after \gr(s)_{*}\big)(\omega)
\\
& = &
\big(\Dst(m\cdot\mu) \after \Nrm \after (\Dst(m\cdot\pi_{2}) \after 
   \gr(h_{1}))_{*} \after \Nrm \after \gr(s)_{*}\big)(\omega)
\\
& \smash{\stackrel{\eqref{OrderThmPrfGivenh}}{=}} &
\big(\Dst(m\cdot\mu) \after \Nrm \after (h\cdot\Dst(A))_{*}
   \after \Nrm \after \gr(s)_{*}\big)(\omega)
\\
& \smash{\stackrel{\eqref{NrmNNatDiag}}{=}} &
\big(\Nrm \after (h\cdot A)_{*} \after \gr(s)_{*}\big)(\omega)
\\
& = &
\big(\Nrm \after ((h\cdot A) \klafter \gr(s))_{*}\big)(\omega)
\\
& \smash{\stackrel{\eqref{OrderThmPrfGivenh}}{=}} &
\big(\Nrm \after \gr(h \klafter s)_{*}\big)(\omega)
\\
& = &
\big(\Nrm \after \gr(t)_{*}\big)(\omega)
\\
& = &
\hypcond{\omega}{t}.
\end{array}$$

In the other direction, let $\hypcond{\omega}{s} \sqsubseteq
\hypcond{\omega}{t}$ via $\Omega \in
\Dst\big(m\cdot\Dst(n\cdot\Dst(A))\big)$, so that
$(\pi_{2})_{*}(\Omega) = \hypcond{\omega}{s}$ and $\Dst(m\cdot
(\pi_{2})_{*})(\Omega) = \hypcond{\omega}{t}$, where
$\omega\in\Dst(A)$ satisfies $\supp(s_{*}(\omega)) = n$ and
$\supp(\omega) = A$. We need to find a map $h\colon n \rightarrow
\Dst(m)$ with $h \klafter s = t$.  Consider the distribution:
\begin{equation}
\label{OrderThmPrfTheta}
\begin{array}{rcl}
\Theta
& = &
\big((\st_{2})_{*} \after \Dst(m\cdot\Dst(\pi_{1}))\big)(\Omega) 
   \;\in\; \Dst(m\cdot n).
\end{array}
\end{equation}

\noindent It is not hard to see that the second marginal
$\Dst(\pi_{2})(\Theta) \in \Dst(n)$ equals $s_{*}(\omega)$. Since the
support of the latter distribution is $n$, by assumption, we may use
Proposition~\ref{PointNormProp}. Hence there is a unique map $h\colon
n \rightarrow \Dst(m)$ with $\Theta = \gr(h)_{*}(s_{*}(\omega))$.

\auxproof{
$$\begin{array}{rcl}
\Dst(\pi_{2})(\Theta)
& = &
\big(\Dst(\pi_{2}) \after (\st_{2})_{*} \after 
   \Dst(m\cdot\Dst(\pi_{1}))\big)(\Omega)
\\
& = &
\big((\Dst(\pi_{2}) \after \st_{2})_{*} \after 
   \Dst(m\cdot\Dst(\pi_{1}))\big)(\Omega)
\\
& \smash{\stackrel{\eqref{StrengthGraphDiag}}{=}} &
\big((\pi_{2})_{*} \after \Dst(m\cdot\Dst(\pi_{1}))\big)(\Omega)
\\
& = &
\big(\mu \after \Dst(\pi_{2}) \after \Dst(m\cdot\Dst(\pi_{1}))\big)(\Omega)
\\
& = &
\big(\mu \after \Dst^{2}(\pi_{1}) \after \Dst(\pi_{2})\big)(\Omega)
\\
& = &
\big(\Dst(\pi_{1}) \after \mu \after \Dst(\pi_{2})\big)(\Omega)
\\
& = &
\Dst(\pi_{1})\big(\hypcond{\omega}{s}\big)
\\
& \smash{\stackrel{\eqref{NrmOutputDiag}}{=}} &
\Dst(\pi_{1})\big(\gr(s)_{*}(\omega)\big)
\\
& = &
\big(\Dst(\pi_{1}) \after \gr(s)\big)_{*}(\omega)
\\
& \smash{\stackrel{\eqref{StrengthGraphDiag}}{=}} &
s_{*}(\omega).
\end{array}$$
}

Our aim is to prove $s \sqsubseteq t$ via $h \klafter s = t$. We shall
switch to a more concrete level. Since the distributions
$\hypcond{\omega}{s} \in \Dst(n\cdot \Dst(A))$ and
$\hypcond{\omega}{t} \in \Dst(m\cdot \Dst(A))$ are normalised, we can
write them as formal convex combinations:
\begin{equation}
\label{OrderThmPrfhyps}
\begin{array}{rclcrcl}
\hypcond{\omega}{s}
& = &
{\displaystyle\sum}_{i\in n} u_{i}\bigket{\kappa_{i}\varphi_{i}}
& \qquad\mbox{and}\qquad &
\hypcond{\omega}{t}
& = &
{\displaystyle\sum}_{j\in m} v_{j}\bigket{\kappa_{j}\psi_{j}},
\end{array}
\end{equation}

\noindent for $\varphi_{i},\psi_{j}\in\Dst(A)$ and $u_{i},v_{j} \in
          [0,1]$ with $\sum_{i}u_{i} = 1 = \sum_{j}v_{j}$. The
equation $\Dst(m\cdot
(\pi_{2})_{*})(\Omega) = \hypcond{\omega}{t}$ means that we can write:
\begin{equation}
\label{OrderThmPrfOmegaOne}
\begin{array}{rclcrcl}
\Omega
& = &
{\displaystyle\sum}_{j\in m} v_{j}\bigket{\kappa_{j}\rho_{j}}
& \qquad\mbox{for $\rho_{j}\in\Dst(n\cdot\Dst(A))$ with}\qquad
(\pi_{2})_{*}(\rho_{j})
& = &
\psi_{j}.
\end{array}
\end{equation}

\noindent The other equation about $\Omega$ gives:
$$\begin{array}{rcccccccl}
\hypcond{\omega}{s}
& = &
(\pi_{2})_{*}(\Omega)
& = &
\mu\big(\Dst(\pi_{2})(\Omega)\big)
& = &
\mu\big(\sum_{j} v_{j}\ket{\rho_{j}}\big)
& = &
{\displaystyle\sum}_{i,\chi} \big(\sum_{j}v_{j}\cdot\rho_{j}(\kappa_{i}\chi)\big)
   \bigket{\kappa_{i}\chi}.
\end{array}$$

\noindent But since this $\hypcond{\omega}{s}$ is normalised, as
described in~\eqref{OrderThmPrfhyps}, the only possible distributions
$\chi\in\Dst(A)$ are $\varphi_{i}$. Hence we can write:
\begin{equation}
\label{OrderThmPrfOmegaTwo}
\begin{array}{rclcrcl}
\rho_{j}
& = &
{\displaystyle\sum}_{i}\,
   \rho_{j}(\kappa_{i}\varphi_{i})\bigket{\kappa_{i}\varphi_{i}}
& \qquad\mbox{with}\qquad &
u_{i}
& = &
\sum_{j}v_{j}\cdot\rho_{j}(\kappa_{i}\varphi_{i}).
\end{array}
\end{equation}

\noindent The second equation in~\eqref{OrderThmPrfOmegaOne} can now
be unfolded to:
\begin{equation}
\label{OrderThmPrfOmegaOneTwo}
\begin{array}{rcl}
\psi_{j}
\hspace*{\arraycolsep}=\hspace*{\arraycolsep}
(\pi_{2})_{*}(\rho_{j})
& \smash{\stackrel{\eqref{OrderThmPrfOmegaTwo}}{=}} &
\mu\big(\Dst(\pi_{2})\big(\sum_{i} \rho_{j}(\kappa_{i}\varphi_{i})
   \bigket{\kappa_{i}\varphi_{i}}\big)\big)
\\
& = &
\mu\big(\sum_{i} \rho_{j}(\kappa_{i}\varphi_{i})
   \bigket{\varphi_{i}}\big)
\\
& = &
{\displaystyle\sum}_{a} 
   \big(\sum_{i} \rho_{j}(\kappa_{i}\varphi_{i})\cdot\varphi_{i}(a)\big)
   \bigket{a}.
\end{array}
\end{equation}

\noindent We can now express the distribution $\Theta\in\Dst(m\cdot
n)$ from~\eqref{OrderThmPrfTheta} as:
$$\begin{array}{rcl}
\Theta
& = &
\big((\st_{2})_{*} \after \Dst(m\cdot\Dst(\pi_{1}))\big)(\Omega) 
\\
& \smash{\stackrel{\eqref{OrderThmPrfOmegaOne}}{=}} &
(\st_{2})_{*}\Big({\displaystyle\sum}_{j} v_{j}
   \ket{\kappa_{j}\Dst(\pi_{1})(\rho_{j})}\Big)
\\
& \smash{\stackrel{\eqref{OrderThmPrfOmegaTwo}}{=}} &
\mu\Big(\Dst(\st_{2})\Big({\displaystyle\sum}_{j} v_{j}
   \ket{\kappa_{j}\big(\sum_{i}\rho_{j}(\kappa_{i}\varphi_{i})\ket{i}\big)}
   \Big)\Big)
\\
& = &
\mu\Big({\displaystyle\sum}_{j} v_{j}
   \ket{\sum_{i}\rho_{j}(\kappa_{i}\varphi_{i})\ket{\kappa_{j}i}}\Big)
\\
& = &
{\displaystyle\sum}_{j,i} v_{j}\cdot \rho_{j}(\kappa_{i}\varphi_{i})
   \ket{\kappa_{j}i}.
\end{array}$$

\noindent According to~\eqref{PointNormDown}, the function $h\colon n
\rightarrow \Dst(m)$ corresponding to this $\Theta\in\Dst(m\cdot n)$ is
given by:
\begin{equation}
\label{OrderThmPrfDefinedh}
\begin{array}{rcccccl}
h(i)
& = &
{\displaystyle\sum}_{j} \displaystyle
   \frac{\Theta(\kappa_{j}i)}{\sum_{j}\Theta(\kappa_{j}i)}\bigket{j}
& = &
{\displaystyle\sum}_{j} \displaystyle
   \frac{v_{j}\cdot \rho_{j}(\kappa_{i}\varphi_{i})}
   {\sum_{j} v_{j}\cdot \rho_{j}(\kappa_{i}\varphi_{i})}\bigket{j}
& \smash{\stackrel{\eqref{OrderThmPrfOmegaTwo}}{=}} &
{\displaystyle\sum}_{j} \displaystyle
   \frac{v_{j}\cdot \rho_{j}(\kappa_{i}\varphi_{i})}{u_{i}}\bigket{j}.
\end{array}
\end{equation}

\noindent We now prove the following equality of distributions in
$\Dst(m\cdot A)$.
$$\begin{array}{rcl}
\gr(h \klafter s)_{*}(\omega)
& \smash{\stackrel{\eqref{OrderThmPrfGivenh}}{=}} &
\big((h\cdot A) \klafter \gr(s)\big)_{*}(\omega)
\\
& \smash{\stackrel{\eqref{NrmMonicDiag}}{=}} &
\big((h\cdot A)_{*} \after (\st_{2})_{*}\big)(\hypcond{\omega}{s})
\\
& \smash{\stackrel{\eqref{OrderThmPrfhyps}}{=}} &
\big((h\cdot A)_{*} \after \mu \after \Dst(\st_{2})\big)
   \big(\sum_{i} u_{i}\bigket{\kappa_{i}\varphi_{i}}\big)
\\
& = &
\big((h\cdot A)_{*} \after \mu\big)\big(\sum_{i} u_{i}
   \bigket{\sum_{a}\varphi_{i}(a)\ket{\kappa_{i}a}}\big)
\\
& = &
\big(\mu \after \Dst(h\cdot A)\big)\big(\sum_{i,a} 
   u_{i}\cdot \varphi_{i}(a)\ket{\kappa_{i}a}\big)
\\
& = &
\mu\big(\sum_{i,a} u_{i} \cdot \varphi_{i}(a) 
   \bigket{\sum_{j} h(i)(j)\ket{\kappa_{j}a}}\big)
\\
& = &
\sum_{j,i,a} u_{i} \cdot \varphi_{i}(a) \cdot h(i)(j) \ket{\kappa_{j}a}
\\
& \smash{\stackrel{\eqref{OrderThmPrfDefinedh}}{=}} &
\sum_{j,i,a} \varphi_{i}(a) \cdot v_{j} \cdot \rho_{j}(\kappa_{i}\varphi_{i})
    \ket{\kappa_{j}a}
\\
& = &
\sum_{j,a} v_{j} \cdot \big(\sum_{i}\varphi_{i}(a) \cdot 
   \rho_{j}(\kappa_{i}\varphi_{i})\big) \ket{\kappa_{j}a}
\\
& \smash{\stackrel{\eqref{OrderThmPrfOmegaOneTwo}}{=}} &
\sum_{j,a} v_{j} \cdot \psi_{j}(a) \ket{\kappa_{j}a}
\\
& = &
\mu\big(\sum_{j}v_{j}\bigket{\sum_{a}\psi_{j}(a)\ket{\kappa_{j}a}}\big)
\\
& = &
\mu\big(\Dst(\st_{2})\big(\sum_{j}v_{j}\bigket{\kappa_{j}\psi_{j}}\big)
\\
& \smash{\stackrel{\eqref{OrderThmPrfhyps}}{=}} &
(\st_{2})_{*}(\hypcond{\omega}{t})
\\
& \smash{\stackrel{\eqref{NrmMonicDiag}}{=}} &
\gr(t)_{*}(\omega).
\end{array}$$

\noindent We have $\Dst(\pi_2)\big(\gr(t)_{*}(\omega)\big) = \omega$,
and $\supp(\omega) = A$ by assumption. Hence we can use uniqueness
from Proposition~\ref{PointNormProp} to obtain the required conclusion
$h\klafter s = t$. \QED
\end{myproof}

We conclude with an example of this refinement theorem, taken from the
unpublished extended version\footnote{Available from
  \url{http://www.cse.unsw.edu.au/~carrollm/probs/Papers/LiCS15.pdf}}
of~\cite{McIverMT15}, building on Example~\ref{HypCondRecoverEx}.

\begin{example}
\label{OrderEx}
We recall the two hyper distributions in $\Dst(2\cdot\Dst(\{H,T\}))$
and $\Dst(3\cdot\Dst(\{H,T\}))$ from Example~\ref{HypCondRecoverEx}:
$$\begin{array}{rcl}
\Phi
& = &
\frac{1}{2}\bigket{\kappa_{0}(\frac{2}{3}\ket{H} + \frac{1}{3}\ket{T})} +
   \frac{1}{2}\bigket{\kappa_{1}(\frac{1}{3}\ket{H} + \frac{2}{3}\ket{T})}
\\
\Psi
& = &
\frac{1}{3}\bigket{\kappa_{0}(\frac{2}{3}\ket{H} + \frac{1}{3}\ket{T})} +
   \frac{1}{3}\bigket{\kappa_{1}(\frac{1}{2}\ket{H} + \frac{1}{2}\ket{T})} +
   \frac{1}{3}\bigket{\kappa_{2}(\frac{1}{3}\ket{H} + \frac{2}{3}\ket{T})}
\end{array}$$

\noindent They satisfy $(\pi_{2})_{*}(\Phi) = \omega =
(\pi_{2})_{*}(\Psi)$, for $\omega = \frac{1}{2}\ket{H} +
\frac{1}{2}\ket{T}$. Moreover, they can be written as $\Phi =
\hypcond{\omega}{s}$ and $\Psi = \hypcond{\omega}{t}$ for tests
$s\colon \{H,T\} \rightarrow \Dst(2)$ and $t\colon \{H,T\} \rightarrow
\Dst(3)$ given by:
$$\left\{\begin{array}{rcl}
s(H)
& = &
\frac{2}{3}\ket{0} + \frac{1}{3}\ket{1}
\\
s(T)
& = &
\frac{1}{3}\ket{0} + \frac{2}{3}\ket{1}
\end{array}\right.
\qquad\mbox{and}\qquad
\left\{\begin{array}{rcl}
t(H)
& = &
\frac{4}{9}\ket{0} + \frac{1}{3}\ket{1} + \frac{2}{9}\ket{2}
\\
t(T)
& = &
\frac{2}{9}\ket{0} + \frac{1}{3}\ket{1} + \frac{4}{9}\ket{2}
\end{array}\right.$$

\noindent We claim $\Phi = \hypcond{\omega}{s} \sqsubseteq
\hypcond{\omega}{t} = \Psi$, via the distribution $\Omega \in
\Dst(3\cdot\Dst(2\cdot\Dst(\{H,T\})))$ given by:
$$\begin{array}{rcl}
\Omega
& = &
\frac{1}{3}\Bigket{\kappa_{0}\big(
   1\bigket{\kappa_{0}(\frac{2}{3}\ket{H}+\frac{1}{3}\ket{T})}\big)} \;+
\\[.5em]
& & \frac{1}{3}\Bigket{\kappa_{1}\big(
   \frac{1}{2}\bigket{\kappa_{0}(\frac{2}{3}\ket{H}+\frac{1}{3}\ket{T})} +
   \frac{1}{2}\bigket{\kappa_{1}(\frac{1}{3}\ket{H}+\frac{2}{3}\ket{T})}\big)}
   \; +
\\[.5em]
& & \frac{1}{3}\Bigket{\kappa_{2}\big(
   1\bigket{\kappa_{1}(\frac{1}{3}\ket{H}+\frac{2}{3}\ket{T})}\big)} 
\end{array}$$

\noindent This $\Omega$ proves the refinement $\Phi \sqsubseteq \Psi$
as in Definition~\ref{OrderDef}, since:
$$\begin{array}{rcl}
(\pi_{2})_{*}(\Omega)
\hspace*{\arraycolsep}=\hspace*{\arraycolsep}
\big(\mu \after \Dst(\pi_{2})\big)(\Omega)
& = &
\frac{1}{3}\bigket{\kappa_{0}(\frac{2}{3}\ket{H}+\frac{1}{3}\ket{T})} \;+
\\
& & \frac{1}{6}\bigket{\kappa_{0}(\frac{2}{3}\ket{H}+\frac{1}{3}\ket{T})} +
   \frac{1}{6}\bigket{\kappa_{1}(\frac{1}{3}\ket{H}+\frac{2}{3}\ket{T})} \;+
\\
& & \frac{1}{3}\bigket{\kappa_{1}(\frac{1}{3}\ket{H}+\frac{2}{3}\ket{T})}
\\
& = &
\frac{1}{2}\bigket{\kappa_{0}(\frac{2}{3}\ket{H}+\frac{1}{3}\ket{T})} +
   \frac{1}{2}\bigket{\kappa_{1}(\frac{1}{3}\ket{H}+\frac{2}{3}\ket{T})}
\\
& = &
\Phi
\\
\Dst(3\cdot(\pi_{2})_{*})(\Omega)
\hspace*{\arraycolsep}=\hspace*{\arraycolsep}
\Dst\big(3\cdot (\mu \after \Dst(\pi_{2}))\big)(\Omega)
& = &
\frac{1}{3}\bigket{\kappa_{0}\big(
   \frac{2}{3}\ket{H}+\frac{1}{3}\ket{T}\big)} \;+
\\
& & \frac{1}{3}\bigket{\kappa_{1}\big(
   \frac{2}{6}\ket{H}+\frac{1}{6}\ket{T} +
   \frac{1}{6}\ket{H}+\frac{2}{6}\ket{T}\big)}
   \; +
\\
& & \frac{1}{3}\bigket{\kappa_{2}\big(
   \frac{1}{3}\ket{H}+\frac{2}{3}\ket{T})}
\\
& = &
\frac{1}{3}\bigket{\kappa_{0}\big(
   \frac{2}{3}\ket{H}+\frac{1}{3}\ket{T}\big)} \;+
\\
& & \frac{1}{3}\bigket{\kappa_{1}\big(
   \frac{1}{2}\ket{H}+\frac{1}{2}\ket{T}\big)}
   \; +
\\
& & \frac{1}{3}\bigket{\kappa_{2}\big(
   \frac{1}{3}\ket{H}+\frac{2}{3}\ket{T})}
\\
& = &
\Psi.    
\end{array}$$

\noindent We illustrate how to obtain from $\Omega$ the map $h\colon 2
\rightarrow \Dst(3)$ that proves the refinement $s \sqsubseteq t$, as
in the proof of Theorem~\ref{OrderThm}, via the distribution $\Theta$
in~\eqref{OrderThmPrfTheta}:
$$\begin{array}{rcl}
\Theta
& = &
\big((\st_{2})_{*} \after \Dst(3\cdot\Dst(\pi_{1}))\big)(\Omega) 
\\
& = &
\mu\Big(\Dst(\st_{2})\Big(
\frac{1}{3}\Bigket{\kappa_{0}\big(1\bigket{0}\big)} +
\frac{1}{3}\Bigket{\kappa_{1}\big(
   \frac{1}{2}\bigket{0} + \frac{1}{2}\bigket{1}\big)} +
\frac{1}{3}\Bigket{\kappa_{2}\big(1\bigket{1}\big)}\Big)\Big)
\\
& = &
\mu\Big(\frac{1}{3}\big(1\bigket{\kappa_{0}0}\big) +
\frac{1}{3}\big(\frac{1}{2}\bigket{\kappa_{1}0} + 
   \frac{1}{2}\bigket{\kappa_{1}1}\big) +
\frac{1}{3}\big(1\bigket{\kappa_{2}1}\big)\Big)
\\
& = &
\frac{1}{3}\bigket{\kappa_{0}0} +
\frac{1}{6}\bigket{\kappa_{1}0} + \frac{1}{6}\bigket{\kappa_{1}1} +
\frac{1}{3}\bigket{\kappa_{2}1}
\end{array}$$

\noindent From this $\Theta$ we obtain the function $h\colon 2
\rightarrow \Dst(3)$ by pointwise normalisation~\eqref{PointNormDown}:
$$\begin{array}{rccclcrcccl}
h(0)
& = &
\frac{\nicefrac{1}{3}}{\nicefrac{1}{2}}\ket{0} + 
   \frac{\nicefrac{1}{6}}{\nicefrac{1}{2}}\ket{1}
& = &
\frac{2}{3}\ket{0} + \frac{1}{3}\ket{1}
& \qquad &
h(1)
& = &
\frac{\nicefrac{1}{6}}{\nicefrac{1}{2}}\ket{1} + 
   \frac{\nicefrac{1}{3}}{\nicefrac{1}{2}}\ket{2}
& = &
\frac{1}{3}\ket{1} + \frac{2}{3}\ket{2}
\end{array}$$

\noindent There is a refinement $s \sqsubseteq t$, as in
Definition~\ref{OrderDef}, since we have $h \klafter s = t$.

\auxproof{
Indeed $t = h \klafter s$ since:
$$\begin{array}{rcl}
(h_{*} \after s)(H)
& = &
\frac{2}{3}\cdot\frac{2}{3}\ket{0} + \frac{2}{3}\cdot\frac{1}{3}\ket{1} + 
   \frac{1}{3}\cdot\frac{1}{3}\ket{1} + \frac{1}{3}\cdot\frac{2}{3}\ket{2} 
\\
& = &
\frac{4}{9}\ket{0} + \frac{1}{3}\ket{1} + \frac{2}{9}\ket{2}
\\
& = &
t(H)
\\
(h_{*} \after s)(T)
& = &
\frac{1}{3}\cdot\frac{2}{3}\ket{0} + \frac{1}{3}\cdot\frac{1}{3}\ket{1} + 
   \frac{2}{3}\cdot\frac{1}{3}\ket{1} + \frac{2}{3}\cdot\frac{2}{3}\ket{2} 
\\
& = &
\frac{2}{9}\ket{0} + \frac{1}{3}\ket{1} + \frac{4}{9}\ket{2}
\\
& = &
t(T)
\end{array}$$
}

In the other direction, given this function $h\colon 2 \rightarrow
\Dst(3)$, one may check that the formula~\eqref{OrderThmPrfOmega}
gives the distribution $\Omega$ that we used above.

\auxproof{
First consider $h_{1} = h\after\pi_{1} \colon 2\cdot\Dst(2)
\rightarrow \Dst(3)$ and its graph $\gr(h_{1}) \colon
2\cdot\Dst(\{H,T\}) \rightarrow \Dst(3\cdot
(2\cdot\Dst(\{H,T\})))$. It is given by:
$$\left\{\begin{array}{rcl}
\gr(h_{1})(\kappa_{0}\psi)
& = &
\frac{2}{3}\bigket{\kappa_{0}\kappa_{0}\psi} +
   \frac{1}{3}\bigket{\kappa_{1}\kappa_{0}\psi}
\\
\gr(h_{1})(\kappa_{1}\psi)
& = &
\frac{1}{3}\bigket{\kappa_{1}\kappa_{1}\psi} +
   \frac{2}{3}\bigket{\kappa_{2}\kappa_{1}\psi}
\end{array}\right.$$

\noindent For readability's sake we abbreviate $\varphi_{0} =
\frac{2}{3}\ket{H}+\frac{1}{3}\ket{T})$ and $\varphi_{1} =
\frac{1}{3}\ket{H}+\frac{2}{3}\ket{T})$ in $\hypcond{\omega}{s} =
\frac{1}{2}\bigket{\kappa_{0}\varphi_{0}} +
\frac{1}{2}\bigket{\kappa_{1}\varphi_{1}} \in
\Dst(2\cdot\Dst(\{H,T\}))$. Then:
$$\begin{array}{rcl}
\gr(h_{1})_{*}(\hypcond{\omega}{s})
& = &
\frac{1}{3}\bigket{\kappa_{0}\kappa_{0}\varphi_{0}} +
   \frac{1}{6}\bigket{\kappa_{1}\kappa_{0}\varphi_{0}} +
   \frac{1}{6}\bigket{\kappa_{1}\kappa_{1}\varphi_{1}} +
   \frac{1}{3}\bigket{\kappa_{2}\kappa_{1}\varphi_{1}}
\end{array}$$

\noindent By normalising we obtain $\Omega$:
$$\begin{array}{rcl}
\hypcond{\big(\hypcond{\omega}{s}\big)}{h_1}
& = &
\Nrm\Big(\gr(h_{1})_{*}(\hypcond{\omega}{s})\Big)
\\
& = &
\Nrm\Big(\frac{1}{3}\bigket{\kappa_{0}\kappa_{0}\varphi_{0}} +
   \frac{1}{6}\bigket{\kappa_{1}\kappa_{0}\varphi_{0}} +
   \frac{1}{6}\bigket{\kappa_{1}\kappa_{1}\varphi_{1}} +
   \frac{1}{3}\bigket{\kappa_{2}\kappa_{1}\varphi_{1}}\Big)
\\
& = &
\frac{1}{3}\Bigket{\kappa_{0}\big(1\bigket{\kappa_{0}\varphi_{0}}\big)} +
   \frac{1}{3}\Bigket{\kappa_{1}\big(\frac{1}{2}\bigket{\kappa_{0}\varphi_{0}} +
      \frac{1}{2}\bigket{\kappa_{1}\varphi_{1}}\big)} +
   \frac{1}{3}\Bigket{\kappa_{2}\big(
     1\bigket{\kappa_{1}\varphi_{1}}\big)} 
\\
& = &
\Omega.
\end{array}$$
}
\end{example}



\section{Concluding remarks}

This paper provides a novel perspective on normalisation of discrete
probability distributions, by presenting it in `hyper' form as a map
$\Nrm \colon \Dst(n\cdot A) \rightarrow \Dst(n\cdot\Dst(A))$ that
satisfies various nice properties. The associated hyper conditioning
operation $\hypcond{\omega}{t}$ performs conditioning for all the
predicates incorporated in the test $t$ in parallel, and is a total
operation too. It has been implemented in the EfProb
tool~\cite{ChoJ17b}, see especially the manual~\cite{JacobsC17}.

Since we deal with finite discrete probability distributions, using
this copower is $n\cdot A$ is quite natural. But one could have
described normalisation also using a cartesian product $B\times A$,
for an arbitrary not necessarily finite set $B$, or as an indexed
coproduct $\coprod_{i\in I}A_{i}$, as in:
$$\xymatrix@C-1pc{
\Dst(B\times A)\ar[rr]^-{\Nrm} & & \Dst(B\times\Dst(A))
\qquad\mbox{or as} \qquad
\Dst(\coprod_{i\in I}A_{i})\ar[rr]^-{\Nrm} & & \Dst(\coprod_{i\in I}\Dst(A_{i}))
}$$

\noindent This does not fundamentally change the theory.

A different dimension of change is to consider other functors than
distribution $\Dst$. First, one could use the multiset functor $\Mlt$
over the non-negative real number, given by:
$$\begin{array}{rcl}
\Mlt(X)
& = &
\set{\varphi\colon X \rightarrow \mathbb{R}_{\geq 0}}{\supp(\varphi)
   \mbox{ is finite}}.
\end{array}$$

\noindent Then one can generalise normalisation from subdistributions
to such multisets (or `scores', as in~\cite{StatonYHKW16}) via a map:
$$\xymatrix@C-1pc{
\Mlt(n\cdot A)\ar[rr] & & \Mlt(n\cdot\Dst(A))
}$$

\noindent One then normalises non-negative real numbers to $1$.

A more drastic step is the move from discrete probability to
continuous probability, by replacing the distribution monad $\Dst$ on
sets with the Giry monad $\Giry$ on measurable
spaces~\cite{Giry82,Panangaden09}. How to best do this will be
explored in later work.

\subsection*{Acknowledgements} The author wishes to thank
Robin Adams, Kenta Cho, Caroll Morgan, Sam Staton, Bram Westerbaan and
Fabio Zanasi for helpful discussions and feedback.


\end{document}

\appendix

\section{Calculations}

\subsection{Example from~\cite{McIverMT15}}

We start with the example from the appendix\footnote{The extended
  version of~\cite{McIverMT15}, with the appendix that we use here, is
  only available online as
  \url{http://www.cse.unsw.edu.au/~carrollm/probs/Papers/LiCS15.pdf}.}
F of~\cite{McIverMT15} and elaborate its details in the current
`tagged' setting. It starts from the hyper distributions in
$\Dst(2\cdot\Dst(2))$ and $\Dst(3\cdot\Dst(2))$:
$$\begin{array}{rcl}
\Phi
& = &
\frac{1}{2}\bigket{\kappa_{0}(\frac{2}{3}\ket{H} + \frac{1}{3}\ket{T})} +
   \frac{1}{2}\bigket{\kappa_{1}(\frac{1}{3}\ket{H} + \frac{2}{3}\ket{T})}
\\
\Psi
& = &
\frac{1}{3}\bigket{\kappa_{0}(\frac{2}{3}\ket{H} + \frac{1}{3}\ket{T})} +
   \frac{1}{3}\bigket{\kappa_{1}(\frac{1}{2}\ket{H} + \frac{1}{2}\ket{T})} +
   \frac{1}{3}\bigket{\kappa_{2}(\frac{1}{3}\ket{H} + \frac{2}{3}\ket{T})}
\end{array}$$

\noindent (The corresponding hypers in~\cite{McIverMT15} are called
$\Delta_{S}$ and $\Delta_{I}$; the letters `$H$' and `$T$' stand for
`head' and `tail' in toin cossing; we treat them as elements of $2 =
\{H, T\}$.)

We first notice that these hyper distributions $\Phi,\Psi$ have the
same underlying state $\omega = \frac{1}{2}\ket{H} +
\frac{1}{2}\ket{T}\in \Dst(2)$ since:
$$\begin{array}{rcl}
\mu\big(\Dst(\nabla)(\Phi)\big)
& = &
(\frac{1}{2}\cdot\frac{2}{3} + \frac{1}{2}\cdot\frac{1}{3})\ket{H} + 
   (\frac{1}{2}\cdot\frac{1}{3} + \frac{1}{2}\cdot\frac{2}{3})\ket{T}
\\
& = &
\frac{1}{2}\ket{H} + \frac{1}{2}\ket{T}
\\
\mu\big(\Dst(\nabla)(\Psi)\big)
& = &
(\frac{1}{3}\cdot\frac{2}{3} + \frac{1}{3}\cdot\frac{1}{2}
    + \frac{1}{3}\cdot\frac{1}{3})\ket{H} + 
   (\frac{1}{3}\cdot\frac{1}{3} + \frac{1}{2}\cdot\frac{1}{2}  
    + \frac{1}{3}\cdot\frac{2}{3})\ket{T}
\\
& = &
\frac{1}{2}\ket{H} + \frac{1}{2}\ket{T}.
\end{array}$$

\noindent We first reconstruct $\Phi$ as hyper conditional $\Phi =
\hypcond{\omega}{s}$ for some test $s\colon 2 \rightarrow \Dst(2)$.
If $\Phi = \hypcond{\omega}{s}$, then, by
Lemma~\ref{HypCondPropLem}~\eqref{HypCondPropLemBang}:
$$\begin{array}{rcccccl}
s_{*}(\omega)
& = &
\Dst(2\cdot\bang)(\hypcond{\omega}{s})
& = &
\Dst(2\cdot\bang)(\Phi)
& = &
\frac{1}{2}\ket{0} + \frac{1}{2}\ket{1}.
\end{array}$$

\noindent This is the first step to find $s$. Write $s(H) = u\ket{0} +
(1-u)\ket{1}$ and $s(T) = v\ket{0} + (1-v)\ket{1}$ for (unknown)
probabilities $u,v\in [0,1]$, then:
$$\begin{array}{rcccl}
s_{*}(\omega)
& = &
\frac{1}{2}u\ket{0} + \frac{1}{2}(1-u)\ket{1} + 
   \frac{1}{2}v\ket{0} + \frac{1}{2}(1-v)\ket{1}
& = &
\frac{u+v}{2}\ket{0} + \frac{2-u-v}{2}\ket{1}
\end{array}$$

\noindent By combining this with the earlier equation we get a system
of two linear equations:
$$\left\{\begin{array}{rcl}
\frac{u+v}{2} & = & \frac{1}{2} \\
\frac{2-u-v}{2} & = & \frac{1}{2}
\end{array}\right.
\qquad\mbox{\it i.e.}\qquad
\left\{\begin{array}{rcl}
u+v & = & 1 \\
-u-v & = & -1
\end{array}\right.$$

\noindent They have many solutions. But we fix $u$, so that $v = 1-u$
and thus $1-v = 1 - (1-u) = u$, and compute the resulting hyper
conditional as second step:
$$\begin{array}{rcl}
\hypcond{\omega}{s}
& = &
\Nrm\big(\gr(s)_{*}(\omega)\big) 
\\
& = &
\Nrm\big(\frac{1}{2}u\ket{\kappa_{0}H} + 
   \frac{1}{2}(1-u)\ket{\kappa_{1}H} + 
   \frac{1}{2}(1-u)\ket{\kappa_{0}T} + 
   \frac{1}{2}u\ket{\kappa_{1}T}\big)
\\
& = &
\frac{1}{2}\bigket{\kappa_{0}(u\ket{H} + (1-u)\ket{T})} +
   \frac{1}{2}\bigket{\kappa_{1}((1-u)\ket{H} + u\ket{T})}
\end{array}$$

\noindent If this hyper equals $\Phi$, then we see that
$u=\frac{2}{3}$ is the only solution, so that $v = 1 - u =
\frac{1}{3}$.

Let's try again for $\Psi$. Let $\Psi = \hypcond{\omega}{t}$, for $t
\colon 2 \rightarrow \Dst(3)$, so that $t_{*}(\omega) =
\Dst(3\cdot\bang)(\Psi) = \frac{1}{3}\ket{0} + \frac{1}{3}\ket{1} +
\frac{1}{3}\ket{2}$. Write $t(H) = u\ket{0}+v\ket{1}+w\ket{2}$ and
$t(H) = x\ket{0}+y\ket{1}+z\ket{2}$ for unkowns $u,v,w$ and $x,y,z$ in
$[0,1]$ with $u+v+w = 1$ and $x+y+z=1$. Then:
$$\begin{array}{rcl}
t_{*}(\omega)
& = &
\frac{1}{2}u\ket{0} + \frac{1}{2}v\ket{1} + \frac{1}{2}w\ket{2} +
   \frac{1}{2}x\ket{0} + \frac{1}{2}y\ket{1} + \frac{1}{2}z\ket{2}
\\
& = &
\frac{1}{2}(u+x)\ket{0} + \frac{1}{2}(v+y)\ket{1} + \frac{1}{2}(w+z)\ket{2}
\end{array}$$

\noindent Hence we can eliminate one half of the variables, namely: $x
= \frac{2}{3}-u$, $y = \frac{2}{3}-v$, $z = \frac{2}{3}-w$. We use
them in the computation of the hyper conditional:
$$\begin{array}{rcl}
\hypcond{\omega}{t}
& = &
\Nrm\big(\gr(t)_{*}(\omega)\big) 
\\
& = &
\Nrm\big(\frac{1}{2} u\ket{\kappa_{0}H} + 
   \frac{1}{2}v\ket{\kappa_{1}H} + \frac{1}{2}w\ket{\kappa_{1}H}
\\
& & \qquad
   \frac{1}{2}\cdot(\frac{2}{3}-u)\ket{\kappa_{0}T} + 
   \frac{1}{2}\cdot(\frac{2}{3}-v)\ket{\kappa_{1}T} + 
   \frac{1}{2}\cdot(\frac{2}{3}-w)\ket{\kappa_{2}T}\big)
\\
& = &
\Nrm\big(\frac{1}{2} u\ket{\kappa_{0}H} + 
   \frac{1}{2}v\ket{\kappa_{1}H} + \frac{1}{2}w\ket{\kappa_{1}H} +
   \frac{2-3u}{6}\ket{\kappa_{0}T}
\\
& & \qquad
    + \frac{2-3v}{6}\ket{\kappa_{1}T} + 
   \frac{2-3w}{6}\ket{\kappa_{2}T}\big)
\\
& = &
\frac{1}{3}\bigket{\kappa_{0}(\frac{3u}{2}\ket{H} + \frac{2-3u}{2}\ket{T})} +
   \frac{1}{3}\bigket{\kappa_{1}(\frac{3v}{2}\ket{H} +
   \frac{2-3v}{2}\ket{T})} \\
& & \qquad + 
   \frac{1}{3}\bigket{\kappa_{2}(\frac{3w}{2}\ket{H} + \frac{2-3w}{2}\ket{T})}
\end{array}$$

\noindent The assumption $\Psi = \hypcond{\omega}{t}$ yields three
equations: $\frac{3u}{2} = \frac{2}{3}$, $\frac{3v}{2} = \frac{1}{2}$,
and $\frac{3w}{2} = \frac{1}{3}$, so that $u = \frac{4}{9}$, $v =
\frac{1}{3}$, $w = \frac{2}{9}$. Hence $x = \frac{2}{3}-u =
\frac{2}{9}$, $y = \frac{2}{3}-v = \frac{1}{3}$, $z = \frac{2}{3}-w =
\frac{4}{9}$.

We summarise what we have found so far: $\Phi = \hypcond{\omega}{s}$
and $\Psi = \hypcond{\omega}{t}$ for state $\omega =
\frac{1}{2}\ket{H} + \frac{1}{2}\ket{T}$ and tests:
$$\left\{\begin{array}{rcl}
s(H)
& = &
\frac{2}{3}\ket{0} + \frac{1}{3}\ket{1}
\\
s(T)
& = &
\frac{1}{3}\ket{0} + \frac{2}{3}\ket{1}
\end{array}\right.
\qquad\mbox{and}\qquad
\left\{\begin{array}{rcl}
t(H)
& = &
\frac{4}{9}\ket{0} + \frac{1}{3}\ket{1} + \frac{2}{9}\ket{2}
\\
t(T)
& = &
\frac{2}{9}\ket{0} + \frac{1}{3}\ket{1} + \frac{4}{9}\ket{2}
\end{array}\right.$$

\auxproof{
\noindent We calculate the associated joint distributions, in
$\Dst(2\cdot 2)$ and $\Dst(3\cdot 2)$.
$$\begin{array}{rcl}
\gr(s)_{*}(\omega)
& = &
\frac{1}{2}\cdot\frac{2}{3}\ket{\kappa_{0}H} + 
   \frac{1}{2}\cdot\frac{1}{3}\ket{\kappa_{1}H} +
   \frac{1}{2}\cdot\frac{1}{3}\ket{\kappa_{0}T} + 
   \frac{1}{2}\cdot\frac{2}{3}\ket{\kappa_{1}T}
\\
& = &
\frac{1}{3}\ket{\kappa_{0}H} + \frac{1}{6}\ket{\kappa_{1}H} +
   \frac{1}{6}\ket{\kappa_{0}T} + \frac{1}{3}\ket{\kappa_{1}T}
\\
\gr(t)_{*}(\omega)
& = &
\frac{1}{2}\cdot\frac{4}{9}\ket{\kappa_{0}H} + 
   \frac{1}{2}\cdot\frac{1}{3}\ket{\kappa_{1}H} + 
   \frac{1}{2}\cdot\frac{2}{9}\ket{\kappa_{2}H}
\\
& & \qquad
\frac{1}{2}\cdot\frac{2}{9}\ket{\kappa_{0}T} + 
   \frac{1}{2}\cdot\frac{1}{3}\ket{\kappa_{1}T} + 
   \frac{1}{2}\cdot\frac{4}{9}\ket{\kappa_{2}T}
\\
& = &
\frac{2}{9}\ket{\kappa_{0}H} + \frac{1}{6}\ket{\kappa_{1}H} + 
   \frac{1}{9}\ket{\kappa_{2}H} +
\frac{1}{9}\ket{\kappa_{0}T} + \frac{1}{6}\ket{\kappa_{1}T} + 
   \frac{2}{9}\ket{\kappa_{2}T}
\end{array}$$

}

We now obtain the tests $s,t$ in a more systematic manner. We first
apply the (Kleisli extension of the) map $[\Dst(\kappa_{i})]_{i}$ from
Lemma~\ref{NrmLem}~\eqref{NrmLemMonic} to $\Phi,\Psi$. This yields
the following distributions in $\Dst(2\cdot 2)$ and $\Dst(3\cdot 2)$.
$$\begin{array}{rcl}
\lefteqn{[\Dst(\kappa_{0}), \Dst(\kappa_{1})]_{*}(\Phi)} \\
& = &
\mu\big(\frac{1}{2}(\frac{2}{3}\ket{\kappa_{0}H} + 
   \frac{1}{3}\ket{\kappa_{0}T}) +
   \frac{1}{2}(\frac{1}{3}\ket{\kappa_{1}H} + \frac{2}{3}\ket{\kappa_{1}T})
\\
& = &
\frac{1}{3}\ket{\kappa_{0}H} + \frac{1}{6}\ket{\kappa_{0}T}) +
   \frac{1}{6}\ket{\kappa_{1}H} + \frac{1}{3}\ket{\kappa_{1}T} 
\\
\lefteqn{[\Dst(\kappa_{0}), \Dst(\kappa_{1}), \Dst(\kappa_{2})]_{*}(\Psi)} \\
& = &
\mu\big(\frac{1}{3}(\frac{2}{3}\ket{\kappa_{0}H} + 
   \frac{1}{3}\ket{\kappa_{0}T}) +
   \frac{1}{3}(\frac{1}{2}\ket{\kappa_{1}H} + \frac{1}{2}\ket{\kappa_{1}T}) +
   \frac{1}{3}(\frac{1}{3}\ket{\kappa_{2}H} + \frac{2}{3}\ket{\kappa_{2}T})
\\
& = &
\frac{2}{9}\ket{\kappa_{0}H} + \frac{1}{9}\ket{\kappa_{0}T} +
   \frac{1}{6}\ket{\kappa_{1}H} + \frac{1}{6}\ket{\kappa_{1}T} +
   \frac{1}{9}\ket{\kappa_{2}H} + \frac{2}{9}\ket{\kappa_{2}T}
\end{array}$$

\noindent We can now extract functions $s\colon 2 \rightarrow \Dst(2)$
and $t\colon 2 \rightarrow \Dst(3)$ via pointwise normalisation as
in~\eqref{??}.
$$\begin{array}{rcl}
s(H)
& = &
\frac{\nicefrac{1}{3}}{\nicefrac{1}{2}}\ket{0} +
   \frac{\nicefrac{1}{6}}{\nicefrac{1}{2}}\ket{1}
\hspace*{\arraycolsep}=\hspace*{\arraycolsep}
\frac{2}{3}\ket{0} + \frac{1}{3}\ket{1}
\\
s(T)
& = &
\frac{\nicefrac{1}{6}}{\nicefrac{1}{2}}\ket{0} +
   \frac{\nicefrac{1}{3}}{\nicefrac{1}{2}}\ket{1}
\hspace*{\arraycolsep}=\hspace*{\arraycolsep}
\frac{1}{3}\ket{0} + \frac{2}{3}\ket{1}
\\
t(H)
& = &
\frac{\nicefrac{2}{9}}{\nicefrac{1}{2}}\ket{0} +
   \frac{\nicefrac{1}{6}}{\nicefrac{1}{2}}\ket{1} +
   \frac{\nicefrac{1}{9}}{\nicefrac{1}{2}}\ket{2}
\hspace*{\arraycolsep}=\hspace*{\arraycolsep}
\frac{4}{9}\ket{0} + \frac{1}{3}\ket{1} + \frac{2}{9}\ket{2}
\\
t(T)
& = &
\frac{\nicefrac{1}{9}}{\nicefrac{1}{2}}\ket{0} +
   \frac{\nicefrac{1}{6}}{\nicefrac{1}{2}}\ket{1} +
   \frac{\nicefrac{2}{9}}{\nicefrac{1}{2}}\ket{2}
\hspace*{\arraycolsep}=\hspace*{\arraycolsep}
\frac{2}{9}\ket{0} + \frac{1}{3}\ket{1} + \frac{4}{9}\ket{2}.
\end{array}$$

We claim $s \sqsubseteq t$ via the map $h$ in a commuting
triangle:
$$\vcenter{\xymatrix@R-1.5pc@C+1pc{
& 2\ar[dd]|-{\bullet}^{h}
\\
2\ar[ur]|-{\bullet}^-{s}\ar[dr]|-{\bullet}_{t} & 
\\
& 3
}}\qquad\mbox{given by}\qquad
\left\{\begin{array}{rcl}
h(0) & = & \frac{2}{3}\ket{0} + \frac{1}{3}\ket{1}
\\
h(1) & = & \frac{1}{3}\ket{1} + \frac{2}{3}\ket{2}.
\end{array}
\right.$$

\noindent Indeed $t = h \klafter s$ since:
$$\begin{array}{rcl}
(h_{*} \after s)(H)
& = &
\frac{2}{3}\cdot\frac{2}{3}\ket{0} + \frac{2}{3}\cdot\frac{1}{3}\ket{1} + 
   \frac{1}{3}\cdot\frac{1}{3}\ket{1} + \frac{1}{3}\cdot\frac{2}{3}\ket{2} 
\\
& = &
\frac{4}{9}\ket{0} + \frac{1}{3}\ket{1} + \frac{2}{9}\ket{2}
\\
& = &
t(H)
\\
(h_{*} \after s)(T)
& = &
\frac{1}{3}\cdot\frac{2}{3}\ket{0} + \frac{1}{3}\cdot\frac{1}{3}\ket{1} + 
   \frac{2}{3}\cdot\frac{1}{3}\ket{1} + \frac{2}{3}\cdot\frac{2}{3}\ket{2} 
\\
& = &
\frac{2}{9}\ket{0} + \frac{1}{3}\ket{1} + \frac{4}{9}\ket{2}
\\
& = &
t(T)
\end{array}$$

We take $\Omega \in \Dst(3\cdot\Dst(2\cdot\Dst(2)))$ as `tagged'
version of the hyper-hyper distribution $\underline{\Delta}$
from~\cite{McIverMT15}:
$$\begin{array}{rcl}
\Omega
& = &
\frac{1}{3}\Bigket{\kappa_{0}\big(
   1\bigket{\kappa_{0}(\frac{2}{3}\ket{H}+\frac{1}{3}\ket{T})}\big)} \;+
\\
& & \frac{1}{3}\Bigket{\kappa_{1}\big(
   \frac{1}{2}\bigket{\kappa_{0}(\frac{2}{3}\ket{H}+\frac{1}{3}\ket{T})} +
   \frac{1}{2}\bigket{\kappa_{1}(\frac{1}{3}\ket{H}+\frac{2}{3}\ket{T})}\big)}
   \; +
\\
& & \frac{1}{3}\Bigket{\kappa_{2}\big(
   1\bigket{\kappa_{1}(\frac{1}{3}\ket{H}+\frac{2}{3}\ket{T})}\big)} 
\end{array}$$

\noindent This $\Omega$ proves $\Phi \sqsubseteq \Psi$ since:
$$\begin{array}{rcl}
\big(\mu \after \Dst(\nabla)\big)(\Omega)
& = &
\frac{1}{3}\bigket{\kappa_{0}(\frac{2}{3}\ket{H}+\frac{1}{3}\ket{T})} \;+
\\
& & \frac{1}{6}\bigket{\kappa_{0}(\frac{2}{3}\ket{H}+\frac{1}{3}\ket{T})} +
   \frac{1}{6}\bigket{\kappa_{1}(\frac{1}{3}\ket{H}+\frac{2}{3}\ket{T})} \;+
\\
& & \frac{1}{3}\bigket{\kappa_{1}(\frac{1}{3}\ket{H}+\frac{2}{3}\ket{T})}
\\
& = &
\frac{1}{2}\bigket{\kappa_{0}(\frac{2}{3}\ket{H}+\frac{1}{3}\ket{T})} +
   \frac{1}{2}\bigket{\kappa_{1}(\frac{1}{3}\ket{H}+\frac{2}{3}\ket{T})}
\\
& = &
\Phi
\\
\Dst\big(3\cdot (\mu \after \Dst(\nabla))\big)(\Omega)
& = &
\frac{1}{3}\Bigket{\kappa_{0}\big(
   \frac{2}{3}\ket{H}+\frac{1}{3}\ket{T}\big)} \;+
\\
& & \frac{1}{3}\Bigket{\kappa_{1}\big(
   \frac{2}{6}\ket{H}+\frac{1}{6}\ket{T} +
   \frac{1}{6}\ket{H}+\frac{2}{6}\ket{T}\big)}
   \; +
\\
& & \frac{1}{3}\Bigket{\kappa_{2}\big(
   1\bigket{\kappa_{1}(\frac{1}{3}\ket{H}+\frac{2}{3}\ket{T})}\big)} 
\\
& = &
\frac{1}{3}\Bigket{\kappa_{0}\big(
   \frac{2}{3}\ket{H}+\frac{1}{3}\ket{T}\big)} \;+
\\
& & \frac{1}{3}\Bigket{\kappa_{1}\big(
   \frac{1}{2}\ket{H}+\frac{1}{2}\ket{T}\big)}
   \; +
\\
& & \frac{1}{3}\Bigket{\kappa_{2}\big(
   1\bigket{\kappa_{1}(\frac{1}{3}\ket{H}+\frac{2}{3}\ket{T})}\big)} 
\\
& = &
\Psi.    
\end{array}$$

Next we illustrate how to obtain $\Omega$ from $h$, following the
recipe of~\cite{McIverMT15}. Consider $h_{1} = h\after\pi_{1} \colon
2\cdot\Dst(2) \rightarrow \Dst(3)$ and its graph $\gr(h_{1}) \colon
2\cdot\Dst(2) \rightarrow \Dst(3\cdot (2\cdot\Dst(2)))$. It is given
by:
$$\left\{\begin{array}{rcl}
\gr(h_{1})(\kappa_{0}\psi)
& = &
\frac{2}{3}\bigket{\kappa_{0}\kappa_{0}\psi} +
   \frac{1}{3}\bigket{\kappa_{1}\kappa_{0}\psi}
\\
\gr(h_{1})(\kappa_{1}\psi)
& = &
\frac{1}{3}\bigket{\kappa_{1}\kappa_{1}\psi} +
   \frac{2}{3}\bigket{\kappa_{2}\kappa_{1}\psi}
\end{array}\right.$$

\noindent For readability's sake we abbreviate $\varphi_{0} =
\frac{2}{3}\ket{H}+\frac{1}{3}\ket{T})$ and $\varphi_{1} =
\frac{1}{3}\ket{H}+\frac{2}{3}\ket{T})$ in $\Phi =
\frac{1}{2}\bigket{\kappa_{0}\varphi_{0}} +
\frac{1}{2}\bigket{\kappa_{1}\varphi_{1}} \in
\Dst(2\cdot\Dst(2))$. Then:
$$\begin{array}{rcl}
\gr(h_{1})_{*}(\Phi)
& = &
\frac{1}{3}\bigket{\kappa_{0}\kappa_{0}\varphi_{0}} +
   \frac{1}{6}\bigket{\kappa_{1}\kappa_{0}\varphi_{0}} +
   \frac{1}{6}\bigket{\kappa_{1}\kappa_{1}\varphi_{1}} +
   \frac{1}{3}\bigket{\kappa_{2}\kappa_{1}\varphi_{1}}
\end{array}$$

\noindent By normalising we obtain $\Omega$:
$$\begin{array}{rcl}
\hypcond{\Phi}{h_1}
& = &
\Nrm\Big(\gr(h_{1})_{*}(\Phi)\Big)
\\
& = &
\Nrm\Big(\frac{1}{3}\bigket{\kappa_{0}\kappa_{0}\varphi_{0}} +
   \frac{1}{6}\bigket{\kappa_{1}\kappa_{0}\varphi_{0}} +
   \frac{1}{6}\bigket{\kappa_{1}\kappa_{1}\varphi_{1}} +
   \frac{1}{3}\bigket{\kappa_{2}\kappa_{1}\varphi_{1}}\Big)
\\
& = &
\frac{1}{3}\Bigket{\kappa_{0}\big(1\bigket{\kappa_{0}\varphi_{0}}\big)} +
   \frac{1}{3}\Bigket{\kappa_{1}\big(\frac{1}{2}\bigket{\kappa_{0}\varphi_{0}} +
      \frac{1}{2}\bigket{\kappa_{1}\varphi_{1}}\big)} +
   \frac{1}{3}\Bigket{\kappa_{2}\big(
     1\bigket{\kappa_{1}\varphi_{1}}\big)} 
\\
& = &
\Omega.
\end{array}$$

\noindent Thus $\Omega \in \Dst(3\cdot\Dst(2\cdot\Dst(2)))$ contains
all the information to reconstruct not only the hyper distributions
$\Psi$ and $\Psi$ with $\Phi \sqsubseteq \Psi$, but also the tests
$s,t$ with $s \sqsubseteq t$ via the map $h$ satisfying $t = h
\klafter s$, and the state $\omega$ such that $\Phi =
\hypcond{\omega}{s}$ and $\Psi = \hypcond{\omega}{t}$.

What we still want to do is see how to obtain $h$ from $\Omega$. We
first compute the following distribution in $\Dst(3\cdot 2)$
$$\begin{array}{rcl}
\lefteqn{\big([\Dst(\kappa_{0}), \Dst(\kappa_{1}), \Dst(\kappa_{2})]_{*}
   \after \Dst(3\cdot\Dst(\pi_{1}))\big)(\Omega)} \\
& = &
[\Dst(\kappa_{0}), \Dst(\kappa_{1}), \Dst(\kappa_{2})]_{*}\Big(
\frac{1}{3}\Bigket{\kappa_{0}\big(1\bigket{0}\big)} +
\frac{1}{3}\Bigket{\kappa_{1}\big(
   \frac{1}{2}\bigket{0} + \frac{1}{2}\bigket{1}\big)} +
\frac{1}{3}\Bigket{\kappa_{2}\big(1\bigket{1}\big)}\Big)
\\
& = &
\mu\Big(\frac{1}{3}\big(1\bigket{\kappa_{0}0}\big) +
\frac{1}{3}\big(\frac{1}{2}\bigket{\kappa_{1}0} + 
   \frac{1}{2}\bigket{\kappa_{1}1}\big) +
\frac{1}{3}\big(1\bigket{\kappa_{2}1}\big)\Big)
\\
& = &
\frac{1}{3}\bigket{\kappa_{0}0} +
\frac{1}{6}\bigket{\kappa_{1}0} + \frac{1}{6}\bigket{\kappa_{1}1} +
\frac{1}{3}\bigket{\kappa_{2}1}
\end{array}$$

\noindent From this we obtain the function $h\colon 2 \rightarrow \Dst(3)$
by pointwise normalisation:
$$\begin{array}{rcccl}
h(0)
& = &
\frac{\nicefrac{1}{3}}{\nicefrac{1}{2}}\ket{0} + 
   \frac{\nicefrac{1}{6}}{\nicefrac{1}{2}}\ket{1}
& = &
\frac{2}{3}\ket{0} + \frac{1}{3}\ket{1}
\\
h(1)
& = &
\frac{\nicefrac{1}{6}}{\nicefrac{1}{2}}\ket{1} + 
   \frac{\nicefrac{1}{3}}{\nicefrac{1}{2}}\ket{2}
& = &
\frac{1}{3}\ket{1} + \frac{2}{3}\ket{2}
\end{array}$$

\subsection{Own example}

We elaborate another, own ad hoc example. Consider tests $t, s$,
related by a Kleisli map $h$ in:
$$\vcenter{\xymatrix@R-1.5pc@C+1pc{
& 3
\\
2\ar[ur]|-{\bullet}^-{s}\ar[dr]|-{\bullet}_{t} & 
\\
& 2\ar[uu]|-{\bullet}_{h}
}}\qquad\mbox{given by}\qquad
\left\{\begin{array}{rcl}
t(0) & = & \frac{1}{3}\ket{0} + \frac{2}{3}\ket{1} 
\\
t(1) & = & 1\ket{1}
\\
s(0) & = & \frac{1}{12}\ket{0} + \frac{11}{18}\ket{1} + \frac{11}{36}\ket{2}
\\
s(1) & = & \frac{2}{3}\ket{1} + \frac{1}{3}\ket{2}
\\
h(0) & = & \frac{1}{4}\ket{0} + \frac{1}{2}\ket{1} + \frac{1}{4}\ket{2}
\\
h(1) & = & \frac{2}{3}\ket{1} + \frac{1}{3}\ket{2}.
\end{array}
\right.$$

\noindent Then indeed $s = h \klafter t$ since:
$$\begin{array}{rcl}
h_{*}(t)(0)
& = &
h_{*}\big(\frac{1}{3}\ket{0} + \frac{2}{3}\ket{1}\big) 
\\
& = &
\frac{1}{3}\cdot\frac{1}{4}\ket{0} + \frac{1}{3}\cdot\frac{1}{2}\ket{1} + 
   \frac{1}{3}\cdot\frac{1}{4}\ket{2} + 
   \frac{2}{3}\cdot\frac{2}{3}\ket{1} + \frac{2}{3}\cdot\frac{1}{3}\ket{2}
\\
& = &
\frac{1}{12}\ket{0} + (\frac{1}{6}+\frac{4}{9})\ket{1} + 
   (\frac{1}{12}+\frac{2}{9})\ket{2}
\\
& = &
\frac{1}{12}\ket{0} + \frac{11}{18}\ket{1} + \frac{11}{36}\ket{2} \\
& = &
s(0)
\\
h_{*}(t)(1)
& = &
h_{*}\big(1\ket{1}\big) 
\\
& = &
\frac{2}{3}\ket{1} + \frac{1}{3}\ket{2}
\\
& = &
s(1).
\end{array}$$

\auxproof{
In general we have:
$$\begin{array}{rcl}
\hypcond{\omega}{s}
& = &
\Nrm\big(\gr(s)_{*}(\omega)\big) 
\\
& = &
\Nrm\big(\sum_{i,x} \omega(x)\cdot s(x)(i)\ket{\kappa_{i}x}\big)
\\
& = &
{\displaystyle\sum}_{i}\big(\sum_{x}\omega(x)\cdot s(x)(i)\big)
   \Bigket{\kappa_{i}\big(\sum_{x}
   \frac{\omega(x)\cdot s(x)(i)}{\sum_{x}\omega(x)\cdot s(x)(i)}\ket{x}\big)}
\\
& = &
{\displaystyle\sum}_{i}\big(\omega\models s_{i}\big)
   \Bigket{\kappa_{i}\big(\sum_{x}
   \frac{\omega(x)\cdot s(x)(i)}{\omega\models s_{i}}\ket{x}\big)}
\end{array}$$

\noindent where $s_{i}$ is the $i$-th predicate in the test $s$, given
by $s_{i}(x) = s(x)(i)$.
}

We consider the state $\omega = \frac{1}{4}\ket{0} +
\frac{3}{4}\ket{1}$. Then $\hypcond{\omega}{s} \in \Dst(2\cdot
\Dst(2))$ and $\hypcond{\omega}{t} \in \Dst(3\cdot \Dst(2))$ are given
by:
$$\begin{array}{rcl}
\hypcond{\omega}{s}
& = &
\Nrm\big(\gr(s)_{*}(\omega)\big)
\\
& = &
\Nrm\big(\frac{1}{4}\cdot\frac{1}{12}\ket{\kappa_{0}0} + 
   \frac{1}{4}\cdot\frac{11}{18}\ket{\kappa_{1}0} +
   \frac{1}{4}\cdot\frac{11}{36}\ket{\kappa_{2}0} +
   \frac{3}{4}\cdot\frac{2}{3}\ket{\kappa_{1}1} +
   \frac{3}{4}\cdot\frac{1}{3}\ket{\kappa_{2}1}\big)
\\
& = &
\Nrm\big(\frac{1}{48}\ket{\kappa_{0}0} + 
   \frac{11}{72}\ket{\kappa_{1}0} +
   \frac{11}{144}\ket{\kappa_{2}0} +
   \frac{1}{2}\ket{\kappa_{1}1} +
   \frac{1}{4}\ket{\kappa_{2}1}\big)
\\
& = &
\frac{1}{48}\bigket{\kappa_{0}\big(1\ket{0}\big)} +
   \frac{47}{72}\bigket{\kappa_{1}\big(
      \frac{11}{47}\ket{0} + \frac{36}{47}\ket{1}\big)} +
   \frac{47}{144}\bigket{\kappa_{2}\big(
      \frac{11}{47}\ket{0} + \frac{36}{47}\ket{1}\big)}
\\
\hypcond{\omega}{t}
& = &
\Nrm(\big(\gr(t)_{*}(\omega)\big)
\\
& = &
\Nrm\big(\frac{1}{4}\cdot\frac{1}{3}\ket{\kappa_{0}0} + 
   \frac{1}{4}\cdot\frac{2}{3}\ket{\kappa_{1}0} +
   \frac{3}{4}\cdot 1\ket{\kappa_{1}1}\big)
\\
& = &
\Nrm\big(\frac{1}{12}\ket{\kappa_{0}0} + \frac{1}{6}\ket{\kappa_{1}0} +
   \frac{3}{4}\ket{\kappa_{1}1}\big)
\\
& = &
\frac{1}{12}\bigket{\kappa_{0}\big(1\ket{0}\big)}
   + \frac{11}{12}\bigket{\kappa_{1}\big(
      \frac{2}{11}\ket{0} + \frac{9}{11}\ket{1}\big)} 
\end{array}$$

\noindent We now take $\Omega \in \Dst(2\cdot\Dst(3\cdot\Dst(2)))$
given by $\Omega = \Nrm\big(\gr(h_{1})_{*}(\hypcond{\omega}{t})\big)$,
as before via $h_{1} = h \after \pi_{1} \colon 2\cdot\Dst(2)
\rightarrow \Dst(3)$. For simplicity we write $\hypcond{\omega}{t} =
\frac{1}{12}\bigket{\kappa_{0}\varphi_{0}} +
\frac{11}{12}\bigket{\kappa_{1}\varphi_{1}}$ where $\varphi_{0} =
1\ket{0}$ and $\varphi_{1} = \frac{2}{11}\ket{0} +
\frac{9}{11}\ket{1}$.  Then:
$$\begin{array}{rcl}
\Omega
& = &
\Nrm\big(\gr(h_{1})_{*}(\hypcond{\omega}{t})\big)
\\
& = &
\Nrm\big(\frac{1}{12}\cdot\frac{1}{4}\ket{\kappa_{0}\kappa_{0}\varphi_{0}} +
\frac{1}{12}\cdot\frac{1}{2}\ket{\kappa_{1}\kappa_{0}\varphi_{0}} +
\frac{1}{12}\cdot\frac{1}{4}\ket{\kappa_{2}\kappa_{0}\varphi_{0}} \;+
\\
& & \qquad
\frac{11}{12}\cdot\frac{2}{3}\ket{\kappa_{1}\kappa_{1}\varphi_{1}} + 
\frac{11}{12}\cdot\frac{1}{3}\ket{\kappa_{2}\kappa_{1}\varphi_{1}}\big)
\\
& = &
\Nrm\big(\frac{1}{48}\ket{\kappa_{0}\kappa_{0}\varphi_{0}} +
\frac{1}{24}\ket{\kappa_{1}\kappa_{0}\varphi_{0}} +
\frac{1}{48}\ket{\kappa_{2}\kappa_{0}\varphi_{0}} \;+
\\
& & \qquad
\frac{11}{18}\ket{\kappa_{1}\kappa_{1}\varphi_{1}} +
\frac{11}{36}\ket{\kappa_{2}\kappa_{1}\varphi_{1}}\big)
\\
& = &
\frac{1}{48}\Bigket{\kappa_{0}\big(1\ket{\kappa_{0}\varphi_{0}}\big)} +
   \frac{47}{72}\Bigket{\kappa_{1}\big(
    \frac{3}{47}\ket{\kappa_{0}\varphi_{0}} + 
    \frac{44}{47}\ket{\kappa_{1}\varphi_{1}}\big)}
  \; +
\\
& & \qquad 
   \frac{47}{144}\Bigket{\kappa_{2}\big(
    \frac{3}{47}\ket{\kappa_{0}\varphi_{0}} + 
    \frac{44}{47}\ket{\kappa_{1}\varphi_{1}}\big)}
\\
& = & 
\frac{1}{48}\Bigket{\kappa_{0}\big(1\ket{\kappa_{0}\big(1\ket{0}\big)}\big)} 
   \;+
\\
& & \frac{47}{72}\Bigket{\kappa_{1}\big(
    \frac{3}{47}\ket{\kappa_{0}\big(1\ket{0}\big)} + 
    \frac{44}{47}\ket{\kappa_{1}\big(
      \frac{2}{11}\ket{0} + \frac{9}{11}\ket{1}\big)}\big)}
  \; +
\\
& & \frac{47}{144}\Bigket{\kappa_{2}\big(
    \frac{3}{47}\ket{\kappa_{0}\big(1\ket{0}\big)} + 
    \frac{44}{47}\ket{\kappa_{1}\big(
      \frac{2}{11}\ket{0} + \frac{9}{11}\ket{1}\big)}\big)}
\end{array}$$

\noindent Then indeed:
$$\begin{array}{rcl}
\big(\mu \after \Dst(\nabla)\big)(\Omega)
& = &
\frac{1}{48}\ket{\kappa_{0}\big(1\ket{0}\big)}
   \;+
\\
& & \frac{3}{72}\ket{\kappa_{0}\big(1\ket{0}\big)} + 
    \frac{11}{18}\ket{\kappa_{1}\big(
      \frac{2}{11}\ket{0} + \frac{9}{11}\ket{1}\big)}
  \; +
\\
& & \frac{1}{48}\ket{\kappa_{0}\big(1\ket{0}\big)} + 
    \frac{11}{36}\ket{\kappa_{1}\big(
      \frac{2}{11}\ket{0} + \frac{9}{11}\ket{1}\big)}
\\
& = &
(\frac{1}{48}+\frac{3}{72}+\frac{1}{48})\ket{\kappa_{0}\big(1\ket{0}\big)} \;+
\\
& & \qquad (\frac{11}{18}+\frac{11}{36})\ket{\kappa_{1}\big(
      \frac{2}{11}\ket{0} + \frac{9}{11}\ket{1}\big)}
\\
& = &
\frac{1}{12}\ket{\kappa_{0}\big(1\ket{0}\big)} +
   \frac{11}{12}\ket{\kappa_{1}\big(
      \frac{2}{11}\ket{0} + \frac{9}{11}\ket{1}\big)}
\\
& = &
\hypcond{\omega}{t}
\\
\Dst(2\cdot (\mu \after \Dst(\nabla)))(\Omega)
& = &
\frac{1}{48}\bigket{\kappa_{0}\big(1\ket{0}\big)} 
   \;+
\\
& & \frac{47}{72}\bigket{\kappa_{1}\big(\frac{3}{47}\ket{0} + 
    \frac{8}{47}\ket{0} + \frac{36}{11}\ket{1}\big)}
  \; +
\\
& & \frac{47}{144}\bigket{\kappa_{2}\big(
    \frac{3}{47}1\ket{0} + \frac{8}{47}\ket{0} + \frac{36}{47}\ket{1}\big)}
\\
& = &
\frac{1}{48}\bigket{\kappa_{0}\big(1\ket{0}\big)} 
   \;+
\\
& & \frac{47}{72}\bigket{\kappa_{1}\big(\frac{11}{47}\ket{0} + 
    \frac{36}{11}\ket{1}\big)}
  \; +
\\
& & \frac{47}{144}\bigket{\kappa_{2}\big(
    \frac{11}{47}\ket{0} + \frac{36}{47}\ket{1}\big)}
\\
& = &
\hypcond{\omega}{s}.
\end{array}$$

\end{document}

Our starting point is an $n$-test on a set $A$, described as a Kleisli
map $p\colon X \klto n$, that is, as a function $p\colon X \rightarrow
\Dst(n)$. The $n$ predicates $p_{i}\in [0,1]^{A}$ involved may be
extracted as $p_{i}(a) = p(a)(i)$. As noted before, we have
$\bigovee_{i}p_{i} = \one$.

Now let's assume we have a state $\omega\in\Dst(A)$, written as map
$1 \klto A$. We can now form the Kleisli composite:
$$\xymatrix{
1\ar[r]|-{\bullet}^-{\omega} & A\ar[r]|-{\bullet}^-{p} & n
}$$

\noindent This composite is the $n$-tuple of validities:
$$\omega\models p_{1}, \;\ldots, \;\omega\models p_{n} \;\in\; [0,1]
\qquad\mbox{with}\qquad
\begin{array}{rcl}
\sum_{i}(\omega\models p_{i})
& = &
1.
\end{array}$$

\noindent More formally, following~\eqref{KlLiftEqn}:
\begin{equation}
\label{DstTestValidityEqn}
\begin{array}{rcl}
p \klafter \omega
\hspace*{\arraycolsep}=\hspace*{\arraycolsep}
p_{*}(\omega)
& = &
\sum_{i}(\sum_{x}\omega(x)\cdot p(x)(i))\ket{i} \\
& = &
\sum_{i}(\sum_{x}\omega(x)\cdot p_{i}(x))\ket{i} \\
& = &
\sum_{i}(\omega \models p_{i})\ket{i}. 
\end{array}
\end{equation}

\begin{definition}
\label{DstHyperCondDef}
In the above situation, with an $n$-test $p\colon A \klto n$ and a
state $\omega\in\Dst(A)$. We use the name \emph{hyper conditional} and
notation $\hypcond{\omega}{p}$ for the following formal convex sum.
\begin{equation}
\label{DstHyperCondEqn}
\begin{array}{rcl}
\hypcond{\omega}{p}
& \;\defn{=}\; &
{\displaystyle\sum}_{i} (\omega\models p_{i})
   \Bigket{\kappa_{i}(\cond{\omega}{p_i})}
   \;\in\; \Dst\big(n\cdot \Dst(A)\big).
\end{array}
\end{equation}

\noindent Thus we have a Kleisli map $\hypcond{(-)}{p} \colon \Dst(A)
\klto \Dst(A) + \cdots + \Dst(A)$.
\end{definition}

A crucial point is that in the definition of the hyper conditional
$\hypcond{\omega}{p}$ we do not have to worry about definedness of the
ordinary conditionals $\cond{\omega}{p_i}$. If the validity
$\omega\models p_{i}$ is zero, then the term $(\omega\models p_{i})
\bigket{\kappa_{i}(\cond{\omega}{p_i})}$ disappears from the convex
sum~\eqref{DstHyperCondEqn}.

\begin{theorem}
\label{DstHyperCondThm}
The hyper conditional $\hypcond{\omega}{p}$ from
Definition~\ref{DstHyperCondDef} is the unique distribution $\varphi
\in \Dst(n\cdot \Dst(A))$ making the following three `Kleisli'
diagrams commute.
\begin{equation}
\label{DstHyperCondDiag}
\vcenter{\xymatrix{
1\ar[r]|-{\bullet}^-{\varphi}\ar[d]|-{\bullet}_{\omega} &
  n\cdot\Dst(A)\ar[d]|-{\bullet}^-{n\cdot\bang}
& 
1\ar[r]|-{\bullet}^-{\varphi}\ar[d]|-{\bullet}_{\omega} &
  n\cdot\Dst(A)\ar[d]|-{\bullet}^-{\nabla}
& 
1\ar[r]|-{\bullet}^-{\varphi}
    \ar[d]|-{\bullet}_{\instr_{p}\klafter\omega} &
  n\cdot\Dst(A)\ar[d]|-{\bullet}^-{[\Dst(\kappa_{i})]_{i}}
\\
A\ar[r]|-{\bullet}_-{p} & n
& 
A & \Dst(A)\ar[l]|-{\bullet}^-{\idmap}
& 
n\cdot A & \Dst(n\cdot A)\ar[l]|-{\bullet}^-{\idmap}
}}
\end{equation}
\end{theorem}

\begin{myproof}
We first show that $\hypcond{\omega}{p}$ satisfies these equations.
The first one in~\eqref{DstHyperCondDiag} says the following: the
outcome $\Dst(n\cdot\bang)(\hypcond{\omega}{p})$ deletes the terms
$\cond{\omega}{p_i}$ from the sum $\hypcond{\omega}{p} = \sum_{i}
(\omega\models p_{i})\bigket{\kappa_{i}(\cond{\omega}{p_i})}$.  Then
we are left with $\sum_{i} (\omega\models p_{i})\ket{i} \in
\Dst(n)$. But this is precisely the outcome of $p\klafter\omega$, as
computed in~\eqref{DstTestValidityEqn}.

In the second equation, the codiagonal $\nabla = [\idmap, \ldots,
  \idmap]$ delets all inner coprojections $\kappa_{i}$ in
$\hypcond{\omega}{p}$. Then:
$$\begin{array}{rcll}
\idmap\klafter \nabla \klafter \hypcond{\omega}{p}
& = &
\idmap[*]\Big(\sum_{i}(\omega\models p_{i})\bigket{\cond{\omega}{p_i}}\Big) \\
& \smash{\stackrel{\eqref{KlLiftEqn}}{=}} &
\sum_{a} (\sum_{i}(\omega\models p_{i})\cdot (\cond{\omega}{p_{i}}(a)))
   \bigket{a} \\
& \smash{\stackrel{\eqref{DstCondDistEqn}}{=}} &
\sum_{a} (\sum_{i}\omega(a)\cdot p_{i}(a)) \bigket{a} \\
& = &
\sum_{a} (\omega(a)\cdot \sum_{i}p(a)(i)) \bigket{a} \\
& = &
\sum_{a}\omega(a)\ket{a} & \mbox{since $p(a)\in\Dst(n)$} \\
& = &
\omega.
\end{array}$$

For the third equation in~\eqref{DstHyperCondDiag} we first notice
that the composite $\instr_{p} \klafter \omega \in \Dst(n\cdot A)$ has
the form:
$$\begin{array}{rcl}
\instr_{p} \klafter \omega
\hspace*{\arraycolsep}=\hspace*{\arraycolsep}
(\instr_{p})_{*}(\omega)
& = &
\sum_{i,a} (\omega(a)\cdot p(a)(i))\bigket{\kappa_{i}a} \\
& = &
\sum_{i,a} (\omega(a)\cdot p_{i}(a))\bigket{\kappa_{i}a}.
\end{array}$$

\noindent The map $[\Dst(\kappa_{i})]_{i}$ pushes the coprojections
$\kappa_{i}$ inside, turning $\kappa_{i}(\cond{\omega}{p_i}) \in
n\cdot\Dst(A)$ into $\sum_{a}\frac{\omega(a)\cdot
  p_{i}(a)}{\omega\models p_{i}} \ket{\kappa_{i}a} \in \Dst(n\cdot
A)$. We then get:
$$\begin{array}{rcl}
\idmap\klafter[\Dst(\kappa_{i})]_{i} \klafter \hypcond{\omega}{p}
& = &
\idmap[*]\Big(\sum_{i}(\omega\models p_{i}) 
   \Bigket{\sum_{a}\frac{\omega(a)\cdot p_{i}(a)}{\omega\models p_{i}} 
      \ket{\kappa_{i}a}}\Big) \\
& = &
\sum_{j,a} \big(\sum_{i}(\omega\models p_{i})\cdot 
   \big(\sum_{a}\frac{\omega(a)\cdot p_{i}(a)}{\omega\models p_{i}} 
      \ket{\kappa_{i}a})(\kappa_{j}a)\big)\bigket{\kappa_{j}a} \\
& = &
\sum_{j,a} \big((\omega\models p_{i})\cdot 
   \big(\frac{\omega(a)\cdot p_{i}(a)}{\omega\models p_{i}}\big)
   \bigket{\kappa_{j}a} \\
& = &
\sum_{j,a} \big(\omega(a)\cdot p_{j}(a)\big)\bigket{\kappa_{j}a}.
\end{array}$$

\noindent This is, up to change of bound variables, indeed the same
outcome as obtained from $\instr_{p} \klafter \omega$.

Now, in the other direction, let $\varphi\in\Dst(n\cdot \Dst(A))$. 
The first diagram in~\eqref{DstHyperCondDiag} says that 
$\varphi$ must be of the form:
$$\begin{array}{rclcrcl}
\varphi
& = &
\sum_{i,j} r_{ij}\ket{\kappa_{i}\psi_{ij}}
& \qquad\mbox{with}\qquad &
\sum_{j}r_{ij}
& = &
(\omega\models p_{i}).
\end{array}$$

\noindent The second equation says that for each $a\in A$,
$$\begin{array}{rcl}
\omega(a)
& = &
\sum_{i,j} r_{ij}\cdot\psi_{ij}(a).
\end{array}$$

\noindent From this we conclude that the support $\supp(\psi_{ij})$ of
each $\psi_{ij}$ is contained in the support $\supp(\omega)$ of
$\omega$. 

The third equation yields for $\ell\leq n$ and and $b\in A$:
$$\begin{array}{rcl}
\omega(b)\cdot p_{\ell}(b)
& = &
\big(\idmap \klafter [\Dst(\kappa_{i})]_{i} \klafter 
   \varphi\big)(\kappa_{\ell}b) \\
& = &
\idmap[*]\Big(\sum_{i,j}r_{ij}\bigket{\sum_{a}\psi_{ij}(a)\ket{\kappa_{i}a}}\Big)
   (\kappa_{\ell}b) \\
& = &
\sum_{i,j}r_{ij}\cdot \big(\sum_{a}\psi_{ij}(a)\ket{\kappa_{i}a}\big)
   (\kappa_{\ell}b) \\
& = &
\sum_{j} r_{\ell j} \cdot \psi_{\ell j}(b).
\end{array}$$

\end{myproof}

\begin{myproof}
Let $h\colon n\cdot A \rightarrow \Dst(m)$ satisfy $h \klafter
\gr(s) = t$. The associated twisted graph map is of the form:
$$\xymatrix{
\gr(h) = \Big(n\cdot A\ar[r]^-{\tuple{h,\idmap}} & 
   \Dst(m)\times (n\cdot A)\ar[r]^-{\st_1} & \Dst(m\cdot (n\cdot A)))
}$$

\noindent Hence we take:
$$\begin{array}{rcl}
\Omega
& \defn{=} &
\gr(h)_{*}\big(\gr(s)_{*}(\omega)\big)
   \;\in\; \Dst\big(m\cdot(n\cdot A)\big).
\end{array}$$

\noindent We first note that:
$$\begin{array}{rcl}
\Dst(m\cdot \pi_{2}) \after \gr(h)_{*} \after \gr(s)
& = &
\big(\Dst(\idmap\times\pi_{2}) \after \st_{1} \after \tuple{h, \idmap}\big)_{*}
   \after \gr(s)
\\
& = &
\mu \after \Dst(\st_{1} \after \tuple{h, \pi_{2}}) \after \st_{1} \after
   \tuple{s, \idmap}
\\
& = &
\mu \after \Dst(\st_{1} \after (h\times\idmap)) \after \st_{1} \after
   (\st_{1}\times\idmap) \after \tuple{\idmap,\pi_{2}} \after \tuple{s,\idmap}
\\
& = &
\mu \after \Dst(\st_{1}) \after \st_{1} \after (\Dst(h)\times\idmap)) \after
   (\st_{1}\times\idmap) \after \tuple{\tuple{s,\idmap},\idmap}
\\
& = &
\st_{1} \after (\mu\times\idmap) \after (\Dst(h)\times\idmap)) \after 
   \tuple{\gr(s), \idmap}
\\
& = &
\gr(h \klafter \gr(s)).
\end{array}$$

\noindent Then:
$$\begin{array}{rcl}
\Dst(\pi_{2})(\Omega)
& = &
\big(\Dst(\pi_{2}) \after \gr(h)_{*} \after \gr(s)_{*}\big)(\omega)
\\
& = &
\big(\big(\Dst(\pi_{2}) \after \gr(h)\big)_{*} \after \gr(s)_{*}\big)(\omega)
\\
& \smash{\stackrel{\eqref{StrengthGraphDiag}}{=}} &
\big(\eta_{*} \after \gr(s)_{*}\big)(\omega)
\\
& = &
\gr(s)_{*}(\omega)
\\
\Dst(m\cdot\pi_{2})(\Omega)
& = &
\big(\Dst(m\cdot\pi_{2}) \after \gr(h)_{*} \after \gr(s)_{*}\big)(\omega)
\\
& = &
\gr(h \klafter \gr(s))_{*}(\omega) \qquad \mbox{as noted above}
\\
& = &
\gr(t)_{*}(\omega).
\end{array}$$

In the other direction, let $\Omega \in \Dst(m\cdot (n\cdot A))$ be
given with $\Dst(\pi_{2})(\Omega) = \gr(s)_{*}(\omega)$ and
$\Dst(m\cdot\pi_{2})(\Omega) = \gr(t)_{*}(\omega)$.
Since, by assumption $\supp\big(\gr(s)_{*}(\omega)\big) = n\cdot A$,
there is a unique function $h\colon n\cdot A \rightarrow \Dst(m)$
with $\Omega = \gr(h)_{*}(\gr(s)_{*}(\omega))$. We obtain $h \klafter
\gr(s) = t$ by uniqueness, from:
$$\begin{array}[b]{rcl}
\gr(h \klafter \gr(s))_{*}(\omega)
& = &
\big(\Dst(m\cdot \pi_{2}) \after \gr(h)_{*} \after \gr(s)\big)_{*}(\omega)
\\
& = &
\big(\Dst(m\cdot \pi_{2}) \after \big(\gr(h)_{*} \after \gr(s)\big)_{*}\big)
   (\omega)
\\
& = &
\big(\Dst(m\cdot \pi_{2}) \after \big(\gr(h) \klafter \gr(s)\big)_{*}\big)
   (\omega)
\\
& = &
\big(\Dst(m\cdot \pi_{2}) \after \gr(h)_{*} \after \gr(s)_{*}\big)
   (\omega)
\\
& = &
\Dst(m\cdot \pi_{2})(\Omega)
\\
& = &
\gr(t)_{*}(\omega)
\end{array}\eqno{\QEDbox}$$

\noindent We first prove $\mu\big(\Dst(\nabla)(\Omega)\big) =
\hypcond{\omega}{s}$ in:
$$\begin{array}{rcl}
\mu\big(\Dst(\nabla)(\Omega)\big)
& = &
\big(\mu \after \Dst(\nabla) \after \Nrm \after 
   \gr(h_{*}\after\st_{2})_{*}\big)(\hypcond{\omega}{s})
\\
& \smash{\stackrel{\eqref{NrmOutputDiag}}{=}} &
\big(\Dst(\pi_{2}) \after \gr(h_{*}\after\st_{2})_{*}\big)(\hypcond{\omega}{s})
\\
& = &
\big(\Dst(\pi_{2}) \after \gr(h_{*}\after\st_{2})\big)_{*}(\hypcond{\omega}{s})
\\
& \smash{\stackrel{\eqref{StrengthGraphDiag}}{=}} &
\eta_{*}(\hypcond{\omega}{s})
\\
& = &
\hypcond{\omega}{s}.
\end{array}$$

\noindent The equation $\Dst\big(m\cdot (\mu
\after\Dst(\nabla))\big)(\Omega) = \hypcond{\omega}{t}$ requires more
work. First, we write $(h,\pi_{2})$ for the map $\st_{1} \after
\tuple{h, \pi_{2}} \colon n\cdot A \rightarrow \Dst(m\cdot A)$. Then
the following diagram commutes.
$$\xymatrix{
\Dst(n\cdot A)\ar[d]_{(h,\pi_{2})_{*}}\ar[r]^-{\Nrm} &
   \Dst(n\cdot\Dst(A))\ar[rr]^-{(h_{*}\after\st_{2},\pi_{2})_{*}} & &
   \Dst(m\cdot\Dst(A))\ar[r]^-{\Nrm} & 
   \Dst(m\cdot\Dst^{2}(A))\ar[d]^{\Dst(m\cdot\mu)}
\\
\Dst(m\cdot A)\ar[rrrr]_-{\Nrm} & & & & \Dst(m\cdot\Dst(A))
}$$

\noindent In order to see that this diagram commute we first elaborate
what these maps of the form $(h, \pi_{2})$ do. First,
$$\begin{array}{rcccl}
(h, \pi_{2})(\kappa_{i}a)
& = &
\st_{2}(h(\kappa_{i}a), a)
& = &
\sum_{j} h(\kappa_{i}a)(j)\ket{\kappa_{j}a}.
\end{array}$$

\noindent Similarly, $(h_{*} \after \st_{2}, \pi_{2}) \colon n\cdot\Dst(A)
\rightarrow \Dst(m\cdot \Dst(A))$ is described by:
$$\begin{array}{rcl}
(h_{*} \after \st_{2}, \pi_{2})(\kappa_{i}\varphi)
& = &
\st_{1}\big(h_{*}(\sum_{a}\varphi(a)\ket{\kappa_{i}a}), \varphi\big)
\\
& = &
\st_{1}\big(\sum_{j}(\sum_{a}\varphi(a)\cdot h(\kappa_{i}a)(j))\ket{j}, 
   \varphi\big)
\\
& = &
\sum_{j}(\sum_{a}\varphi(a)\cdot h(\kappa_{i}a)(j))\ket{\kappa_{j}\varphi}.
\end{array}$$

\noindent Hence:
$$\begin{array}{rcl}
\lefteqn{\big(\Dst(m\cdot\mu) \after \Nrm \after 
   (h_{*} \after \st_{2}, \pi_{2})_{*} \after \Nrm\big)(\omega)}
\\
& = &
\big(\Dst(m\cdot\mu) \after \Nrm \after 
   (h_{*} \after \st_{2}, \pi_{2})_{*}\big)\Big(
   {\displaystyle\sum}_{i} \omega[i]\bigket{\kappa_{i}\big(
   \sum_{a} \frac{\omega(\kappa_{i}a)}{\omega[i]}\ket{a}\big)}\Big)
\\
& = &
\big(\Dst(m\cdot\mu) \after \Nrm\big)\Big(
   {\displaystyle\sum}_{j,i} \big(\omega[i]\cdot 
   (\sum_{a} \frac{\omega(\kappa_{i}a)}{\omega[i]}\cdot h(\kappa_{i}a)(j))\big)
   \bigket{\kappa_{j}\big(
   \sum_{a} \frac{\omega(\kappa_{i}a)}{\omega[i]}\ket{a}\big)}\Big)
\\
& = &
\big(\Dst(m\cdot\mu) \after \Nrm\big)\Big(
   {\displaystyle\sum}_{j,i} 
   \big(\sum_{a} \omega(\kappa_{i}a)\cdot h(\kappa_{i}a)(j))\big)
   \bigket{\kappa_{j}\big(
   \sum_{a} \frac{\omega(\kappa_{i}a)}{\omega[i]}\ket{a}\big)}\Big)
\\
& = &
\Dst(m\cdot\mu)\Big({\displaystyle\sum}_{j} 
   \big(\sum_{i,a} \omega(\kappa_{i}a)\cdot h(\kappa_{i}a)(j))\big)
   \bigket{\kappa_{j}\big(\sum_{i}
   \frac{\sum_{a} \omega(\kappa_{i}a)\cdot h(\kappa_{i}a)(j))}
        {\sum_{i,a} \omega(\kappa_{i}a)\cdot h(\kappa_{i}a)(j))}
   \bigket{\sum_{a} \frac{\omega(\kappa_{i}a)}{\omega[i]}\ket{a}}\big)}\Big)
\\
& = &
\\
& = &
{\displaystyle\sum}_{j} \big(\sum_{i,a} \omega(\kappa_{i}a) \cdot
   h(\kappa_{i}a)(j)\big)\bigket{\kappa_{j}\big(
   \sum_{a} \frac{\sum_{i} \omega(\kappa_{i}a) \cdot h(\kappa_{i}a)(j)}
   {\sum_{i,a} \omega(\kappa_{i}a) \cdot h(\kappa_{i}a)(j)}\ket{a}\big)}
\\
& = &
\Nrm\big({\displaystyle\sum}_{j,a} \big(\sum_{i} \omega(\kappa_{i}a) \cdot
   h(\kappa_{i}a)(j)\big)\bigket{\kappa_{j}a}\big)
\\
& = &
\big(\Nrm \after (h, \pi_{2})_{*}\big)(\omega).
\end{array}$$

We can then compute:
$$\begin{array}{rcl}
\lefteqn{\Dst\big(m\cdot (\mu \after\Dst(\nabla))\big)(\Omega)} \\
& = &
\big(\Dst(m\cdot\mu) \after \Dst(m\cdot\Dst(\nabla)) \after \Nrm \after
   \gr(h_{*}\after\st_{2})_{*} \after \Nrm \after \gr(s)_{*}\big)(\omega)
\\
& \smash{\stackrel{\eqref{NrmNatDiag}}{=}} &
\big(\Dst(m\cdot\mu) \after \Nrm \after \Dst(m\cdot\nabla) \after 
   \gr(h_{*}\after\st_{2})_{*} \after \Nrm \after \gr(s)_{*}\big)(\omega)
\\
& \smash{\stackrel{(*)}{=}} &
\big(\Dst(m\cdot\mu) \after \Nrm \after 
   (h_{*}\after\st_{2},\pi_{2})_{*} \after \Nrm \after \gr(s)_{*}\big)(\omega)
\\
& = &
\big(\Nrm \after (h,\pi_{2})_{*} \after \gr(s)_{*}\big)(\omega)
\\
& \smash{\stackrel{(**)}{=}} &
\big(\Nrm \after \gr(t)_{*}\big)(\omega)
\\
& = &
\hypcond{\omega}{t}.   
\end{array}$$

The equation $\smash{\stackrel{(*)}{=}}$ follows from:
$$\begin{array}{rcl}
\Dst(m\cdot\nabla) \after \gr(h_{*}\after\st_{2})
& = &
\Dst(\idmap\times\pi_{2}) \after \st_{1} \after 
   \tuple{h_{*} \after \st_{2}, \idmap}
\\
& = &
\st_{1} \after (\idmap\times\pi_{2}) \after \tuple{h_{*} \after \st_{2}, \idmap}
\\
& = &
\st_{1} \after \tuple{h_{*} \after \st_{2}, \pi_{2}}
\\
& = &
(h_{*}\after\st_{2},\pi_{2})
\end{array}$$

The second marked equation $\smash{\stackrel{(**)}{=}}$ is obtained via:
$$\begin{array}{rcl}
(h,\pi_{2})_{*} \after \gr(s)
& = &
\mu \after \Dst(\st_{1} \after \tuple{h, \pi_{2}}) \after \st_{1} \after
   \tuple{s,\idmap}
\\
& = &
\mu \after \Dst(\st_{1} \after (h\times\idmap) \after \tuple{\idmap, \pi_{2}}) 
   \after \st_{1} \after \tuple{s,\idmap}
\\
& = &
\mu \after \Dst(\st_{1} \after (h\times\idmap) \after \alpha \after 
   \idmap\times\Delta) \after \st_{1} \after \tuple{s,\idmap}
\\
& = &
\mu \after \Dst(\st_{1} \after (h\times\idmap)) \after \Dst(\alpha) \after 
   \st_{1} \after (\idmap\times\Delta) \after \tuple{s,\idmap}
\\
& = &
\mu \after \Dst(\st_{1} \after (h\times\idmap)) \after \st_{1} \after 
   (\st_{1}\times\idmap) \after \alpha \after \tuple{s,\Delta}
\\
& = &
\mu \after \Dst(\st_{1}) \after \st_{1} \after (\Dst(h)\times\idmap)) \after 
   (\st_{1}\times\idmap) \after \tuple{\tuple{s,\idmap},\idmap}
\\
& = &
\st_{1} \after (\mu\times\idmap) \after (\Dst(h)\times\idmap)) \after 
   \tuple{\gr(s), \idmap}
\\
& = &
\st_{1} \after \tuple{h \klafter \gr(s), \idmap}
\\
& = &
\st_{1} \after \tuple{t, \idmap}
\\
& = &
\gr(t).
\end{array}$$

In the other direction, let $\hypcond{\omega}{s} \sqsubseteq
\hypcond{\omega}{t}$ via $\Omega \in
\Dst\big(m\cdot\Dst(n\cdot\Dst(A))\big)$, so that
$\mu\big(\Dst(\nabla)\big)\big(\Omega\big) = \hypcond{\omega}{s}$ and
$\Dst\big(m\cdot (\mu \after\Dst(\nabla))\big)(\Omega) =
\hypcond{\omega}{t}$. We need to find a map $h\colon n\cdot A
\rightarrow \Dst(m)$ with $h \klafter \gr(s) = t$. Consider the
distribution $\Psi \in \Dst(m\cdot (n\cdot A))$ defined as:
$$\begin{array}{rcl}
\Psi
& = &
\big((\st_{2})_{*} \after \Dst(m\cdot (\st_{2})_{*})\big)(\Omega).
\end{array}$$

\noindent Then $\Dst(\pi_{2})(\Psi) = \gr(s)_{*}(\omega) \in
\Dst(n\cdot A)$ as computed in:
$$\begin{array}{rcl}
\Dst(\pi_{2})(\Psi)
& = &
\big(\Dst(\pi_{2}) \after (\st_{2})_{*} \after 
   \Dst(m\cdot (\st_{2})_{*})\big)(\Omega)
\\
& = &
\big(((\Dst(\pi_{2}) \after \st_{2})_{*} \after 
   \Dst(m\cdot (\st_{2})_{*})\big)(\Omega)
\\
& \smash{\stackrel{\eqref{StrengthGraphDiag}}{=}} &
\big((\pi_{2})_{*} \after \Dst(m\cdot (\st_{2})_{*})\big)(\Omega)
\\
& = &
\big(\mu \after \Dst(\pi_{2}) \after \Dst(m\cdot (\st_{2})_{*})\big)(\Omega)
\\
& = &
\big(\mu \after \Dst((\st_{2})_{*}) \after \Dst(\pi_{2})\big)(\Omega)
\\
& = &
\big(\mu \after \Dst(\mu \after \Dst(\st_{2})) \after
   \Dst(\nabla)\big)(\Omega)
\\
& = &
\big(\mu \after \mu \after \Dst^{2}(\st_{2}) \after
   \Dst(\nabla)\big)(\Omega)
\\
& = &
\big(\mu \after \Dst(\st_{2}) \after \mu \after 
   \Dst(\nabla)\big)(\Omega)
\\
& = &
(\st_{2})_{*}(\hypcond{\omega}{s})
\\
& \smash{\stackrel{\eqref{NrmMonicDiag}}{=}} &
\gr(s)_{*}(\omega).
\end{array}$$

\noindent Hence there is a unique map $h\colon n\cdot A \rightarrow \Dst(m)$
with $\Psi = \gr(h)_{*}(\gr(s)_{*}(\omega))$. We first note that:
$$\begin{array}{rcl}
\Dst(m\cdot \pi_{2}) \after \gr(h)_{*} \after \gr(s)
& = &
\big(\Dst(\idmap\times\pi_{2}) \after \st_{1} \after \tuple{h, \idmap}\big)_{*}
   \after \gr(s)
\\
& = &
\mu \after \Dst(\st_{1} \after \tuple{h, \pi_{2}}) \after \st_{1} \after
   \tuple{s, \idmap}
\\
& = &
\mu \after \Dst(\st_{1} \after (h\times\idmap)) \after \st_{1} \after
   (\st_{1}\times\idmap) \after \tuple{\idmap,\pi_{2}} \after \tuple{s,\idmap}
\\
& = &
\mu \after \Dst(\st_{1}) \after \st_{1} \after (\Dst(h)\times\idmap)) \after
   (\st_{1}\times\idmap) \after \tuple{\tuple{s,\idmap},\idmap}
\\
& = &
\st_{1} \after (\mu\times\idmap) \after (\Dst(h)\times\idmap)) \after 
   \tuple{\gr(s), \idmap}
\\
& = &
\gr(h \klafter \gr(s)).
\end{array}$$

\noindent Hence:
$$\begin{array}{rcl}
\gr(h \klafter \gr(s))_{*}(\omega)
& = &
\big(\Dst(m\cdot \pi_{2}) \after \gr(h)_{*} \after \gr(s)\big)_{*}(\omega)
\\
& = &
\big(\Dst(m\cdot \pi_{2}) \after \big(\gr(h)_{*} \after \gr(s)\big)_{*}\big)
   (\omega)
\\
& = &
\big(\Dst(m\cdot \pi_{2}) \after \big(\gr(h) \klafter \gr(s)\big)_{*}\big)
   (\omega)
\\
& = &
\big(\Dst(m\cdot \pi_{2}) \after \gr(h)_{*} \after \gr(s)_{*}\big)
   (\omega)
\\
& = &
\Dst(m\cdot \pi_{2})(\Psi)
\\
& = &
\big(\Dst(m\cdot\pi_{2}) \after (\st_{2})_{*} \after 
   \Dst(m\cdot (\st_{2})_{*})\big)(\Omega)
\\
& = &
\big((\Dst(\idmap\times\pi_{2}) \after \st_{2})_{*} \after 
   \Dst(m\cdot (\st_{2})_{*})\big)(\Omega)
\\
& = &
\big((\st_{2} \after (\idmap\times\Dst(\pi_{2})))_{*} \after 
   \Dst(m\cdot (\st_{2})_{*})\big)(\Omega)
\\
& = &
\big((\st_{2})_{*} \after \Dst(\idmap\times\Dst(\pi_{2})) \after 
   \Dst(m\cdot (\st_{2})_{*})\big)(\Omega)
\\
& = &
\big((\st_{2})_{*} \after 
   \Dst(m\cdot (\Dst(\pi_{2}) \after \st_{2})_{*})\big)(\Omega)
\\
& = &
\big((\st_{2})_{*} \after \Dst(m\cdot (\pi_{2})_{*})\big)(\Omega)
\\
& = &
(\st_{2})_{*}(\hypcond{\omega}{t})
\\
& \smash{\stackrel{\eqref{NrmMonicDiag}}{=}} &
\gr(t)_{*}(\omega)
\end{array}$$

\noindent Hence the bijective correspondence of
Proposition~\ref{PointNormProp} gives us the required equation
$h \klafter \gr(s) = t$. \QED
\end{myproof}

\end{document}

%% file: prooftree.tex
\newdimen\proofrulebreadth \proofrulebreadth=.05em
\newdimen\proofdotseparation \proofdotseparation=1.25ex
\newdimen\proofrulebaseline \proofrulebaseline=2ex
\newcount\proofdotnumber \proofdotnumber=3
\let\then\relax
\def\hfi{\hskip0pt plus.0001fil}
\mathchardef\squigto="3A3B
\newif\ifinsideprooftree\insideprooftreefalse
\newif\ifonleftofproofrule\onleftofproofrulefalse
\newif\ifproofdots\proofdotsfalse
\newif\ifdoubleproof\doubleprooffalse
\let\wereinproofbit\relax
\newdimen\shortenproofleft
\newdimen\shortenproofright
\newdimen\proofbelowshift
\newbox\proofabove
\newbox\proofbelow
\newbox\proofrulename
\def\shiftproofbelow{\let\next\relax\afterassignment\setshiftproofbelow\dimen0 }
\def\shiftproofbelowneg{\def\next{\multiply\dimen0 by-1 }%
\afterassignment\setshiftproofbelow\dimen0 }
\def\setshiftproofbelow{\next\proofbelowshift=\dimen0 }
\def\setproofrulebreadth{\proofrulebreadth}

\def\prooftree{
\ifnum  \lastpenalty=1
\then   \unpenalty
\else   \onleftofproofrulefalse
\fi
\ifonleftofproofrule
\else   \ifinsideprooftree
        \then   \hskip.5em plus1fil
        \fi
\fi
\bgroup
\setbox\proofbelow=\hbox{}\setbox\proofrulename=\hbox{}%
\let\justifies\proofover\let\leadsto\proofoverdots\let\Justifies\proofoverdbl
\let\using\proofusing\let\[\prooftree
\ifinsideprooftree\let\]\endprooftree\fi
\proofdotsfalse\doubleprooffalse
\let\thickness\setproofrulebreadth
\let\shiftright\shiftproofbelow \let\shift\shiftproofbelow
\let\shiftleft\shiftproofbelowneg
\let\ifwasinsideprooftree\ifinsideprooftree
\insideprooftreetrue
\setbox\proofabove=\hbox\bgroup$\displaystyle 
\let\wereinproofbit\prooftree
\shortenproofleft=0pt \shortenproofright=0pt \proofbelowshift=0pt
\onleftofproofruletrue\penalty1
}

\def\eproofbit{
\ifx    \wereinproofbit\prooftree
\then   \ifcase \lastpenalty
        \then   \shortenproofright=0pt  
        \or     \unpenalty\hfil         
        \or     \unpenalty\unskip       
        \else   \shortenproofright=0pt  
        \fi
\fi
\global\dimen0=\shortenproofleft
\global\dimen1=\shortenproofright
\global\dimen2=\proofrulebreadth
\global\dimen3=\proofbelowshift
\global\dimen4=\proofdotseparation
\global\count255=\proofdotnumber
$\egroup  
\shortenproofleft=\dimen0
\shortenproofright=\dimen1
\proofrulebreadth=\dimen2
\proofbelowshift=\dimen3
\proofdotseparation=\dimen4
\proofdotnumber=\count255
}

\def\proofover{
\eproofbit 
\setbox\proofbelow=\hbox\bgroup 
\let\wereinproofbit\proofover
$\displaystyle
}%
\def\proofoverdbl{
\eproofbit 
\doubleprooftrue
\setbox\proofbelow=\hbox\bgroup 
\let\wereinproofbit\proofoverdbl
$\displaystyle
}%
\def\proofoverdots{
\eproofbit 
\proofdotstrue
\setbox\proofbelow=\hbox\bgroup 
\let\wereinproofbit\proofoverdots
$\displaystyle
}%
\def\proofusing{
\eproofbit 
\setbox\proofrulename=\hbox\bgroup 
\let\wereinproofbit\proofusing
\kern0.3em$
}

\def\endprooftree{
\eproofbit 
  \dimen5 =0pt
\dimen0=\wd\proofabove \advance\dimen0-\shortenproofleft
\advance\dimen0-\shortenproofright
\dimen1=.5\dimen0 \advance\dimen1-.5\wd\proofbelow
\dimen4=\dimen1
\advance\dimen1\proofbelowshift \advance\dimen4-\proofbelowshift
\ifdim  \dimen1<0pt
\then   \advance\shortenproofleft\dimen1
        \advance\dimen0-\dimen1
        \dimen1=0pt
        \ifdim  \shortenproofleft<0pt
        \then   \setbox\proofabove=\hbox{%
                        \kern-\shortenproofleft\unhbox\proofabove}%
                \shortenproofleft=0pt
        \fi
\fi
\ifdim  \dimen4<0pt
\then   \advance\shortenproofright\dimen4
        \advance\dimen0-\dimen4
        \dimen4=0pt
\fi
\ifdim  \shortenproofright<\wd\proofrulename
\then   \shortenproofright=\wd\proofrulename
\fi
\dimen2=\shortenproofleft \advance\dimen2 by\dimen1
\dimen3=\shortenproofright\advance\dimen3 by\dimen4
\ifproofdots
\then
        \dimen6=\shortenproofleft \advance\dimen6 .5\dimen0
        \setbox1=\vbox to\proofdotseparation{\vss\hbox{$\cdot$}\vss}%
        \setbox0=\hbox{%
                \advance\dimen6-.5\wd1
                \kern\dimen6
                $\vcenter to\proofdotnumber\proofdotseparation
                        {\leaders\box1\vfill}$%
                \unhbox\proofrulename}%
\else   \dimen6=\fontdimen22\the\textfont2 
        \dimen7=\dimen6
        \advance\dimen6by.5\proofrulebreadth
        \advance\dimen7by-.5\proofrulebreadth
        \setbox0=\hbox{%
                \kern\shortenproofleft
                \ifdoubleproof
                \then   \hbox to\dimen0{%
                        $\mathsurround0pt\mathord=\mkern-6mu%
                        \cleaders\hbox{$\mkern-2mu=\mkern-2mu$}\hfill
                        \mkern-6mu\mathord=$}%
                \else   \vrule height\dimen6 depth-\dimen7 width\dimen0
                \fi
                \unhbox\proofrulename}%
        \ht0=\dimen6 \dp0=-\dimen7
\fi
\let\doll\relax
\ifwasinsideprooftree
\then   \let\VBOX\vbox
\else   \ifmmode\else$\let\doll=$\fi
        \let\VBOX\vcenter
\fi
\VBOX   {\baselineskip\proofrulebaseline \lineskip.2ex
        \expandafter\lineskiplimit\ifproofdots0ex\else-0.6ex\fi
        \hbox   spread\dimen5   {\hfi\unhbox\proofabove\hfi}%
        \hbox{\box0}%
        \hbox   {\kern\dimen2 \box\proofbelow}}\doll%
\global\dimen2=\dimen2
\global\dimen3=\dimen3
\egroup 
\ifonleftofproofrule
\then   \shortenproofleft=\dimen2
\fi
\shortenproofright=\dimen3
\onleftofproofrulefalse
\ifinsideprooftree
\then   \hskip.5em plus 1fil \penalty2
\fi
}
